\RequirePackage{fix-cm}
\documentclass[smallextended]{svjour3}       
\smartqed  
\usepackage{graphicx}
\usepackage{subfigure}
\usepackage{color}
\usepackage{natbib} 
\usepackage{bm}
\usepackage{epstopdf}
\usepackage{enumerate}
 \usepackage{mathptmx}      
\usepackage{amsfonts}
\usepackage{amsmath,amssymb,bm}
\usepackage{bm}
\usepackage{caption}
\usepackage{multirow}
\usepackage{tabularx} 

\newcommand{\argmin}{\mathop{\rm arg\,min}\limits}

\begin{document}
\title{Quantitative Evaluation of Predictability of Linear Reduced-order Model based on Particle-image-velocimetry Data of Separated Flow Field around Airfoil}
\titlerunning{Quantitative Evaluation of Linear ROM based on PIV Data}        

\author{Taku Nonomura, Koki Nankai, Yuto Iwasaki, Atsushi Komuro, and Keisuke Asai }

\authorrunning{T. Nonomura et al.} 

\institute{T. Nonomura \at
              Department of Aerospace Engineering, Tohoku University/ Institute of Fluid Science, Tohoku University\\
              Aoba 6-6-01, Aramaki, Aoba-ku, Sendai, Miyagi, Japan
              Tel.: +81-22-795-4075\\
              Fax: +81-22-678910\\
              \email{nonomura@aero.mech.tohoku.ac.jp}           
}

\date{Received: date / Accepted: date}

\maketitle
\begin{abstract}
A quantitative evaluation method for a reduced-order model of the flow field around an NACA0015 airfoil based on particle image velocimetry (PIV) data is proposed in the present paper. The velocity field data obtained by the time-resolved PIV measurement were decomposed into significant modes by a proper orthogonal decomposition (POD) technique, and a linear reduced-order model was then constructed by the linear regression of the time advancement of the POD modes or the sparsity promoting dynamic mode decomposition (DMD). The present evaluation method can be used for the evaluation of the estimation error and the model predictability. The model was constructed using different numbers of POD or DMD modes for order reduction of the fluid data and different methods of estimating the linear coefficients, and the effects of these conditions on the model performance were quantitatively evaluated. The results illustrates that forward (standard) model works the best with two to ten significant DMD modes selected by sparsity promoting DMD. 
\end{abstract}
\section{Introduction}
\label{sec:intro}
Flow separation control by active flow control devices has recently attracted a great deal of attention, and the control performance of such devices, represented by dielectric barrier discharge plasma actuators, has been investigated in many studies \citep{corke2007sdbd,little2010high,sekimoto2017burst,komuro2018multiple,sato2015multifactorial,sato2015mechanisms,aono2017plasma,sato2019mechanisms,sato2020unified}. It has been demonstrated that the control input (e.g., the burst-mode frequency) influences the effectiveness of the control and the input should be adapted to the flow configuration. In other words, for the effective control of an unsteady flow, such as a flow field on an airfoil, the control input should be determined based on the state of the flow field. Therefore, we are aiming to construct an optimal feedback flow control system and to determine the control input by taking the system output into consideration. The construction of the system requires an observer with a model which estimates the state of the flow field from the limited system outputs.

Recently, the reduced-order modeling has got attention for this purpose. Proper orthogonal decomposition (POD) introduced by \citet{lumley1967structure} is an essential and popular tool for the reduction of both the computational cost of the estimation and the noise of the data. POD is a well-developed technique for the order reduction of the data and used for extracting coherent structures of fluid flows as orthogonal modes called ``POD modes.'' Fluid data can be reconstructed with a minimum number of orthogonal basis of energetic POD modes, and thus, it is often used for reduced-order modeling of flow fields.  \citep{rowley2004model,semeraro2012analysis,cammilleri2013pod,suzuki2014pod,suzuki2020few}.

With regard to the modeling of the dynamics, the Galerkin projection using governing equations and the regression to the simplified model are often used, where the former is deductive and applicable to computation, and the latter is inductive and applicable to both computations and experiments. The popular method of the latter inductive model is the dynamic mode decomposition (DMD) proposed by \citet{schmid2010dynamic}, which approximately corresponds to the linear model. Here, DMD considers that the fluid dynamics is explained by a linear system and investigates the linear operator obtained by the experimental or numerical fluid data set. A lot of studies used a DMD analysis for investigations of various fluid flows, such as jets \citep{rowley2009spectral}, cavity flows \citep{seena2011dynamic} or flows around an airfoil \citep{pan2011dynamical}. Now, the combination of POD and the linear model reconstruction are expected to be used for the short-time flow field prediction towards the feedback control. It should be noted that several online implementations have been also proposed for the real-time applications \citep{zhang2019online,nonomura2018dynamic,nonomura2019extended,matsumoto2017on-the-fly}

In those approach, several methods have been proposed, not for the prediction, but for the more accurate reconstruction. The efficient selection methods of sparse DMD modes which represents the original data were proposed based on a compress sensing idea by \citet{jovanovic2014sparsity-promoting}. The simultaneous estimation of spatial modes, eigenvalues and initial values was proposed as optimized DMD by \citet{tu2014dynamic} and practically implemented by \citet{askham2018variable}. The optimized mode decomposition, the formulation of which naturally includes the low-rank constraint, was also proposed for the better choice of relevant modes by \citet{wynn2013optimal}. However, these methods have only been used for the model reconstruction and it has not been evaluated in terms of prediction of the flow fields.

In our previous study, we have constructed a linear reduced-order model which estimates the time advancement of low-dimensionalized flow fields around an NACA0015 airfoil based on particle image velocimetry (PIV) data and qualitatively investigated the predictability of the model \citep{nankai2019linear} (``reproductivity'' in that study). In this previous study, the low-dimensional description of the flow field data were obtained by the POD technique. In addition, the model described by a linear equation similar to DMD was straightforwardly adapted to the linear control theory. It has been shown that the linear reduced-order model predicts the original data near the initial condition. However, this evaluation was performed only by qualitative observation in the previous work.\citep{nankai2019linear}

In the present study, the estimation error is focused on and a quantitative evaluation method of the predictability is proposed. The wind tunnel experiments of the improved quality were newly conducted for the evaluation of the model. The two different order-reduction methods, i.e. POD-based and sparsity-promoting-DMD(DMDsp)-based linear reduced-order models are considered, where the former is the same method as that used by \citet{nankai2019linear} and the latter is newly considered in the present study. The effects of parameters, i.e. the number of POD modes used in the model and the method for computing the coefficient matrix of the linear state-space model equation on the predictability  were investigated for two different order-reduction methods (POD-based or DMDsp-based) based on the cross-validation using this evaluation index. Here, the number of POD mode is an important parameter of the model and this point is discussed in detail. Finally, the performances of two different order-reduction methods (POD-based or DMDsp-based) are addressed. This study shows the guideline for choices of parameters and methodology when building the linear reduced-order model based on experimental data. 

\section{Linear Reduced-order Model}
\label{sec:LROM}
\begin{figure}[h]
\captionsetup{justification=raggedright}
    \centering
    \includegraphics[width=80mm]{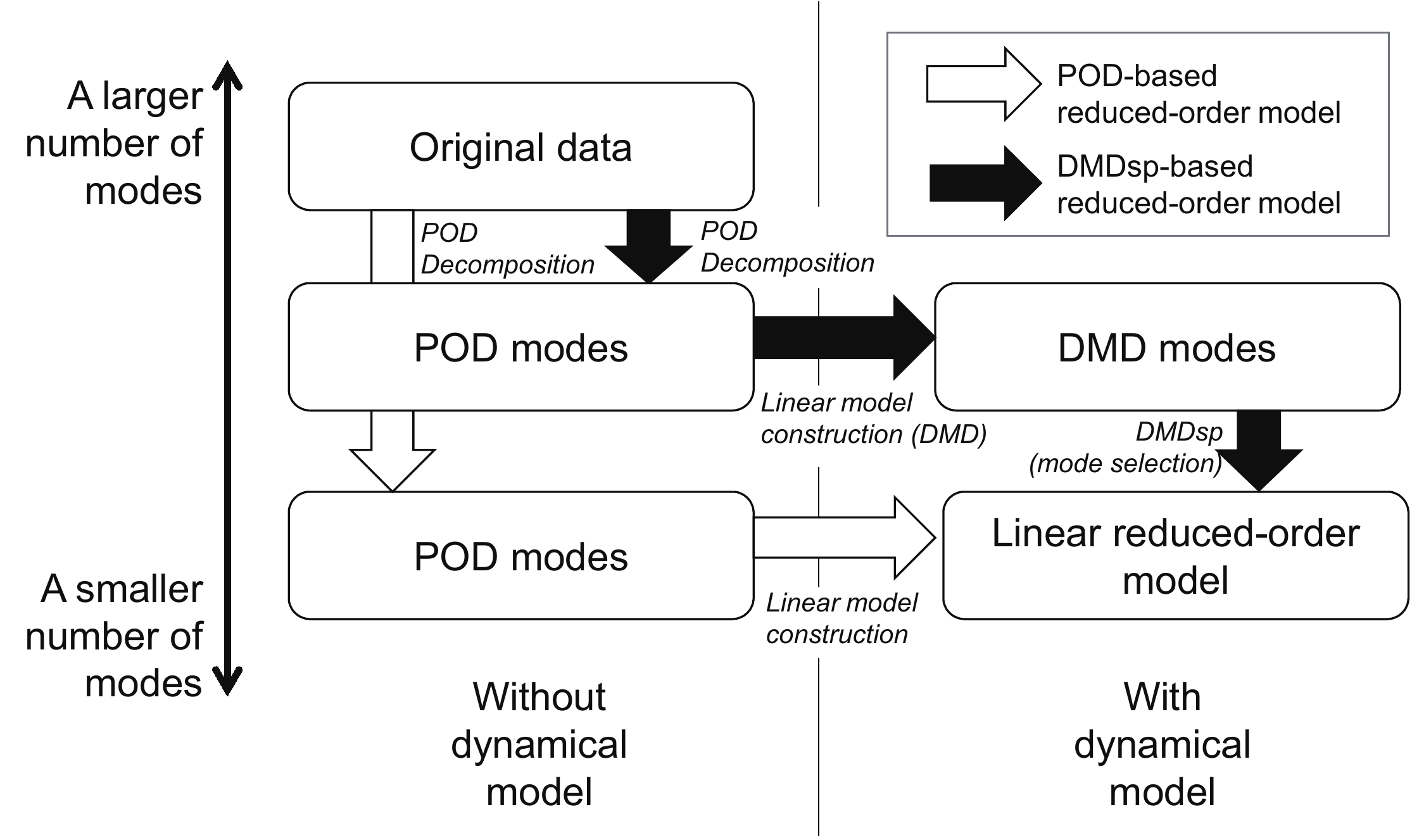}
    \caption{Schematic of processes of two different reduced-order modeling investigated in the present study.}
    \label{fig:Model}
\end{figure}

In the present paper, two kinds of order-reduction methods are considered for a linear low-dimensional model. One is an order-reduction method based on only POD. In this case, a linear model is constructed for the remaining temporal coefficients of POD modes. Any further order reductions are not conducted in the model construction. The other is an order-reduction method based on POD and sparsity promoting DMD (DMDsp). First, the linear model based on DMD is constructed with a certain number of POD modes, and the active modes are selected by using DMDsp. Figure~\ref{fig:Model} illustrates the schematic of two kinds of order-reduction methods. 

\subsection{POD-based reduced-order models}\label{sec:LROM1}
Here, the POD-based reduced-order models are described which was previously shown in the white arrow process in Fig.~\ref{fig:Model}. The construction of the linear reduced-order model starts with the derivation of a low-dimensional description of the velocity field data acquired by the time-resolved PIV measurement. The fluctuation of velocity fields $\mathbf{u}(\mathbf{x}, t)$ are decomposed into a set of temporal and spatial modes $z(t)$ and $\Phi(\mathbf{x})$ by POD:
\begin{equation}
\label{eq:POD}
    \mathbf{u}(\mathbf{x}, t)=\sum_{i} z_{i}(t) \Phi_i(\mathbf{x}),
\end{equation}
where the temporal coefficient of the $i$th POD modes $z_{i}(t)$ contains the POD-mode energy $\sigma_{i}^{2}$ corresponding to the amount of information the $i$th POD mode holds. It should be noted that the temporal coefficients $z$ of POD modes are not normalized, but the POD spatial modes $\Phi$ are normalized. The low-rank approximation of the velocity field data is acquired from the first $r_\textrm{POD}$ energetic POD modes:
\begin{equation}
\label{eq:POD_low}
    \mathbf{u}_{\mathrm{low}}(\mathbf{x}, t)=\sum_{i=1}^{r_{\textrm{POD}}} z_{i}(t) \Phi_i(\mathbf{x}).
\end{equation}

The time advancement of the POD-mode coefficients is approximated using a discrete linear state-space system based on the concept of DMD:
\begin{equation}
\label{eq:model}
    \mathbf{z}(t_{n+1})=\mathbf{A} \mathbf{z}(t_{n}),
\end{equation}
where $\mathbf{z}(t_{n})$ is the state variable vector that contains  $r_\textrm{POD}$ POD-mode coefficients at the $n$th time step. The state-space model is constructed by computing the coefficient matrix $\mathbf{A}$ from the training data set where it is called a coefficient matrix according to the modern control theory \citep{williams2007linear}.  {The coefficient matrix $\mathbf{A}$ can also be viewed as a Koopman matrix \citep{rowley2009spectral} or a DMD operator in the POD subspace \citep{schmid2010dynamic}, and also called a POD spectral operator by \citet{cammilleri2013pod} when applied it to the POD subspace.}

The temporal estimation of each POD-mode coefficient by the linear state-space model at an arbitrary time step is performed recursively based on the original data at the first time step $n_{0}$. {The estimated temporal coefficients $\hat{\mathbf{z}}$ of POD modes  at the $n$th time step are calculated as:
\begin{equation}
\label{eq:model_estimation}
    \hat{\mathbf{z}}(t_{n})=\mathbf{A}^{n-n_{0}} \mathbf{z}(t_{n_{0}}).
\end{equation}}

In the construction of the present model, there are some parameters that are considered to affect the model performance, such as a number $r_{\textrm{POD}}$ of POD modes and the coefficient matrix $\mathbf{A}$. It is also called a DMD matrix in the following discussion of DMDsp. With regard to the coefficient matrix, the standard model is constructed by computing it using the least-squares method corresponding to the exact DMD method by \citet{tu2013dynamic} as follows:
\begin{equation}
\label{eq:exactDMD}
    \mathbf{A}=\mathbf{Z}_{2:N} \mathbf{Z}_{1:N-1}^{+},
\end{equation}
where the matrix $\mathbf{Z}_{i:j}$ contains the POD-mode coefficients $\mathbf{z}(t_{n})$ at $i$-to-$j$th time steps, and the $\mathbf{Z}_{1:N-1}^{+}$ is the Moore-Penrose pseudo inverse matrix of $\mathbf{Z}_{1:N-1}$. Here, $N$ is a number of training data. This method is hereafter referred to as the ``forward (standard) method.'' In the present paper, three additional noise-robust methods were employed for the calculation of the coefficient matrix in addition to the forward (standard) method. 

The first one is the forward-backward method proposed by \citet{dawson2016characterizing}. This method considers the following forward and backward dynamical systems:
\begin{equation}
\label{eq:forward}
    \mathbf{z}(t_{n})\approx \mathbf{A}_{\mathrm{f}} \mathbf{z}(t_{n-1}),
\end{equation}
\begin{equation}
\label{eq:backward}
    \mathbf{z}(t_{n-1})\approx \mathbf{A}_{\mathrm{b}} \mathbf{z}(t_{n}).
\end{equation}
The two matrices $\mathbf{A}_{\mathrm{f}}$ and $\mathbf{A}_{\mathrm{b}}$ are computed by the least square method. Note that $\mathbf{A}_{\mathrm{f}}$ corresponds to the standard coefficient matrix $\mathbf{A}$; that is, $\mathbf{A}=\mathbf{A}_{\mathrm{f}}$.  If these matrices are computed from a linear dynamical system, the forward propagator matrix should be the inverse of the backward matrix. In reality, they have the same type of eigenvalue bias and are only approximate inverses. \citet{dawson2016characterizing} have shown that the corresponding de-biased matrix can be estimated by combining them as
\begin{equation}
\label{eq:A_forward-backward}
    \mathbf{A}_{\mathrm{fb}}=\left(\mathbf{A}_{\mathrm{f}}\left(\mathbf{A}_{\mathrm{b}}\right)^{-1}\right)^{0.5},
\end{equation}
in the DMD framework.

The second method is the total least-squares (total-least-squares) method developed for DMD by \citet{hemati2017de-biasing}. The forward (standard) least squares method minimizes the error concerning time-shifted data $\mathbf{Z}_{2:N}$; i.e., it does not assume noise on $\mathbf{Z}_{1:N-1}$. On the other hand, the total-least-squares method assumes noise on both matrices. The new coefficient matrix $\mathbf{A}_{\mathrm{tls}}$ is computed by performing a linear fitting in which the Frobenius norms of the errors on $\mathbf{Z}_{1:N-1}$ and $\mathbf{Z}_{2:N}$ are minimized, namely solving the following problem:
\begin{equation}
\label{eq:A_total-least-squares}
    \min _{\mathbf{A}_{\mathrm{tls}}, \Delta \mathbf{Z}_{2:N}, \Delta \mathbf{Z}_{1:N-1}}\left\|\left[\begin{array}{c}
{\Delta \mathbf{Z}_{1:N-1}} \\
{\Delta \mathbf{Z}_{2:N}}
\end{array}\right]\right\|_{\mathrm{F}}, \text { subject to }\left(\mathbf{Z}_{2:N}+\Delta \mathbf{Z}_{2:N}\right)=\mathbf{A}_{\mathrm{tls}}\left(\mathbf{Z}_{1:N-1}+\Delta \mathbf{Z}_{1: N-1}\right),
\end{equation}
where $\Delta$ is the error component of each data matrix.

These two DMD-based methods have been shown to be effective for de-biasing the eigenvalues of the propagator matrix against the effects of the observation noise. In addition to these methods, the coefficient matrix was also computed based on the approach taken by \citet{perret2006polynomial}. The following ordinary differential equation (ODE) was assumed:
\begin{equation}
\label{eq:ODE}
    \mathbf{\dot{z}}(t)=\mathbf{D} \mathbf{z}(t),
\end{equation}
where $\mathbf{D}$ is the coefficient matrix of the linear term. A second-order finite difference scheme was adopted, and the time derivatives were estimated with the approach by \citet{perret2006polynomial}:
\begin{equation}
\label{eq:second-order_zdot}
    \mathbf{\dot{z}}(t+\Delta t / 2) = \frac{\mathbf{z}(t+\Delta t)-\mathbf{z}(t)}{\Delta t} +\mathcal{O}(\Delta t^2).
\end{equation}
In addition, the POD-mode coefficients are modified and both amplitude of the POD-mode coefficients and their time derivatives are simultaneously evaluated, as
\begin{equation}
\label{eq:second-order_z}
    \mathbf{z}(t+\Delta t / 2) = \frac{\mathbf{z}(t+\Delta t)+\mathbf{z}(t)}{2} + \mathcal{O}(\Delta t^2).
\end{equation}
Equation \ref{eq:ODE} can then be modified as
\begin{eqnarray}
    \mathbf{\dot{z}}(t+\Delta t / 2)&=&\mathbf{D} \mathbf{z}(t+\Delta t / 2) \nonumber\\
    \frac{\mathbf{z}(t+\Delta t)-\mathbf{z}(t)}{\Delta t}&=&\mathbf{D} \frac{\mathbf{z}(t+\Delta t)+\mathbf{z}(t)}{2} + \mathcal{O}(\Delta t^2),\\
     \label{eq:ODE_mod2}
    \mathbf{z}(t+\Delta t)-\mathbf{z}(t)& \approx & \mathbf{D} \Delta t \frac{\mathbf{z}(t+\Delta t)+\mathbf{z}(t)}{2},\\
    \delta \mathbf{z}(t + \Delta t / 2) &=& \mathbf{D} \Delta t \tilde{\mathbf{z}}(t+\Delta t / 2),
\end{eqnarray}
where the second-order average and difference of $\mathbf{z}(t+\Delta t)$ are defined to be $\tilde{\mathbf{z}}(t+\Delta t / 2)$ and $\delta \mathbf{z}(t+\Delta t / 2)$, respectively. The matrix $\mathbf{D} \Delta t$ is computed by the least squares method:
\begin{equation}
\delta \mathbf{Z}=\mathbf{D} \Delta t \tilde{\mathbf{Z}},
\end{equation}
where
\begin{eqnarray}
        \tilde{\mathbf{Z}}&=&\left[\tilde{\mathbf{z}}\left(t_{1}+\Delta t / 2\right) \quad \tilde{\mathbf{z}}\left(t_{2}+\Delta t / 2\right) \quad \cdots \quad \tilde{\mathbf{z}}\left(t_{N-1}+\Delta t / 2\right)\right],\\
        \delta \mathbf{Z}&=&\left[\delta \mathbf{z}\left(t_{1}+\Delta t / 2\right) \quad \delta \mathbf{z}\left(t_{2}+\Delta t / 2\right) \quad \cdots \quad \delta \mathbf{z}\left(t_{N-1}+\Delta t / 2\right)\right].
\end{eqnarray}
The following equation is obtained by integrating Eq.~\ref{eq:ODE}:
\begin{equation}
\label{eq:ODE_integrated}
    \mathbf{z}(t)=e^{\mathbf{D} t} \mathbf{\zeta},
\end{equation}
where $\mathbf{\zeta}$ is a constant and $e$ is the base of natural logarithm. Equation \ref{eq:ODE_integrated} indicates that the history of the temporal coefficients of the POD modes can be acquired as follows:
\begin{equation}
\label{eq:ODE_estimate1}
    \begin{aligned}
\mathbf{z}(t+\Delta t) &=e^{\mathbf{D}(t+\Delta t)} \mathbf{\zeta} \\
&=e^{\mathbf{D} \Delta t} e^{\mathbf{D} t} \mathbf{\zeta} \\
&=e^{\mathbf{D} \Delta t} \mathbf{z}(t),
    \end{aligned}
\end{equation}
\begin{equation}
\label{eq:ODE_estimate2}
    \Rightarrow \mathbf{z}\left(t_{1}+n \Delta t\right)=\left(e^{\mathbf{D} \Delta t}\right)^{n} \mathbf{z}\left(t_{1}\right).
\end{equation}
Therefore, the new coefficient matrix based on the ODE (ODE-based method) corresponds to the time evolution operator in Eq.~\ref{eq:ODE_estimate2}:
\begin{equation}
    \mathbf{A}_{\mathrm{ode}}=e^{\mathbf{D} \Delta t}.
\end{equation}

\subsection{DMDsp-based reduced-order model}\label{sec:LROMDMDsp}
Here, the DMDsp-based reduced-order models are described which was previously shown in the black arrow processes in Fig.~\ref{fig:Model}. The model reduction is also conducted after linear state-space model construction in these models. 

The POD low-dimensionalization with $r_\textrm{POD}$ of a large number is firstly conducted similar to the previous POD-based reduced-order models. In this study, $r_\textrm{POD}=N/10$ is selected for saving the computational costs. After that, linear state-space models are constructed based on the methods previously introduced. Then, the eigenvalue decomposition of the coefficient matrix $\mathbf{A}$ is conducted, and the eigenvalues $\lambda_{i}$ and the DMD modes ${\bm{\Psi}}_i$ in the POD space are obtained. After that, the most relevant DMD modes in the POD space are selected by DMDsp for the reconstruction of the test data. 

The objective function of DMDsp is as follows:
\begin{equation}
\label{eq:DMDsp}
\argmin \|\mathbf{Z}-\bm{\Psi}\mathbf{D_\alpha}\mathbf{V_\textrm{and}}\|^2_2 + \gamma \|\bm{\alpha}(t_{0})\|_1,
\end{equation}
where 
\begin{eqnarray}
\bm{\Psi}&=&[\begin{array}{cccc} \bm{\Psi}_1 & \bm{\Psi_2}& \dots & \bm{\Psi}_{r_{\rm{POD}}}\end{array}], \\
\mathbf{D_\alpha}(t_0)&=&\textrm{diag}[\begin{array}{cccc} \alpha_1(t_0) & \alpha_2(t_0)& \dots & \alpha_{r_{\rm{POD}}}(t_0)\end{array}] , \\
\bm{\alpha}(t_0)&=&[\begin{array}{cccc} \alpha_1(t_0) & \alpha_2(t_0)& \dots & \alpha_{r_{\rm{POD}}}(t_0)\end{array}]^{\mathsf{T}},\\
\mathbf{V_{\textrm{and}}} &=& 
\left[ \begin{array}{cccc}
1 & \lambda_1 & \cdots & \lambda_1^{m-1} \\
1 & \lambda_2 & \cdots & \lambda_2^{m-1} \\
\vdots & \vdots & \ddots & \vdots \\
1 & \lambda_{r_{\rm{POD}}} & \cdots & \lambda_{r_{\rm{POD}}}^{m-1} 
\end{array} \right].
\end{eqnarray}
Here, $\bm{\alpha}$ is the time-varying mode amplitude, $\mathbf{V_{\textrm{and}}}$ is the Vandermonde matrix, and $\gamma$ is the weight for the sparsity promoting term (the second term of Eq.~\ref{eq:DMDsp}). The time-varying mode amplitudes $\bm{\alpha}$ are not normalized, but the DMD modes ${\bm{\Psi}}$ are normalized in the present study, similar to POD modes. The subscript $i$ denotes the quantity related with $i$th DMD mode and $\alpha_i(t_0)$  represents the initial values of the DMD mode amplitude while they are complex values. This implicates that the DMDsp gives us the suboptimal sparse initial values for the reconstruction of the data in the framework of the initial value problem of the linear system. 

Because the objective function of DMDsp is designed for the reconstruction of the data, its use for the prediction model is not straightforward. Therefore, the activated mode in the DMDsp-based reconstruction are only used for the model prediction in the present study. Although the different definition of a sparsity promoting term might works better for the mode selection of the prediction than the straightforward implementation, it seems to require careful discussions and detailed algorithm developments. Therefore, those issues are left for the future study.

In the present study, DMDsp is applied to test data $\mathbf{Z}$ and the mode with nonzero components of the initial values are only activated and then used for the model prediction. Given the DMD modes with nonzero initial values in the test data $\bm{\Psi}_{\textrm{sp}}=[\begin{array}{cccc}\psi_{\textrm{sp},1}&\psi_{\textrm{sp},2}&\cdots&\psi_{\textrm{sp},r_{\textrm{DMDsp}}}\end{array}]$, the pseudo-inverse operation gives us the estimated DMD amplitude from the full observation as follows:
\begin{equation}
\bm{\alpha}_{\textrm{sp}}(t_{n_0})=\bm{\Psi}_{\textrm{sp}}^{+} \mathbf{z} (t_{n_0}),
\end{equation}
where $\bm{\alpha}_\textrm{sp}=[\begin{array}{cccc} \alpha_{\textrm{sp}1}(t_0) & \alpha_{\textrm{sp}2}(t_0)& \dots & \alpha_{r_{\textrm{DMDsp}}}(t_0)\end{array}]^{\mathsf{T}}$ is the amplitude of the DMD modes selected by DMDsp. Once the estimated mode amplitude is obtained from the full observation, the time advancement of the DMD mode is given as follows: 
\begin{equation}
\bm{\alpha}_{\textrm{sp}}(t_n) = \bm{\Lambda}^{n-{n_0}}_{\textrm{sp}} \bm{\alpha}_\textrm{sp} (t_{n_0}), 
\end{equation}
where $\bm{\Lambda}_\textrm{sp}=\textrm{diag}[\begin{array}{cccc} \lambda_{\textrm{sp}1} & \lambda_{\textrm{sp}2}& \dots & \lambda_{r_{\textrm{DMDsp}}}\end{array}]$ are the DMD eigenvalues of the DMD modes selected by DMDsp. With using these amplitudes, the POD modes and low-dimensional flow fields can be approximated as follows:
\begin{eqnarray}
\mathbf{z}_{\textrm{sp}}(t) &=& \bm{\Psi}_{\textrm{sp}} \bm{\alpha}_{\textrm{sp}}(t),\\
u(\mathbf{x},t) &\approx& \sum_{i=1}^{r_\textrm{POD}} {z_{\textrm{sp}i}} \Phi_i(t),
\end{eqnarray}
where $\mathbf{z}_\textrm{sp}$ is the converted POD modes from the DMD modes selected by DMDsp, and $z_{\textrm{sp}i}$ is the $i$th component of $\mathbf{z}_\textrm{sp}$.

{Finally, Table \ref{tab:models} summarize the methods introduced in the previous and present subsections. Here, the DMDsp-based reduced-order models are only constructed with the forward (standard) and ODE-based methods, as discussed later.}
\begin{table}[ht]
\centering
\caption{Summary of the models and methods evaluated in the present study. The models are constructed in the order of steps 1 to 3.}
\label{tab:models}
\begin{tabular}{lllll}
\hline
            &           & step 1 :        & step 2 :              & step 3 : \\
reduced-order &method    & Mode       & $\mathbf{A}$ matrix                     & Mode       \\
model       &            & reduction  & construction  & reduction  \\ \hline
\multirow{4}{*}{POD-based} & forward(standard)   & \multirow{4}{*}{POD} & $\mathbf{A}_f$ & \multirow{4}{*}{Not conducted} \\
                        & forward-backward    &  & $(\mathbf{A}_f\mathbf{A}_b^{-1})^{1/2}$ & \\
                        & total-least-squares &  &  $    \argmin_{\mathbf{A}_{\mathrm{tls}}}\left\|\left[\begin{array}{c}
{\Delta \mathbf{Z}_{1:N-1}} \\
{\Delta \mathbf{Z}_{2:N}}
\end{array}\right]\right\|_{\mathrm{F}}
                                                                           $&  \\
                        & ODE-based           &  & $e^{\mathbf{D}t}$ &  \\\hline
\multirow{2}{*}{DMDsp-based} &forward(standard)  & \multirow{2}{*}{POD} & $\mathbf{A}_f$ & \multirow{2}{*}{DMDsp} \\
                             &ODE-based          &  & $e^{\mathbf{D}t}$ &  \\ \hline
\end{tabular}
\end{table}

\section{Quantitative evaluation of the model performance}
\label{sec:evaluationmethod}
The estimation error of the model is focused on for the quantitative evaluation of the model predictability. The new evaluation method enables us to the specification of the best set of parameters for the construction of the model ($r$ and $\mathbf{A}$, as described in Sec. \ref{sec:LROM}) for the highest predictability. The estimation results of time histories of the POD-mode coefficients were obtained by the present model as described in Eq.~\ref{eq:model_estimation}. 
Although the unified evaluation formulation can be obtained for the POD-based and DMDsp-based reduced-order modelings, each evaluation formulation is introduced for simplicity. 

\begin{figure}[h]
\captionsetup{justification=raggedright}
    \centering
    \includegraphics[width=80mm]{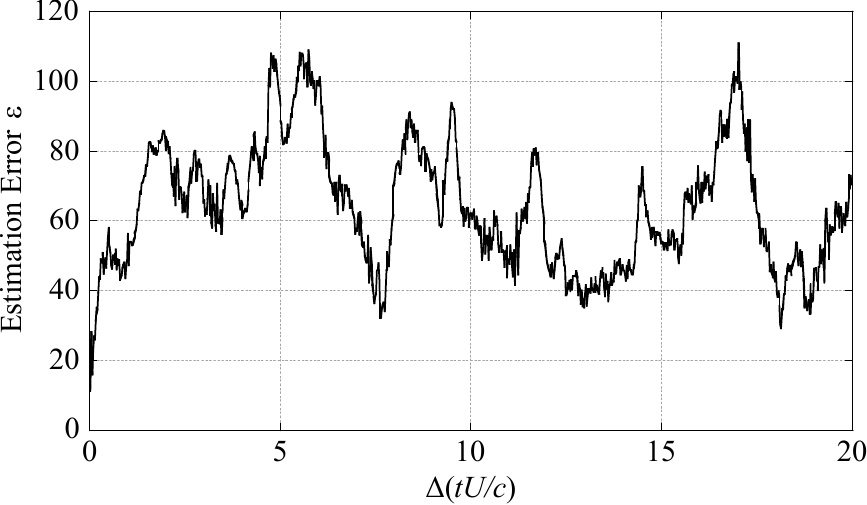}
    \caption{Schematic of the estimation error defined by Eq.~\ref{eq:error_allmodes_POD-based}.}
    \label{fig:estimation_error_schematic}
\end{figure}

\begin{figure}[h]
\captionsetup{justification=raggedright}
    \centering
    \includegraphics[width=110mm]{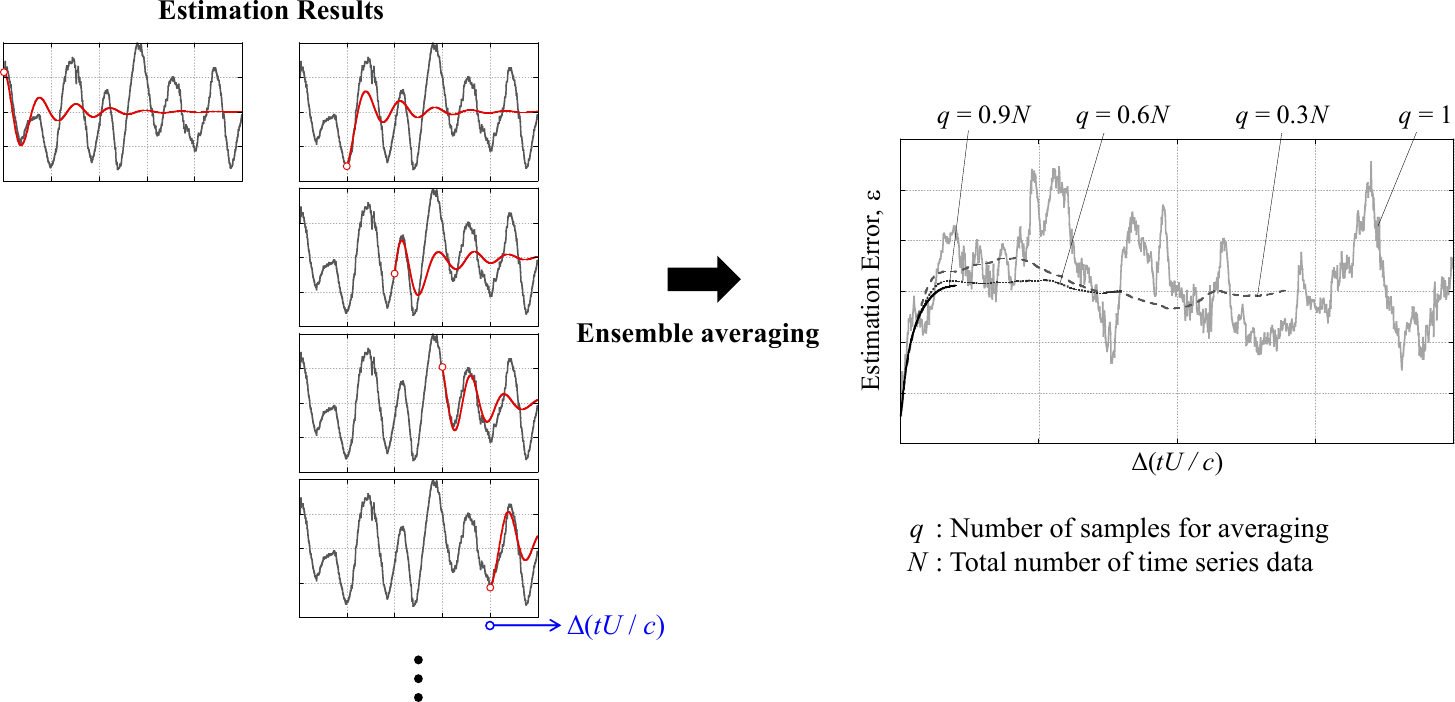}
    \caption{Schematic of the ensemble averaging procedure for the estimation error investigation.}
    \label{fig:error_averaging_schematic}
\end{figure}

First, the error estimation for the POD-based reduced-order model is introduced. The model prediction error vector $\bm{\epsilon}$ between the original and estimated temporal coefficient vectors of POD modes is calculated at each time step as
\begin{equation}
\label{eq:error}
    \bm{\epsilon}(t_{n})=\mathbf{z}(t_{n})-\hat{\mathbf{z}}(t_{n}).
\end{equation}
The instantaneous error $\epsilon$ can be defined as the root sum of squares of each component of $\bm{\epsilon}$ because of the normality of POD modes:
\begin{equation}
\label{eq:error_allmodes_POD-based}
    \epsilon(t_{n})=\| \bm{\epsilon}(t_{n}) \|_2.
\end{equation}
The temporal evolution of the error described by Eq.~\ref{eq:error_allmodes_POD-based} can then be plotted in a graph, as shown in Fig.~\ref{fig:estimation_error_schematic}. The vertical axis represents the estimation error ${\epsilon}$, and the horizontal axis represents the nondimensionalized time from the initial time step, which is the time step at which the original POD-mode coefficient is given to the model. However, in fact, the instantaneous error varies over a wide range as shown in Fig.~\ref{fig:estimation_error_schematic}, and is difficult to investigate accurately. Therefore, the ensemble average of the estimation results was taken, and a smooth curve of the temporal evolution of the error was produced. Equation \ref{eq:model_estimation} shows that multiple estimation results can be obtained by changing the time step for the initial value $n_{0}$. Accordingly, estimation error curves were generated as many as possible in the range of the test data and the ensemble average of these curves was taken for the smooth estimation error curve $\bar{\epsilon}(t_{n})$ (shown in Fig.~\ref{fig:error_averaging_schematic}).

\begin{figure}[h]
\captionsetup{justification=raggedright}
    \centering
    \includegraphics[width=80mm]{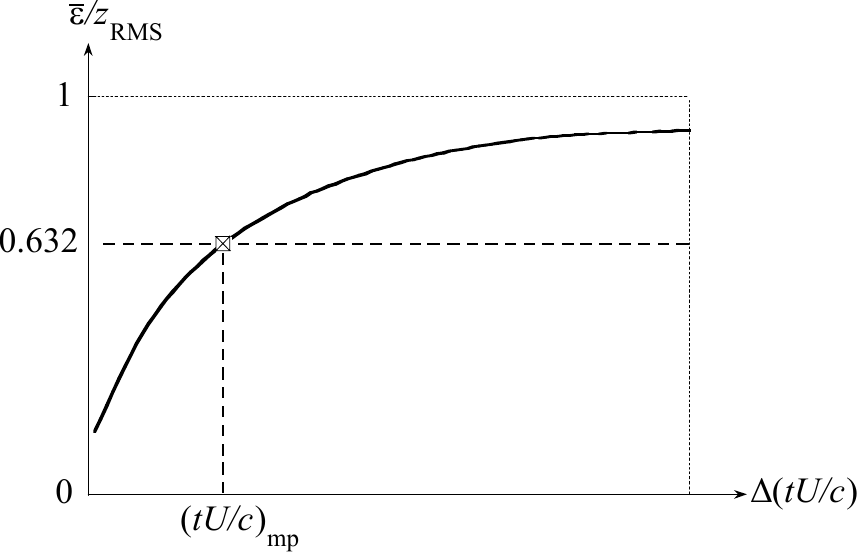}
    \caption{Schematic of the derivation of the model predictability.}
    \label{fig:evaluate_schematic}
\end{figure}

The model predictability was quantitatively evaluated based on the estimation error curve. First, the forward (standard) model was considered. The estimation error at after sufficiently long time was expected to be determined by the original POD-mode coefficients because the estimated temporal coefficients of POD modes finally approach zero as shown in the previous work \citep{nankai2019linear} and the error becomes equal to the deviation of the original mode coefficients around zero:
\begin{equation}
\label{eq:error_convergence}
    \begin{aligned}
\lim _{n \rightarrow \infty} {\epsilon}(t_{n}) &=\lim _{n \rightarrow \infty} \|{\bm{\epsilon}(t_{n})}\|_2 \\
&=\lim _{n \rightarrow \infty} \|{\mathbf{z}(t_{n}) - \hat{\mathbf{z}}(t_{n})}\|_2 \\
&=\lim _{n \rightarrow \infty} \|{\mathbf{z}(t_{n}) - \mathbf{0}}\|_2 \\
&=\lim _{n \rightarrow \infty} \|{\mathbf{z}(t_{n})}\|_2.
    \end{aligned}
\end{equation}
Therefore, the asymptotic value of the error was defined by the root mean square (RMS) of the original POD-mode coefficients $z_{\mathrm{RMS}}$, as given by
\begin{equation}
\label{eq:zrms}
    z_{\mathrm{RMS}}=\sqrt{\frac{\sum_{n=1}^{N} \| \mathbf{z}(t_{n}) \|^2_2}{N}} .
\end{equation}
The error for the model prediction seems to behave similar to a first-order lag system \citep{love2007first} as $\bar{\epsilon}(t)/z_{\mathrm{RMS}} \approx (1-e^{-t/\tau})$ in the forward (standard) model, where $\tau$ is the time constant for the first-order lag system. Therefore, the ``model prediction permissive time range'' $t_{\mathrm{mp}}$ is defined to be the time corresponding to $t_{\mathrm{mp}}=\tau$ at which the error reaches $(1-e^{-1}) \approx 0.632$ of $z_{\mathrm{RMS}}$, i.e., $\bar{\epsilon}({t_{\mathrm{mp}}}) \approx 0.632z_{\mathrm{RMS}}$, under the assumption of a first-order lag system for simplicity, as illustrated in Fig.~\ref{fig:evaluate_schematic}.  The model predictability is nondimensionalized as $(tU/c)_{\mathrm{mp}}$, where $U$ and $c$ are the freestream velocity and the chord length. 

\begin{figure}[h]
\captionsetup{justification=raggedright}
    \centering
    \includegraphics[width=120mm]{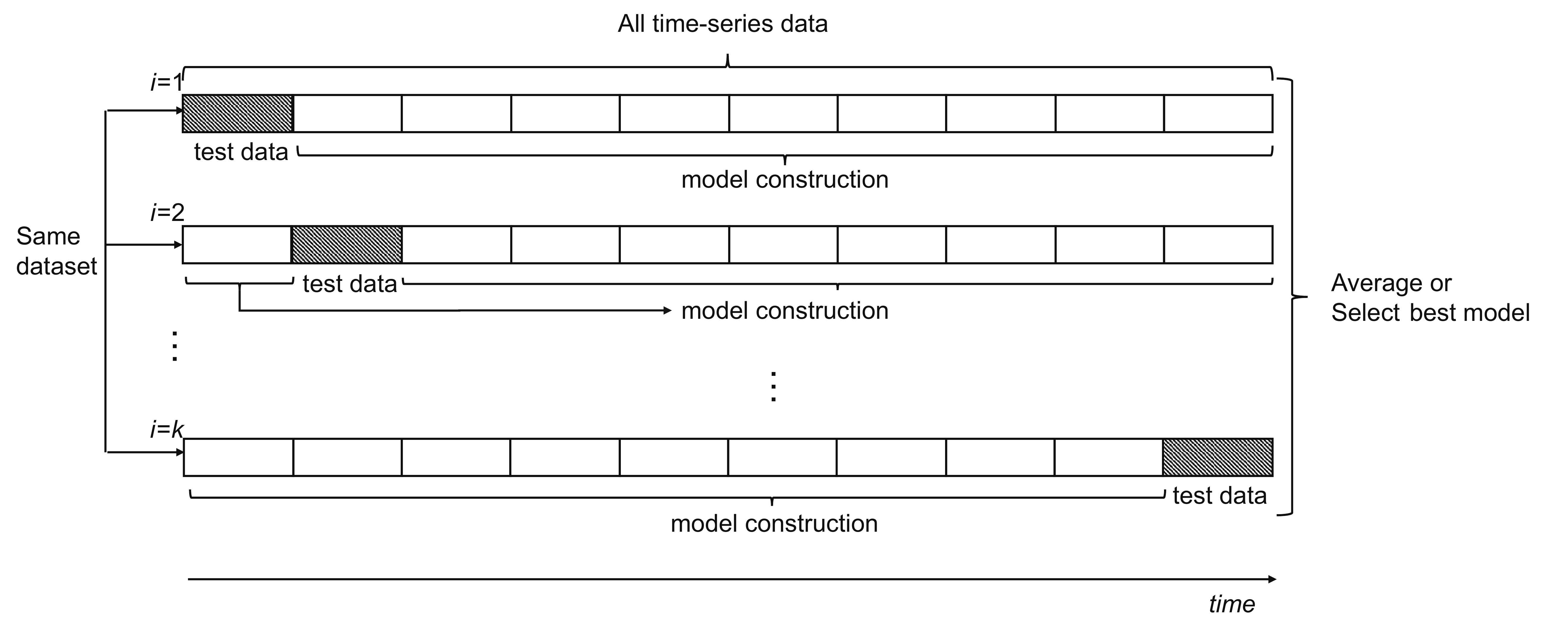}
    \caption{Schematic of cross-validation for POD-based reduced-order model.}
    \label{fig:CVPOD}
\end{figure}

The $k$-fold cross-validation technique was used for the evaluation of the model predictability of the linear model, while the POD analysis was only once applied to whole data and unchanged. The time-series data of temporal coefficients of POD modes was partitioned into ten ($k = 10$) blocks of the dataset where each block has equal sizes. A single block of the dataset was used as test data for the estimation by the model which was kept separated from any other data sets, and the remaining data sets were used as training data for the construction of the model. The training and testing procedure was repeated ten times and an average of the results of predictability was determined as the final results for the evaluation \citep{brunton2019data}. The schematic of the procedure of the $k$-fold cross-validation is shown in Fig.~\ref{fig:CVPOD}.



Similarly, the error estimation for DMD-based reduced-order modeling is presented. 
In the present study, the model is only used for the model evaluation, and initial value is assumed to be obtained from the full observation of the data. For this purpose, the projection matrix for the selected DMD modes is created and multiplied to the initial and test dataset. Although the time history of the DMD mode amplitude could be estimated from the eigenvalues and the initial values estimated by multiplying the pseudo inverse matrix of selected DMD modes to the full observation of the data when we apply it to the practical problem, the results should be again projected onto the POD space and the effects of the nonnormality of the DMD modes on the error norm should be carefully eliminated in that case. For these extra processes are avoided, the projection matrix to the subspace of selected DMD modes is only multiplied and the error is evaluated in the present study. 

The projection matrix onto the subspace of the selected DMD modes in the original POD space can be described as follows:
\begin{equation}
\label{eq:proj}
\mathbb{P}_{\textrm{sp}}=\bm{\Psi}_{\textrm{sp}}(\bm{\Psi}_{\textrm{sp}}^{\mathsf{T}}\bm{\Psi}_{\textrm{sp}})^{-1}\bm{\Psi}_{\textrm{sp}}^{\mathsf{T}}.
\end{equation}
In this case, the solution vector $z$ should be projected onto the subspace and time advancement of the estimated solution vector $\hat{\mathbf{z}}$ in the DMDsp subspace can be written, respectively, as follows:
\begin{eqnarray}
\mathbf{z}_{\textrm{sp}}(t_n)&=&\mathbb{P}_{\textrm{sp}}\mathbf{z}(t_{n}),\\
\hat{\mathbf{z}}_{\textrm{sp}}(t_n)&=&\mathbf{A}^{n-n_{0}}\mathbb{P}_{\textrm{sp}}\mathbf{z}(t_{n0}).
\end{eqnarray}
The model error vector and the model instantaneous error  of the DMDsp-based reduced-order model can be
\begin{equation}
\label{eq:error}
    {\bm{\epsilon}}_{\textrm{sp}}(t_{n})={\mathbf{z}}_{\textrm{sp}}(t_{n})-{\hat{\mathbf{z}}_{\textrm{sp}}}(t_{n}),
\end{equation}
and
\begin{equation}
\label{eq:error_allmodes}
    \bar{\epsilon}_{\textrm{sp}}(t_{n})=\overline{\|{\bm{\epsilon}}(t_{n}) \|}_2,
\end{equation}
respectively. Also, the asymptotic value of the model error was similarly defined as 
\begin{equation}
\label{eq:zrms}
    {z}_{\mathrm{spRMS}}=\sqrt{\frac{\sum_{n=1}^{N} \| {\mathbf{z}_{\textrm{sp}}}(t_{n}) \|^2_2.}{N}} 
\end{equation}
The model prediction permissive time range $t_{\mathrm{mp}}$ for the DMDsp-based reduced-order model is again defined as the time at which the error reaches $1-e^{-1}$ of $\bar{z}_{\mathrm{RMS}}$, i.e., $\bar{\epsilon}_{\rm{sp}}({t_{\mathrm{mp}}}) \approx 0.632\bar{z}_{\mathrm{RMS}}$.

\begin{figure}[h]
\captionsetup{justification=raggedright}
    \centering
    \includegraphics[width=120mm]{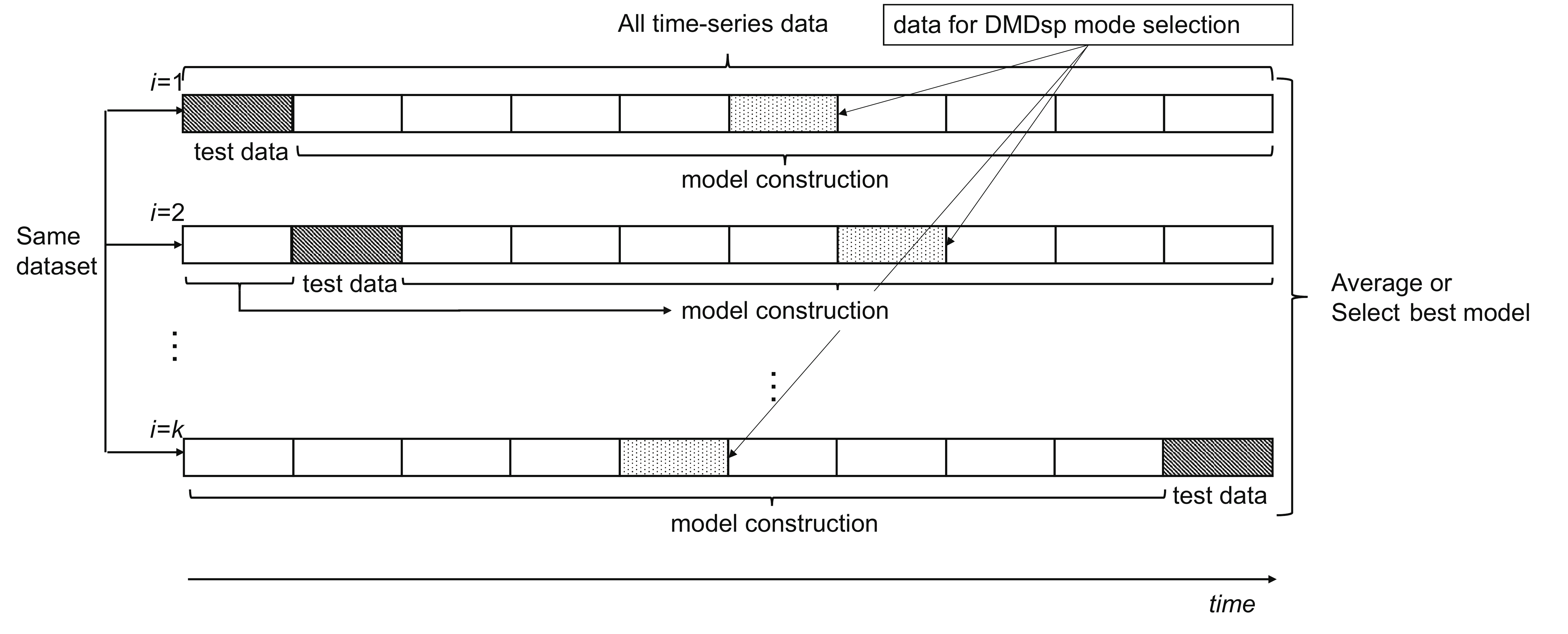}
    \caption{Schematic of cross-validation for DMDSp-based reduced-order model.}
    \label{fig:CVDMDsp}
\end{figure}

The $k$-fold cross-validation technique was also used for the evaluation of the model predictability \citep{brunton2019data}. In addition to the previous cross-validation procedure, the mode selection for the $i$th block of dataset is conducted for the ($\textrm{mod}(i+k/2-1,k)+1$)th block of dataset, here mod represents the modulo function. The number of data in one block seems to be sufficiently large for DMDsp. The schematic of the procedure of the $k$-fold cross-validation for DMDsp-based reduced-order modeling is shown in Fig.~\ref{fig:CVDMDsp}.

\section{Experimental Setup}
The wind tunnel testing was conducted in the small low-turbulence wind tunnel (SLTWT) at the Institute of Fluid Science, Tohoku University, similar to the previous study for the flow control using a plasma actuator \citep{komuro2018multiple}. Here, SLTWT has an open-type test section with an octagonal cross-section of 293~mm diagonal distance. The airfoil of the test model has an NACA0015 profile with a chord length $c$ of 100~mm and a span width of 300~mm. The model was fabricated using stereolithography, which is a high-precision three-dimensional printing method. The freestream velocity $U$ was set to 10~m/s, corresponding to the chord Reynolds number of approximately $6.4\times10^4$. The angle of attack $\alpha$ was changed from 14~deg to 22~deg, which are post-stall angles. The stall angle in this test condition is $\alpha=13$~deg.

\begin{figure}[h]
\captionsetup{justification=raggedright}
    \centering
    \includegraphics[width=70mm]{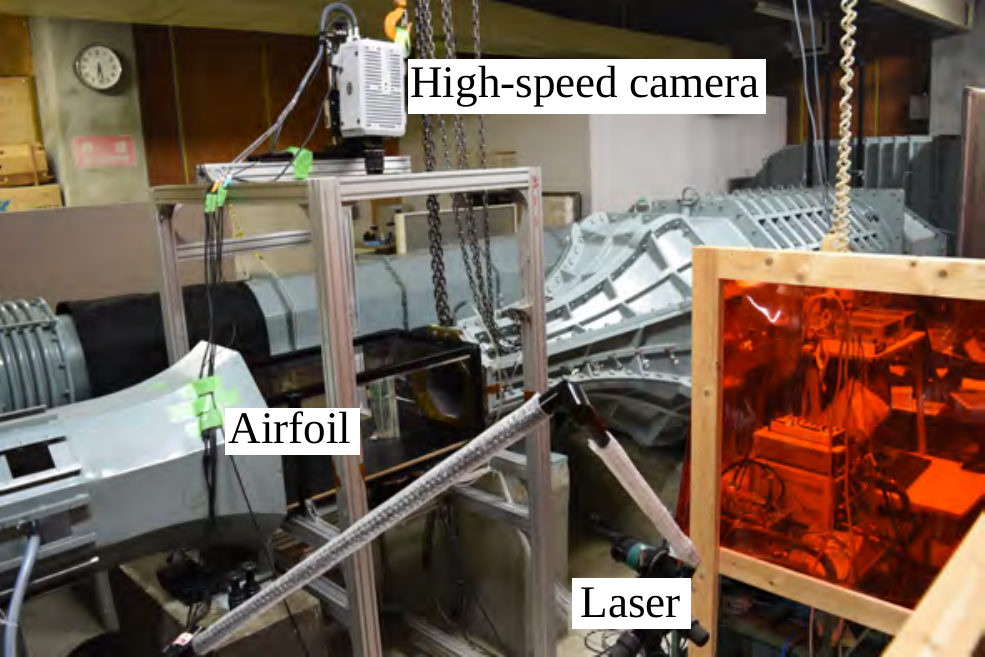}
    \caption{Experimental setup of the PIV measurement.}
    \label{fig:setup}
\end{figure}

Time-resolved PIV measurement was conducted and the unsteady flow-field data were acquired according to the test conditions given above. Figure \ref{fig:setup} shows a schematic of the PIV measurement system. Dioctyl sebacate was used as tracer particles. The particle images were acquired using a double pulse laser (DM30-527, Photonics Industries) and a high-speed camera (SA-X2, Photron) that were synchronized with each other. The laser light sheet was inserted from a side direction to the airfoil model mounted vertically on the test section. The spatial resolution, the sampling rate and the total number of particle images were set to be $1024 \times 1024$~pixels, 5~kHz and 10,000~pairs, respectively. The sampling time 2~s corresponds to a nondimensionalized time of $tU/c=200$ based on the freestream velocity and the chord length. Dynamic Studio 5.1 (Dantec Dynamics) was used as the post-processing software of the PIV. The time-resolved data of two-dimensional velocity vectors were computed using an adaptive PIV algorithm with $8 \times 8$ pixels interrogation windows at a minimum size. Moving average validation algorithm was employed to smooth out each vector using $3 \times 3$ vectors around itself.

Here, the dataset employed in the present study is available in Reference \citep{nonomura2021github}. The authors hope that the dataset helps further developments of reduced-order models of complex flow fields and related technologies. 

\section{Results and Discussions}
\subsection{Flow Fields around Airfoil}
\begin{figure}[h]
\captionsetup{justification=raggedright}
    \centering
    \includegraphics[width=120mm]{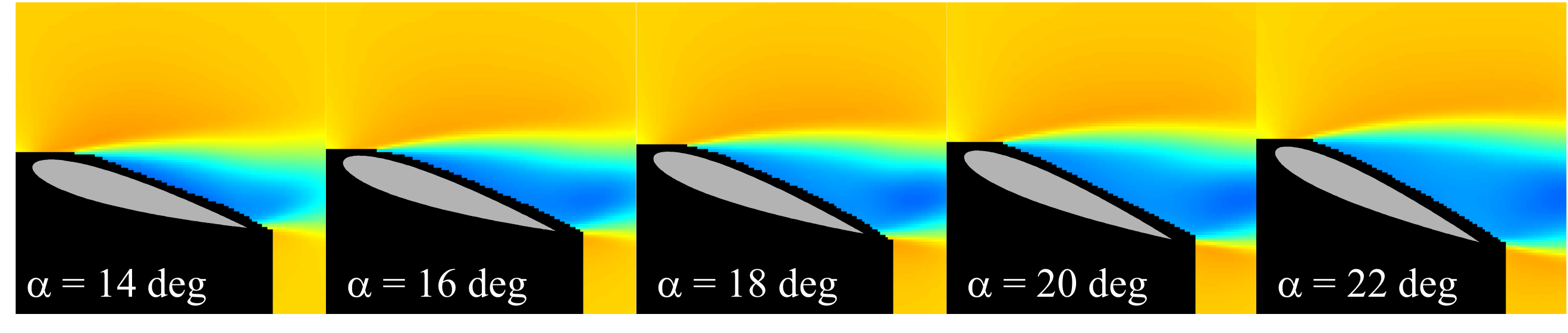}
    \subfigure[streamwise velocity]{\includegraphics[width=60mm]{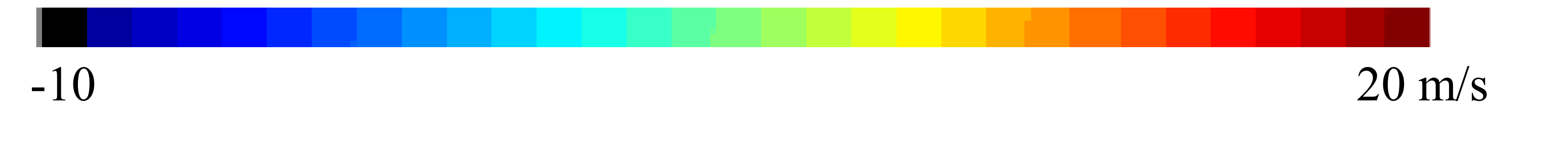}}
    \includegraphics[width=120mm]{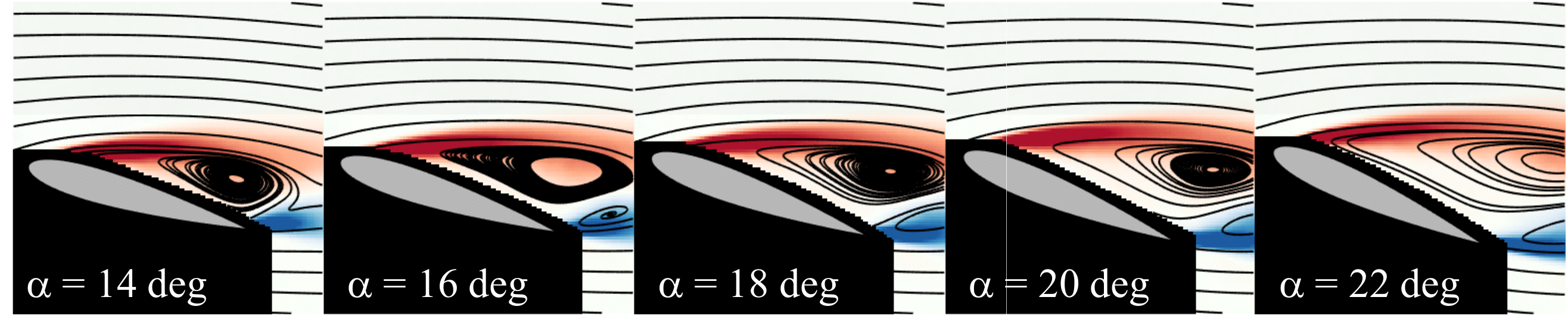}
    \subfigure[vorticity and streamlines]{\includegraphics[width=60mm]{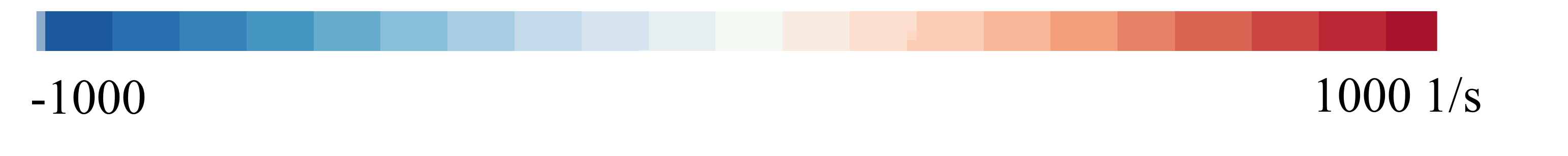}}
    \caption{Time-averaged flow fields.}
    \label{fig:u_ave}
\end{figure}

\begin{figure}[h]
\captionsetup{justification=raggedright}
    \centering
    \includegraphics[width=72mm]{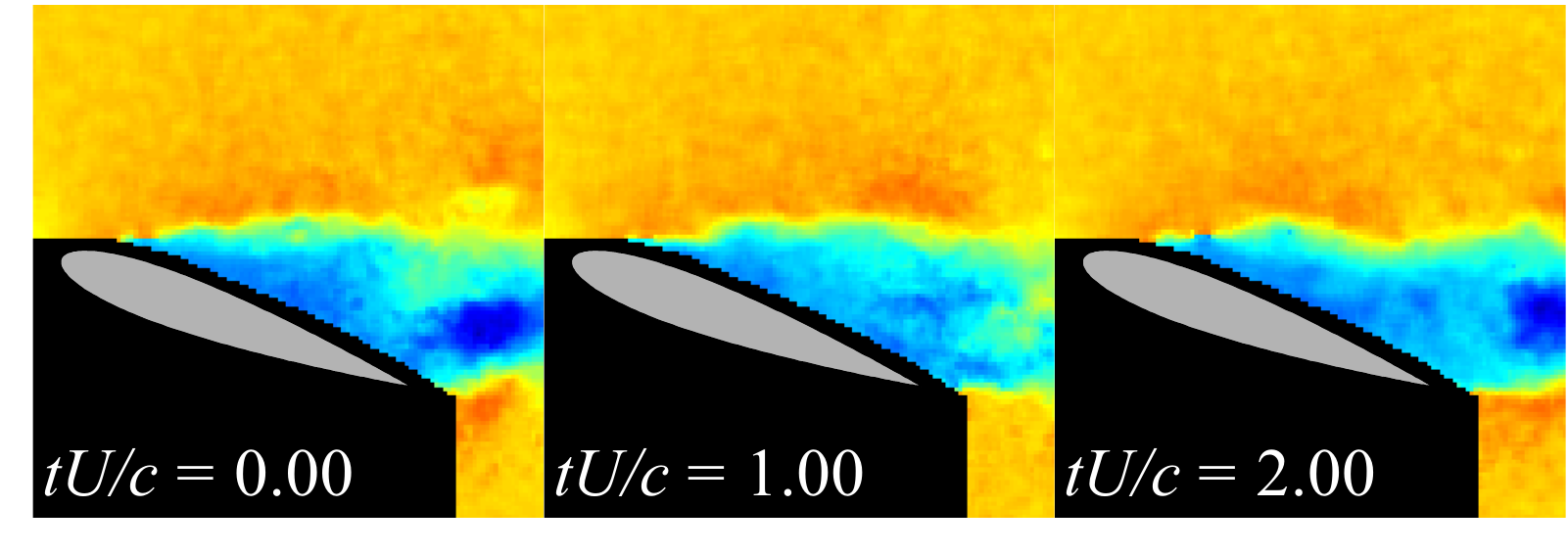}
    \subfigure[streamwise velocity fluctuation ]{\includegraphics[width=60mm]{colorbar_u-eps-converted-to.pdf}}
    \includegraphics[width=72mm]{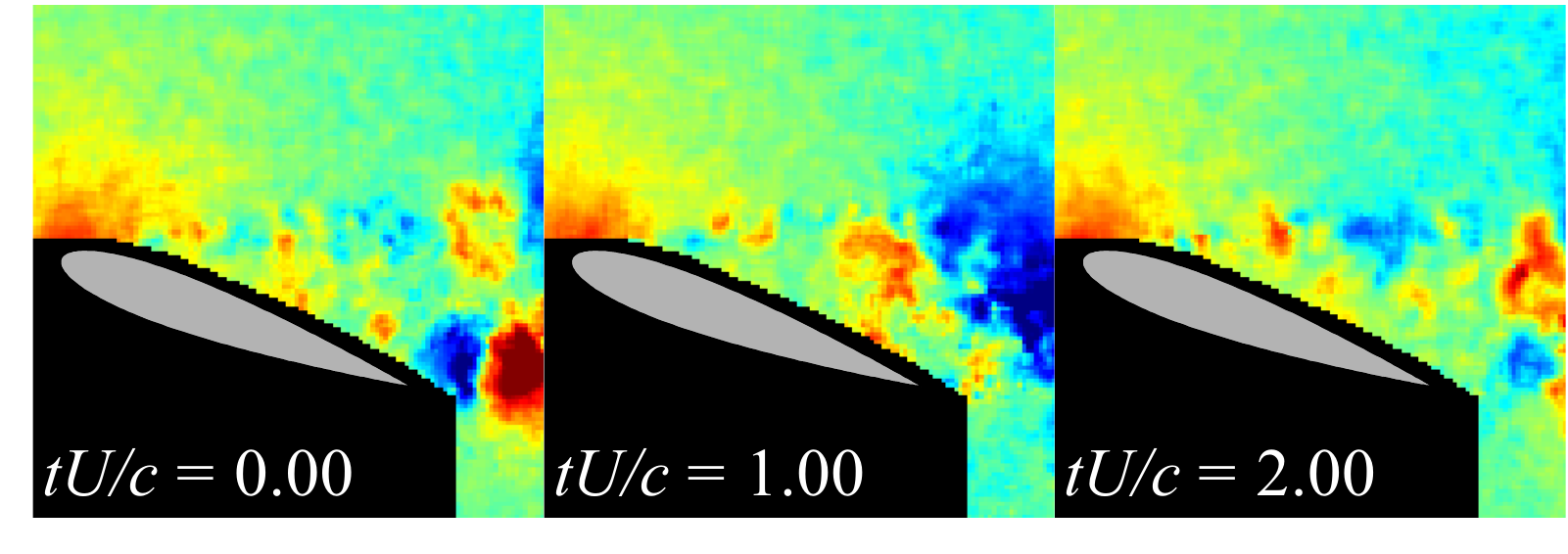}
    \subfigure[transverse velocity fluctuation ]{\includegraphics[width=60mm]{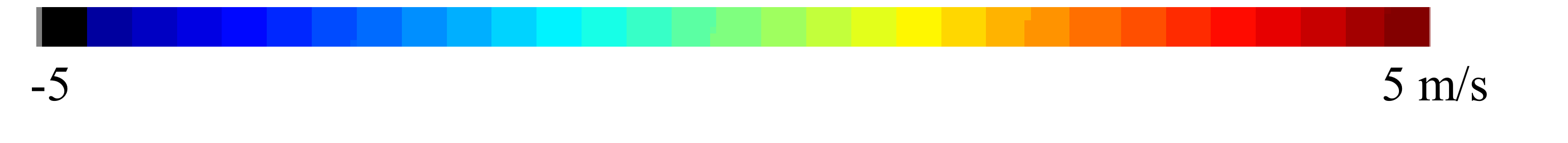}}
    \includegraphics[width=72mm]{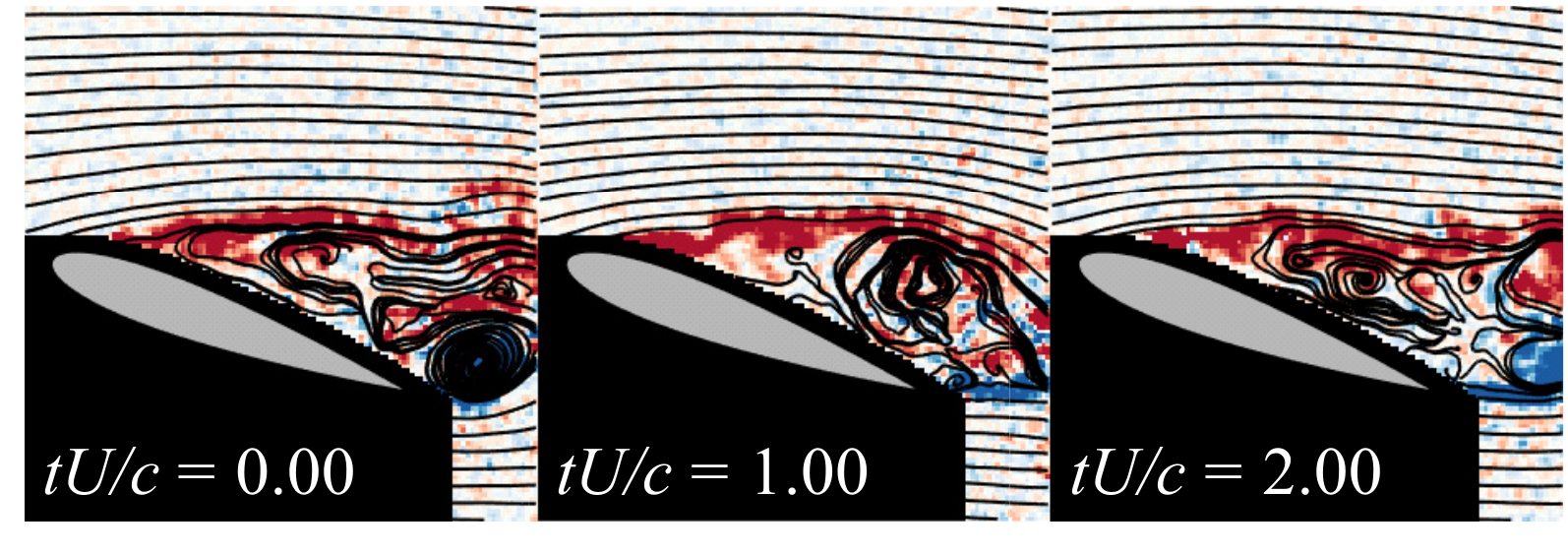}
    \subfigure[vorticity and streamlines]{\includegraphics[width=60mm]{colorbar_omega-eps-converted-to.pdf}}
    \caption{Snapshots of velocity fluctuation fields at $\alpha=18 \mathrm{deg}$}
    \label{fig:tr_18deg}
\end{figure}

Figure \ref{fig:u_ave} shows the streamwise velocity and vorticity fields with streamlines of the time-averaged flow fields. In this study, the calculated velocity data near the airfoil and behind the laser light were not used because their reliability is reduced by the presence of reflections and a lack of tracer particles \citep{theunissen2008improvement}. The black and gray regions in the figures represent the masked region and the position of the airfoil, respectively. Hereafter, we refer to the results of $\alpha=18$~deg because the quantitative results of linear models based on the cross-validation does not significantly change with the angle of attack in the range we investigated. It should be noted that the absence of the sensitivity of an angle of attack for the results are different from the qualitative observation of the presence of sensitivity of an angle of attack for the results in the previous work by \citet{nankai2019linear}. We consider that such a presence of the sensitivity of an angle of attack was accidentally and qualitatively observed in the previous work \citep{nankai2019linear} depending on a choice of the test and training data, and the absence of the sensitivity of an angle of attack might be a more reliable result according to the quantitative evaluation using cross-validation, although the experimental condition itself was similar, but different (the different wind tunnel was used) for the previous and present studies. This implies that the quantitative evaluation of the linear model using cross-validation is important for the detailed discussion of the model.  Figure \ref{fig:tr_18deg} shows the snapshots of the velocity and vorticity fields with streamlines. The figures demonstrate that the velocity fluctuations due to the flow separation that are the target for the linear reduced-order model were clearly acquired by the PIV measurement.

\begin{figure}[h]
\captionsetup{justification=raggedright}
    \centering
    \includegraphics[width=110mm]{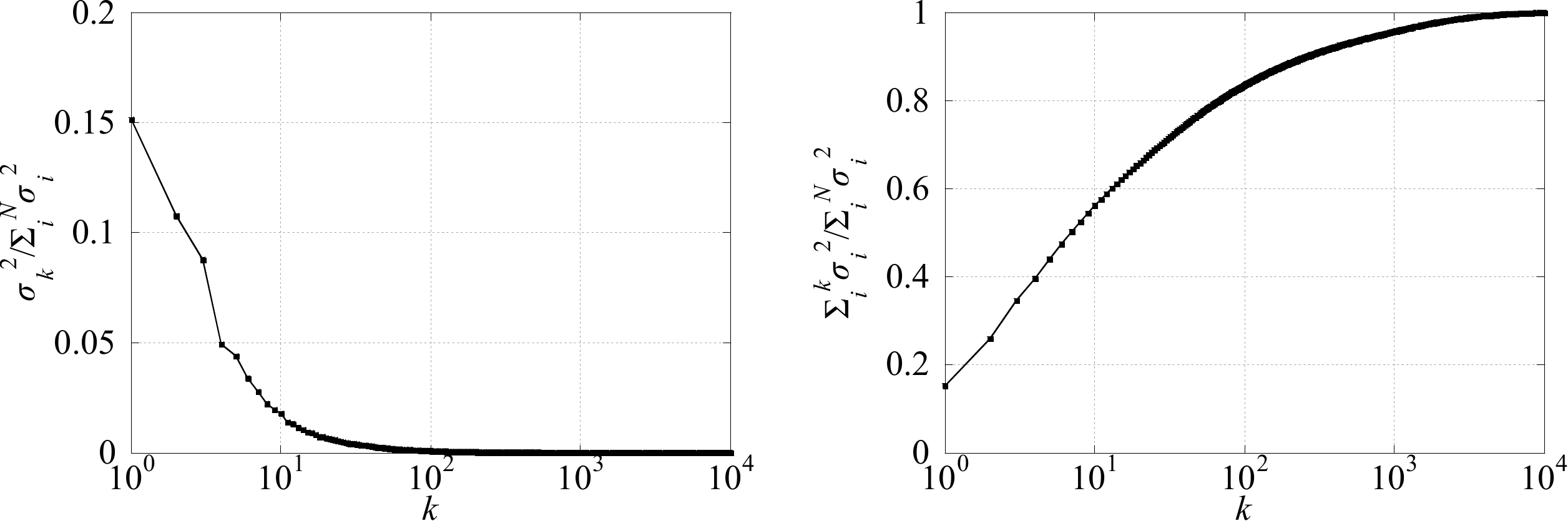}
    \caption{POD-mode energy distributions: (a) normalized POD-eigenvalues spectrum; (b) partial amount of the energy contained in the first $k$ POD modes.}
    \label{fig:mode_energy}
\end{figure}

\begin{figure}[h]
\captionsetup{justification=raggedright}
    \centering
    \subfigure[streamwise velocity component]{\includegraphics[width=110mm]{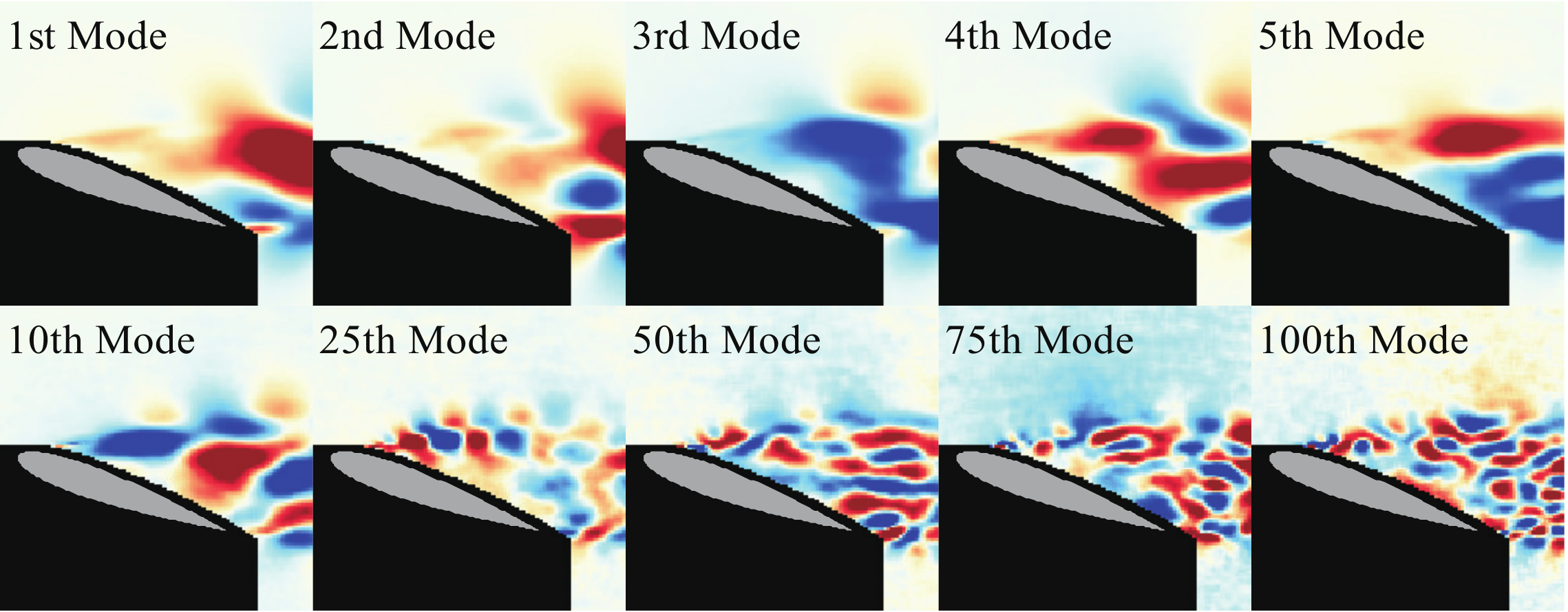}}
    \subfigure[transverse velocity component]{\includegraphics[width=110mm]{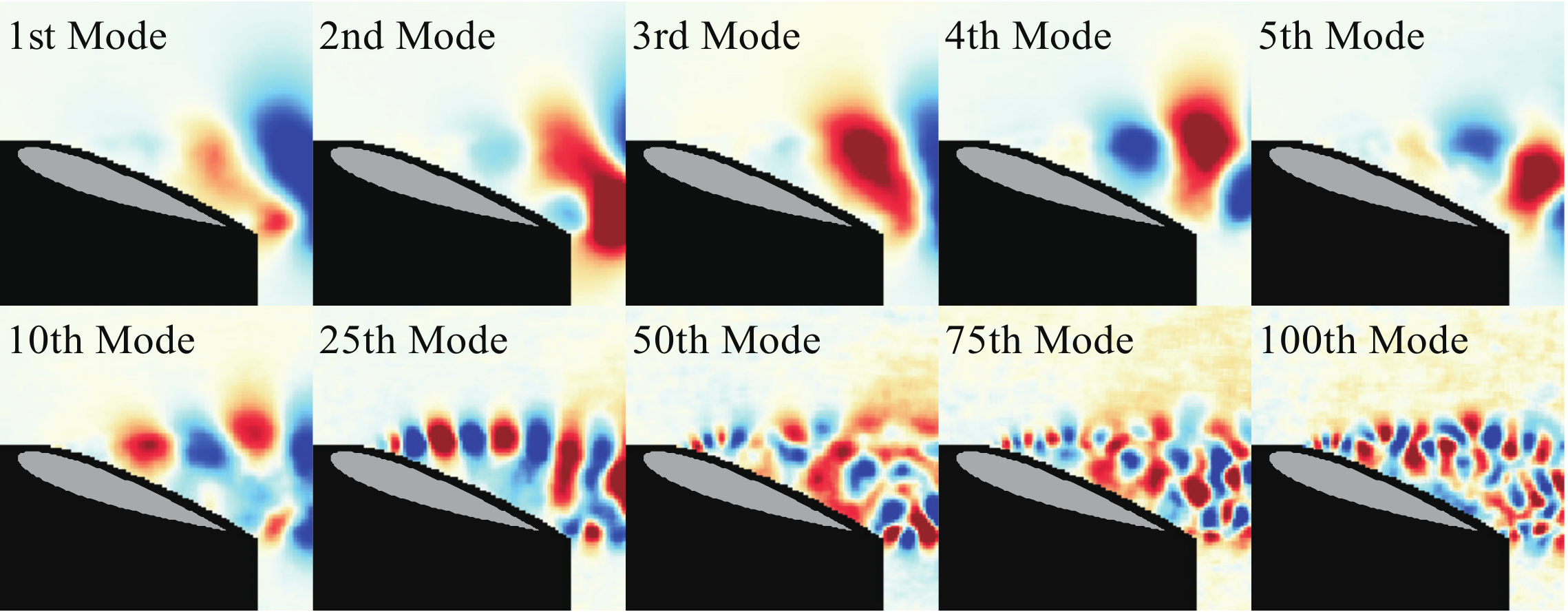}}
    \caption{Velocity fields of POD modes. Here, ranges of the contours are set to be the same for all the mode, while the color bar is not shown because the POD mode is normalized.}
    \label{fig:U}
\end{figure}

Figure \ref{fig:mode_energy} illustrates the POD-mode energy distributions. Figures~\ref{fig:mode_energy}(a) and (b) represents the energy ratio of each POD mode and the amount of energy contained in the first $k$ POD modes, respectively. The first ten POD modes represent approximately $60\%$ of the total energy. Figure \ref{fig:U} displays the velocity-fluctuation fields of several of the POD modes. Here, ranges of the contours are set to be the same for all the mode, while the color bar is not shown because the POD mode is normalized. In addition, POD modes represent the fluctuation components and their representation using arrows or streamlines are not considered to be so appropriate. Therefore, only the streamwise and transverse component of velocity fluctuations are presented in the present study. These results illustrate the principal flow structures extracted from the complex flow fields as POD modes and also indicate that more energetic (low-order) POD modes express larger flow structures than less energetic (high-order) POD modes. See our previous report \citep{nankai2019linear} for more detailed discussion of the POD analysis.

\subsection{POD-based Linear Reduced-order Model}
\label{subsec:pod}
The linear reduced-order model was constructed with different sets of the parameters $r$ and $\mathbf{A}$, and their effects on the model performance were investigated. The number of POD modes was varied from $r=2$ to $1,000$, and the coefficient matrix was computed using the four methods described in Sec. \ref{sec:LROM}.

\begin{figure}[h]
\captionsetup{justification=raggedright}
    \subfigure[$r_{\textrm{POD}}=10$, mode 1]{\includegraphics[width=60mm]{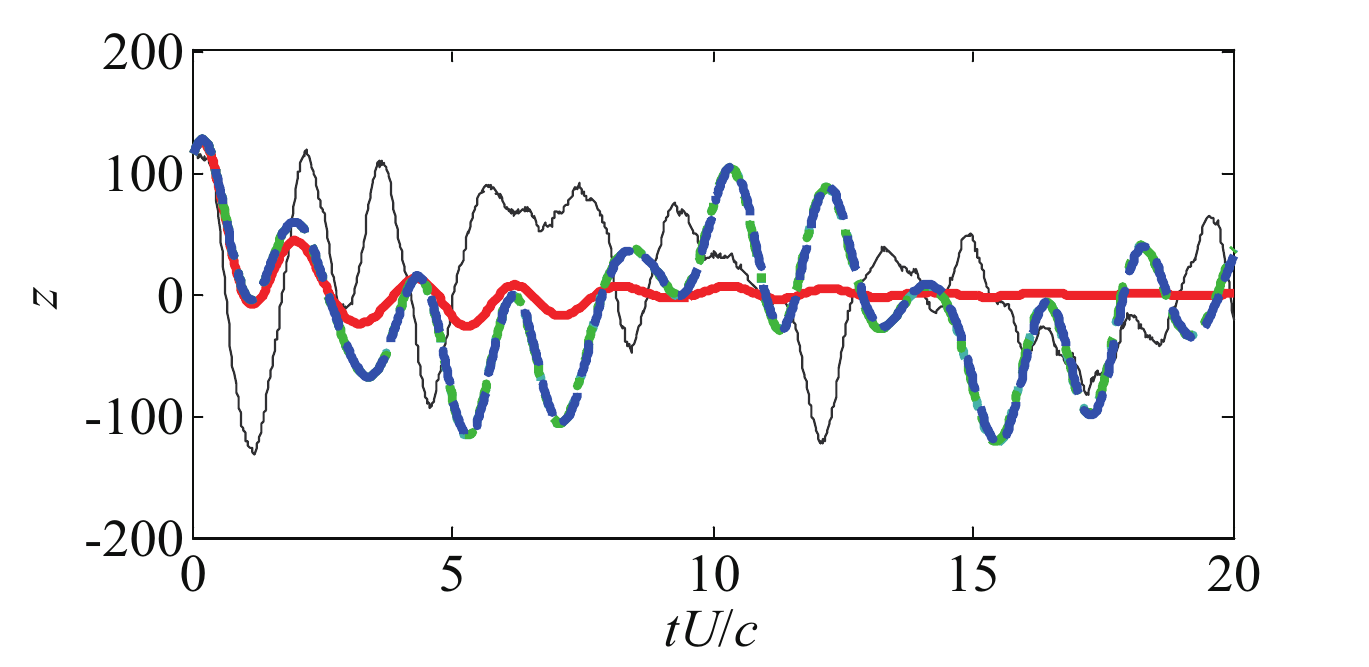}}
    \subfigure[$r_{\textrm{POD}}=10$, mode 2]{\includegraphics[width=60mm]{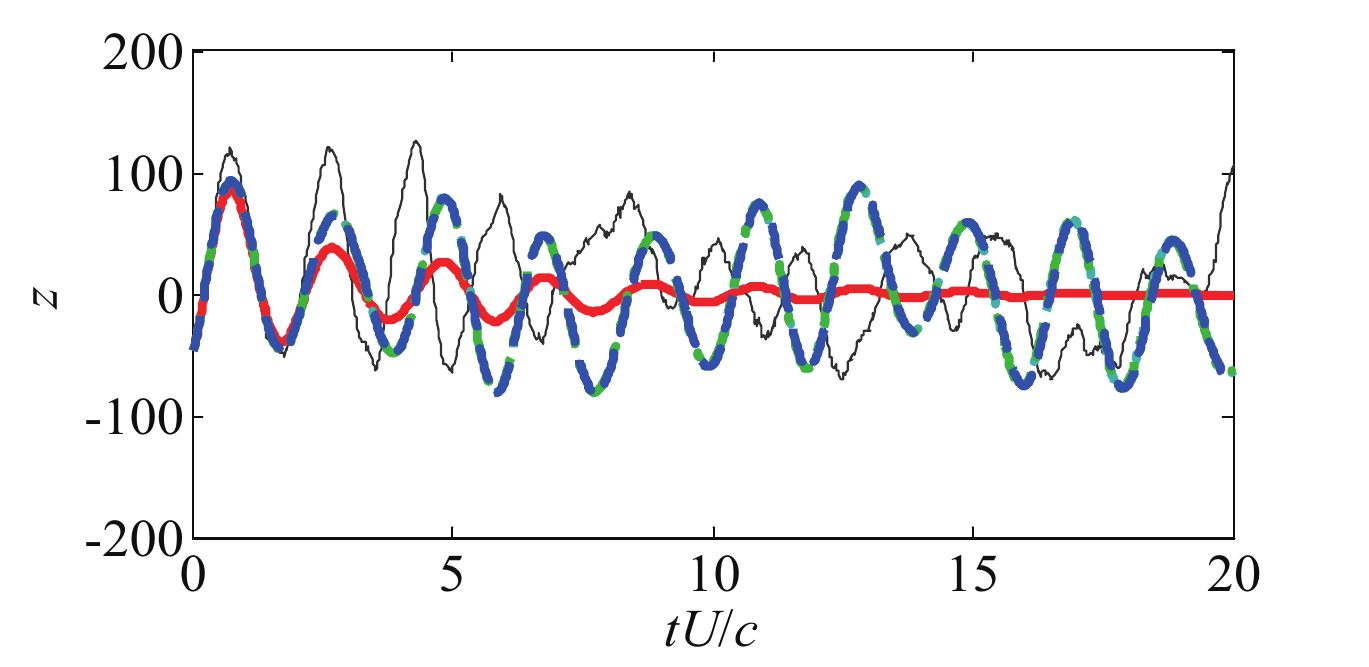}}\\
    \subfigure[$r_{\textrm{POD}}=100$, mode 1]{\includegraphics[width=60mm]{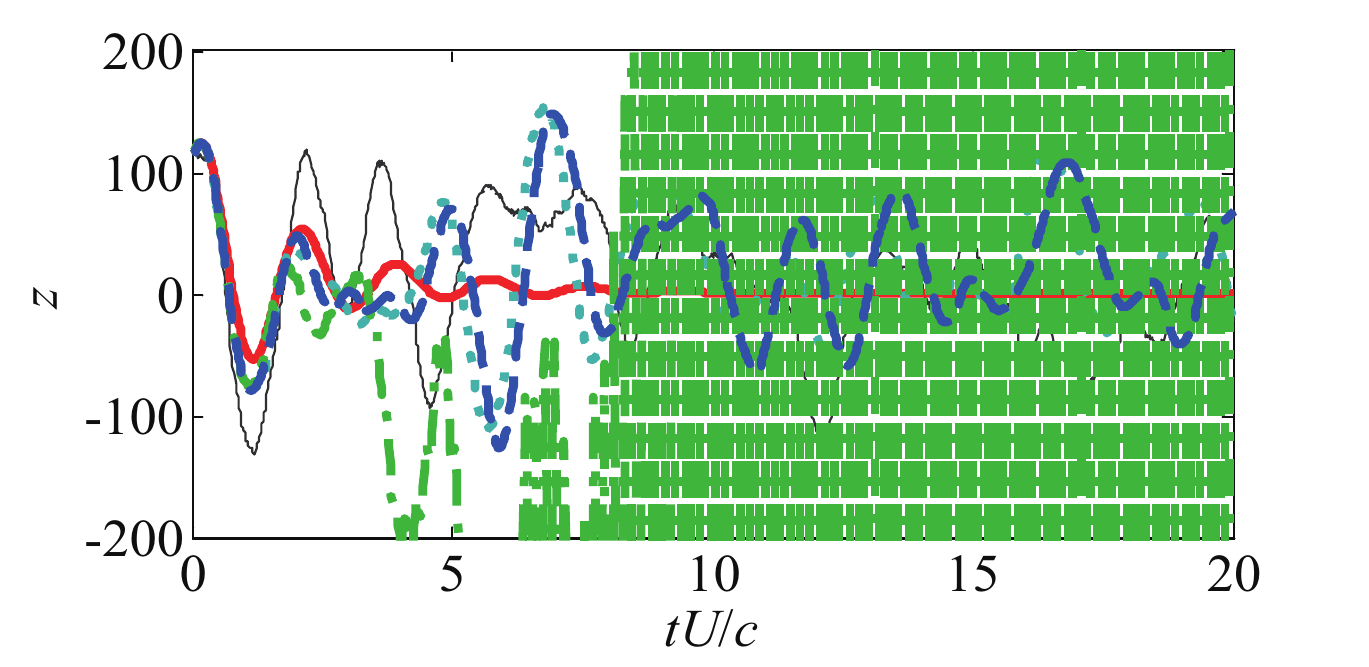}}
    \subfigure[$r_{\textrm{POD}}=100$, mode 2]{\includegraphics[width=60mm]{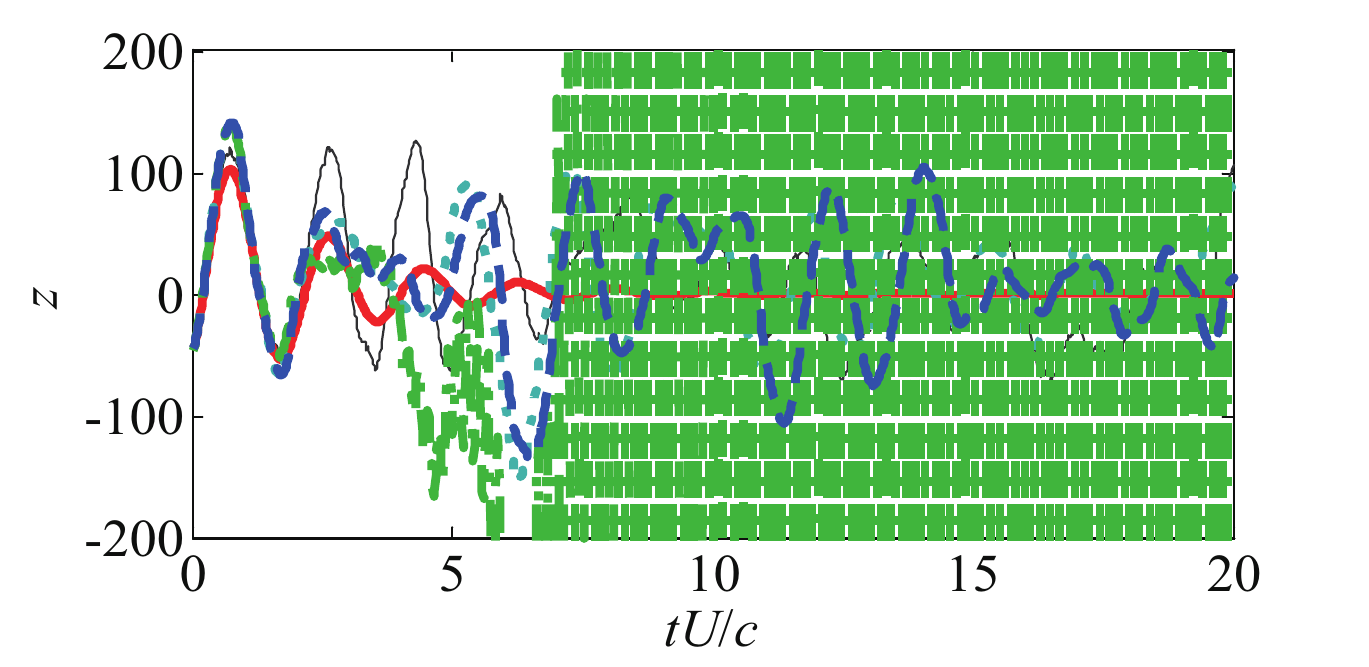}}\\
    \subfigure[$r_{\textrm{POD}}=1000$, mode 1]{\includegraphics[width=60mm]{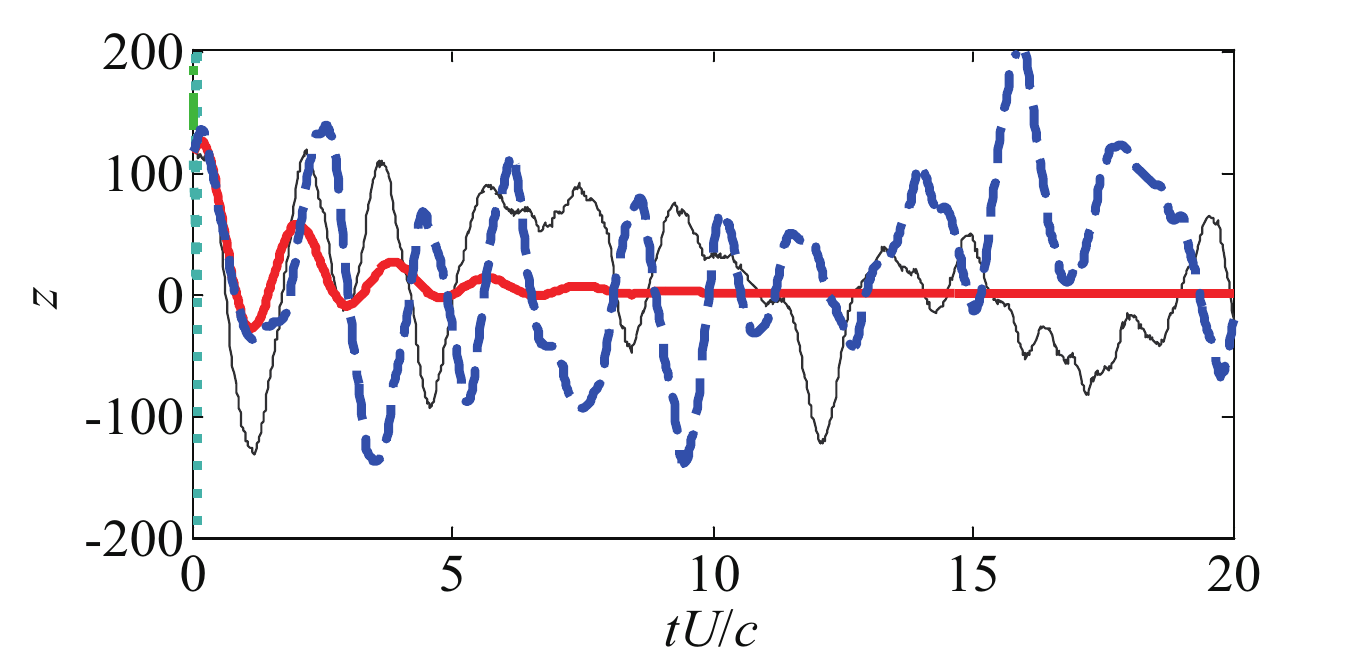}}
    \subfigure[$r_{\textrm{POD}}=1000$, mode 2]{\includegraphics[width=60mm]{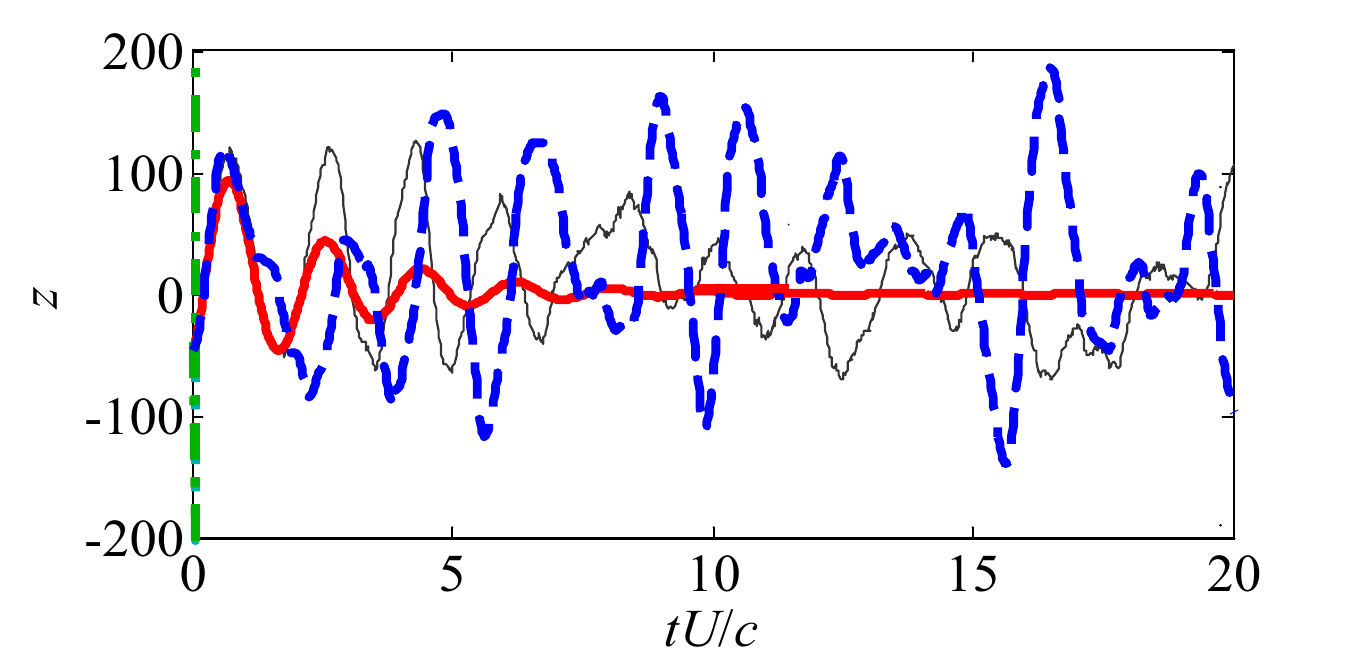}}\\
    \begin{center}
    \includegraphics[width=60mm]{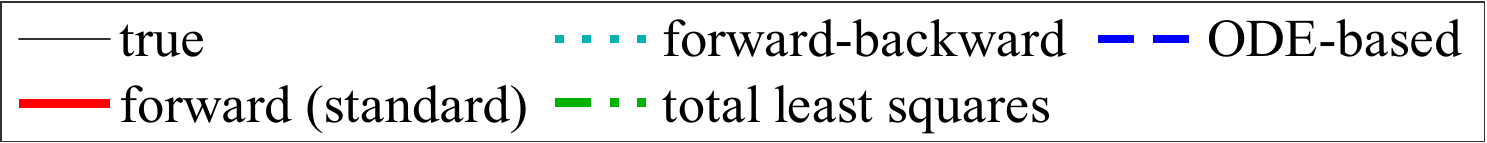}\\    
    \end{center}
    \caption{Time histories of estimated POD modes using the model. Results of the forward-backward and total-least-squares models immediately go outside the plot range because they drastically diverge.}
    \label{fig:SV_estimate}
\end{figure}

\begin{figure}[h]
\captionsetup{justification=raggedright}
    \subfigure[$r=10$~modes]{\includegraphics[width=60mm]{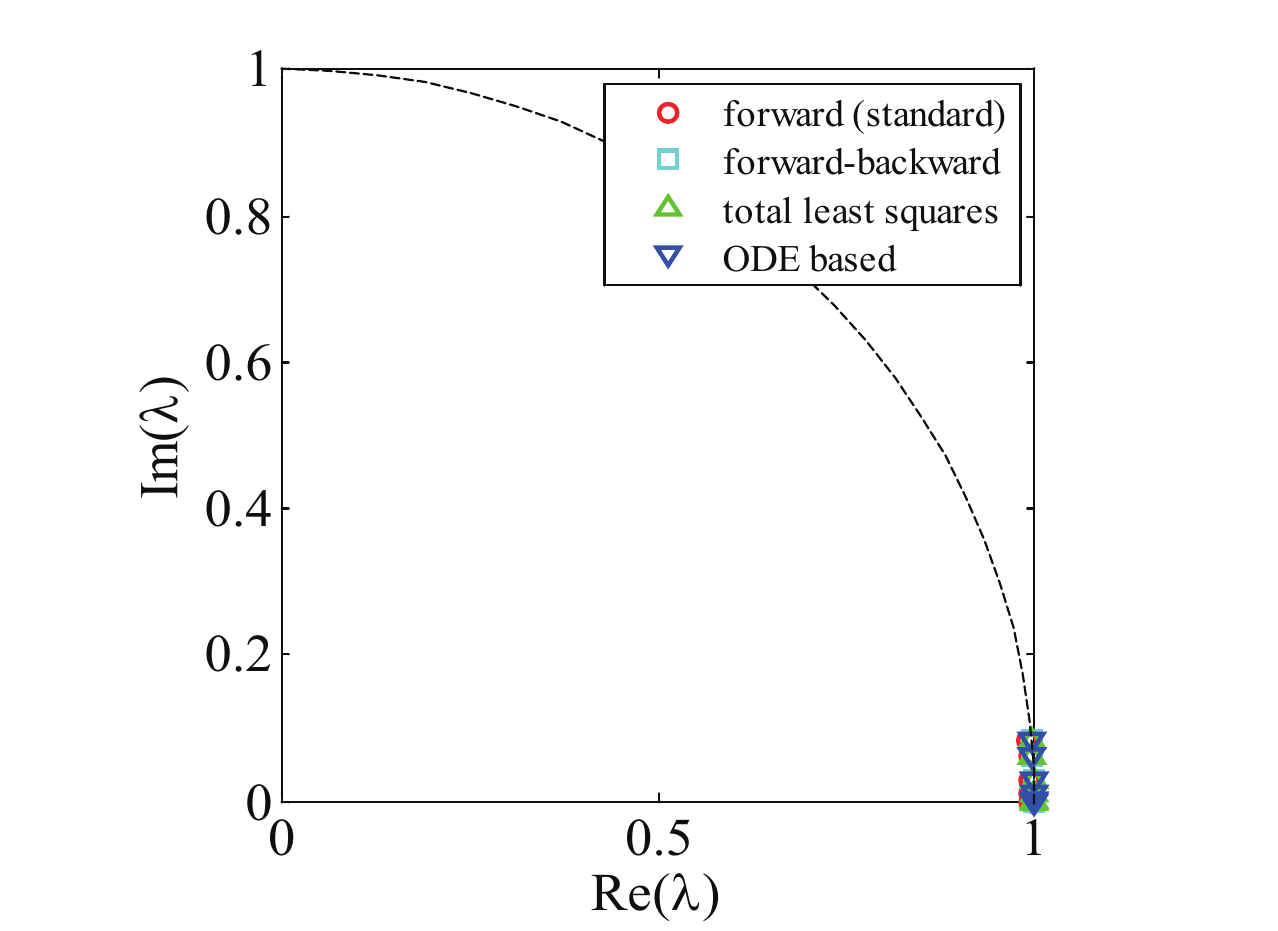}}
    \subfigure[$r=100$~modes]{\includegraphics[width=60mm]{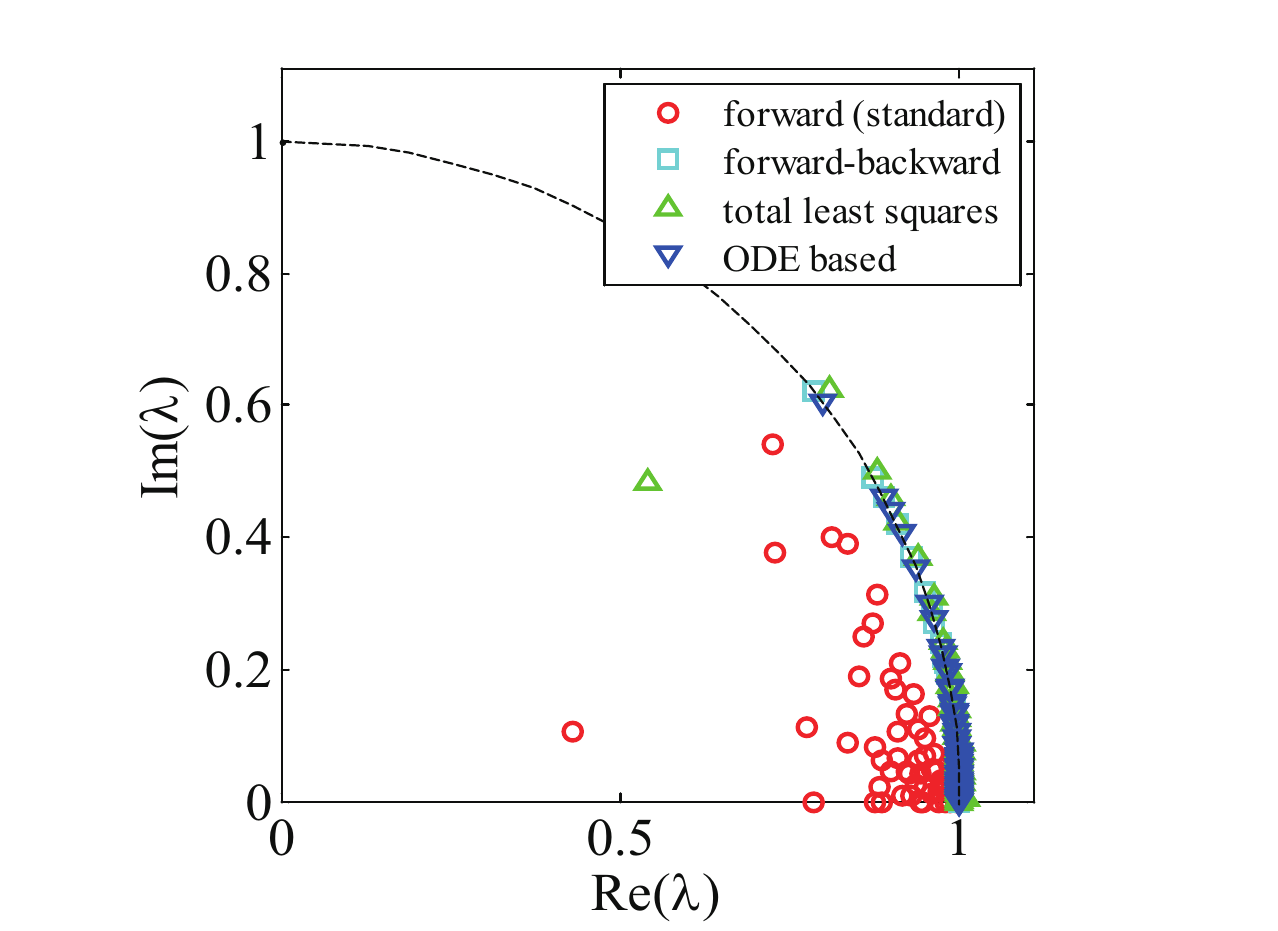}}\\
    \subfigure[$r=1000$~modes]{\includegraphics[width=60mm]{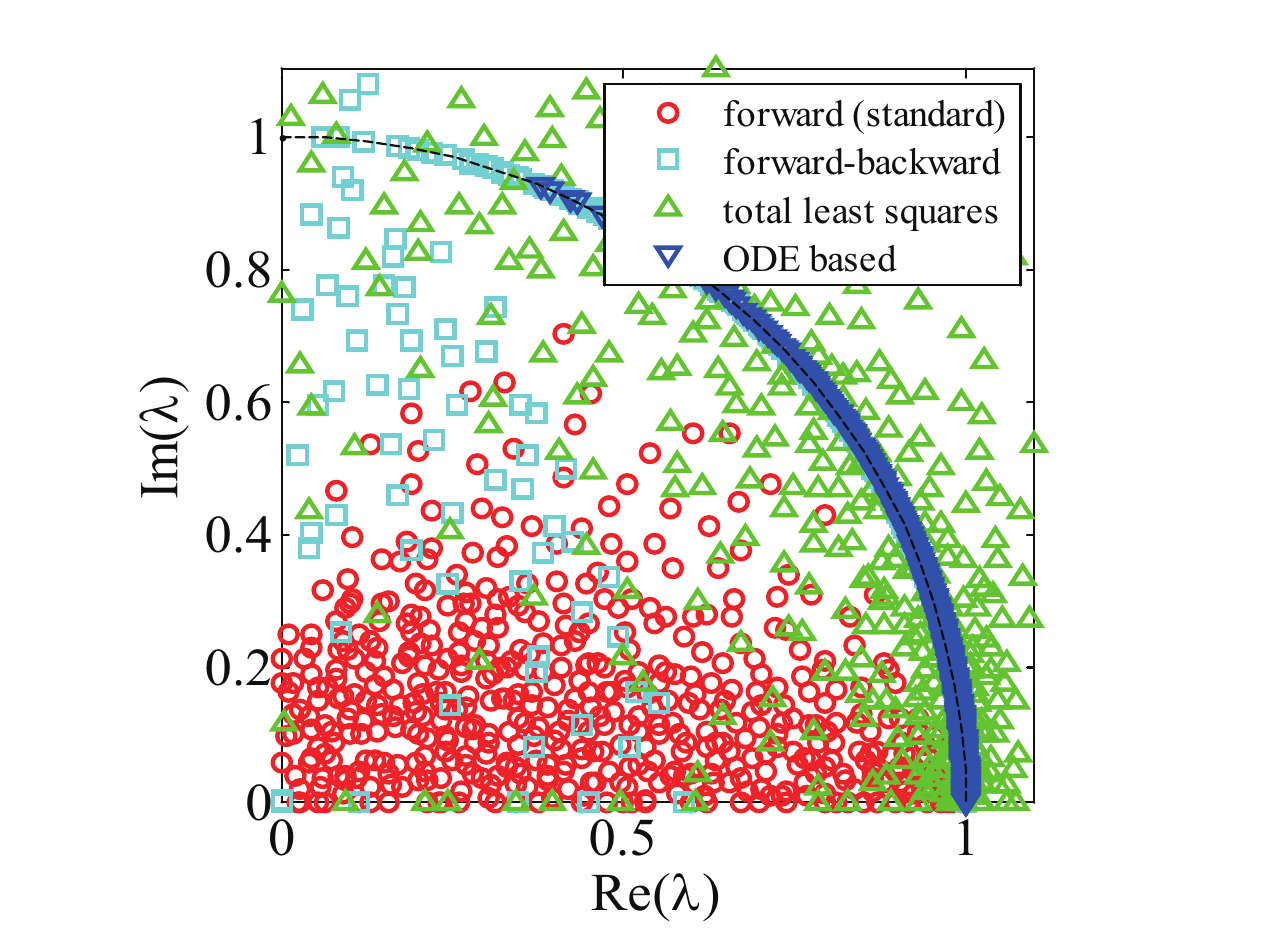}}
    \caption{Eigenvalue distributions of the coefficient matrices.}
    \label{fig:eigenvalues}
\end{figure}

\begin{figure}[h]
\captionsetup{justification=raggedright}
    \subfigure[$r=10$~modes]{\includegraphics[width=60mm]{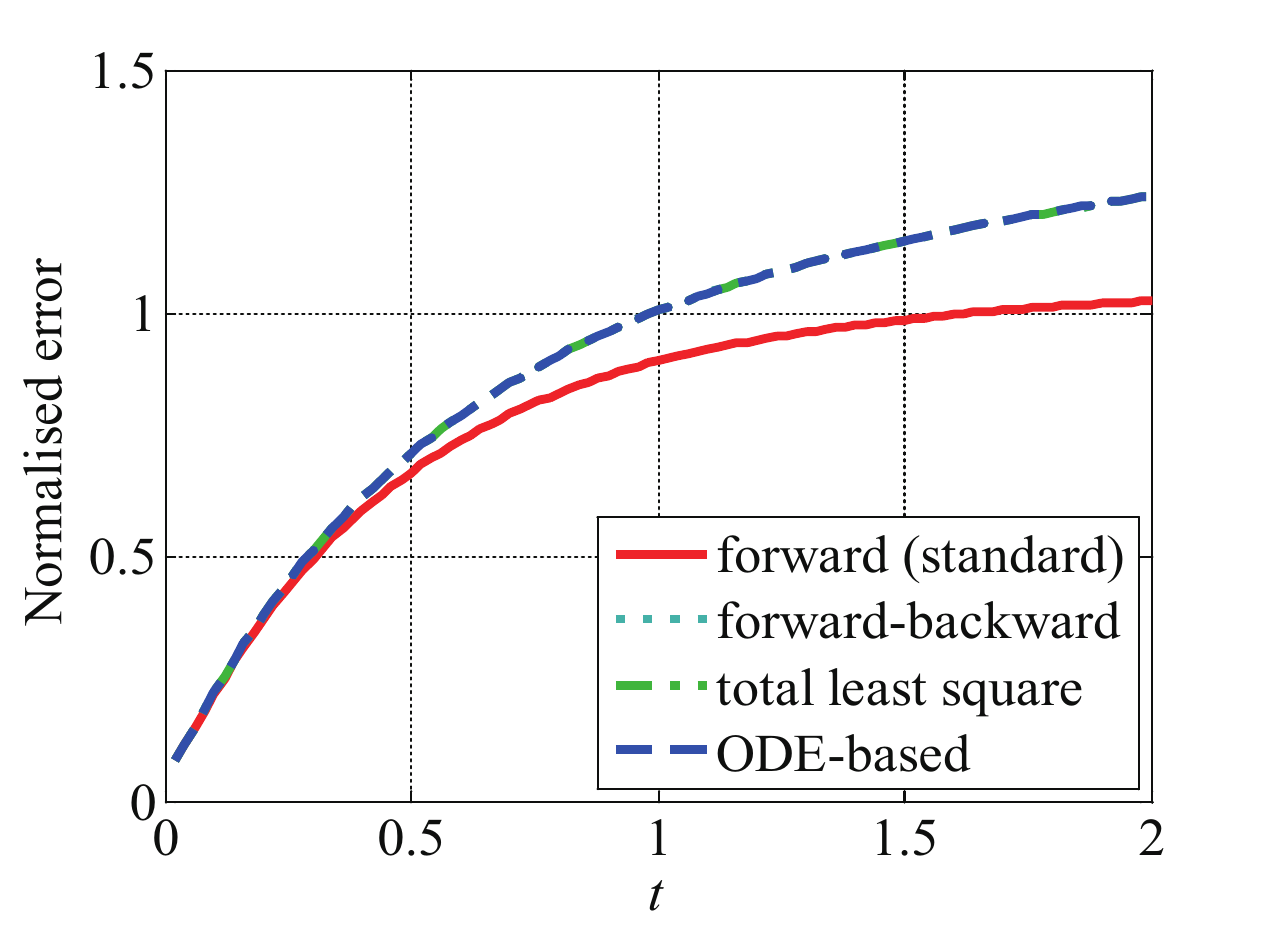}}
    \subfigure[$r=100$~modes]{\includegraphics[width=60mm]{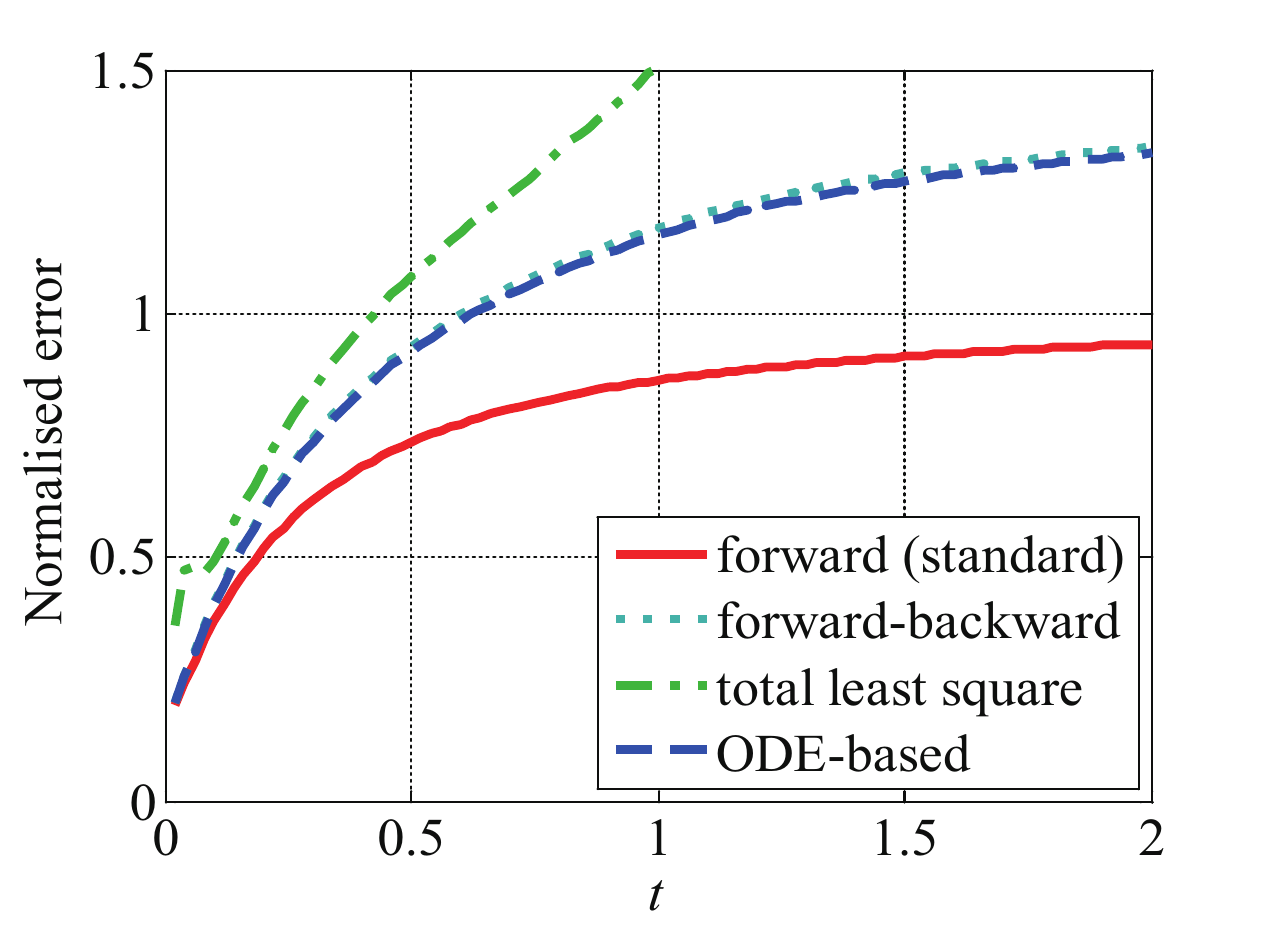}}
    \subfigure[$r=1000$~modes]{\includegraphics[width=60mm]{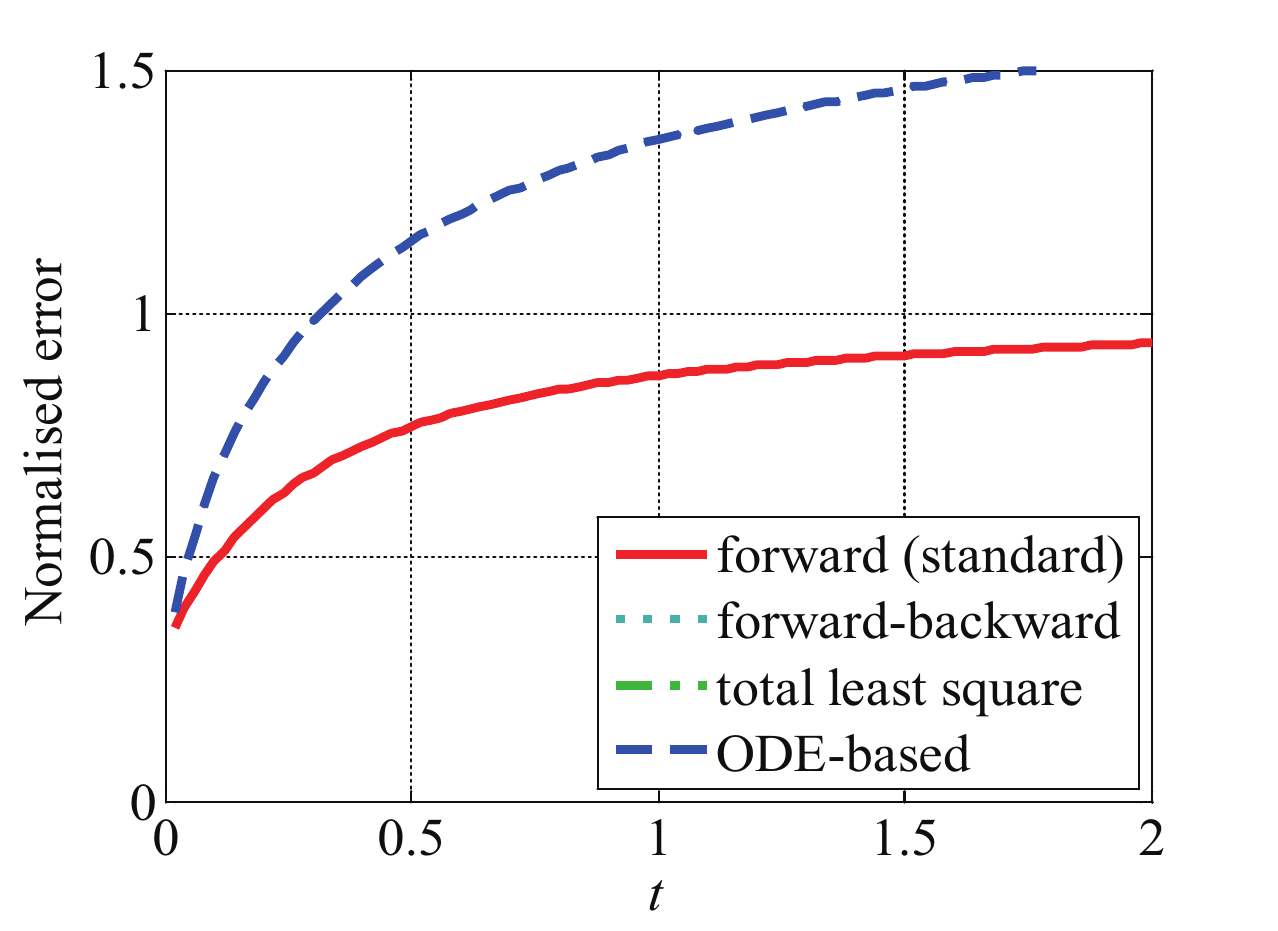}}
    \caption{Estimation error of the model. The results of the forward-backward and total-least-squares models in (c) are not included in the range of the graph because the error of these methods is very large from the beginning.}
    \label{fig:model_error}
\end{figure}

Figures~\ref{fig:SV_estimate}, \ref{fig:eigenvalues} and \ref{fig:model_error} depict the histories of the estimated temporal coefficients of POD modes, eigenvalue distributions of the coefficient matrices, and the estimation error of the model, respectively. The behavior of the estimated temporal coefficients of POD modes is determined by the eigenvalues of the coefficient matrix of the linear state space model. 
This is because Eq. \ref{eq:model_estimation} can be written as
\begin{eqnarray}
\label{eq:model_estimation_eigen}
    \mathbf{z}(t_{n})&=&\mathbf{A}^{n-n_{0}} \mathbf{z}(t_{n_{0}}) \nonumber \\
        &=& \bm{\Psi \Lambda}^{n-n_{0}} \bm{\Psi}^{-1}\mathbf{z}(t_{n_{0}}).
\end{eqnarray}
where $\bm{\Lambda}$ is the diagonal matrix of eigenvalues of $\lambda_1 \dots, \lambda_r$ and  $\mathbf{A}=\bm{\Psi\Lambda \Psi}^{-1}$.
The amplification factor, which shows how the mode evolves in time, corresponds to the magnitude of the eigenvalues $\lambda$ and the frequency of the time fluctuation of each mode corresponds to the argument of $\lambda$ \citep{taira2017modal}.

For all values of $r$, the estimated temporal coefficients of POD modes by the forward (standard) model attenuate and approach to zero, as shown in Fig.~\ref{fig:SV_estimate}. This is also demonstrated by the eigenvalues of $\mathbf{A}_\mathrm{f}$ (i.e., the magnitudes of all eigenvalues is less than unity) as shown in Fig.~\ref{fig:eigenvalues}. In contrast, the estimated temporal coefficients of POD modes by the other three models do not attenuate, and they seem to reproduce the low-frequency component of the time fluctuation of the original POD modes better than those obtained by the forward (standard) method, as shown in Fig~\ref{fig:SV_estimate}. However, the amplitude is not perfectly consistent with that of the original POD modes, and the phase of the fluctuations shifts gradually. They are presumed to cause the very poor predictability at some time steps. In addition, the estimation error curve shown in Fig~\ref{fig:model_error}  indicates that the error increases with time. This means that the predictability of all models decreases as time goes, although the performances of the forward (standard) model and the additional noise-robust models differ greatly in terms of their attenuation behaviors. The three noise-robust models show similar behavior in the case of low $r$, as shown in Figs.~\ref{fig:SV_estimate}(a) and (b), and they display increasingly different behaviors as $r$ becomes large, as illustrated in Figs.~\ref{fig:SV_estimate}(c) to (f). In particular, the total-least-squares model with large $r$ has a large amplification factor as shown in Figs.~\ref{fig:SV_estimate}(c) and (d). Figure \ref{fig:eigenvalues}(c) illustrates that the eigenvalues of $\mathbf{A}_\mathrm{tls}$ are scattered around the unit circle. The results demonstrate that $\mathbf{A}_\mathrm{tls}$ includes unstable eigenvalues and the estimated temporal coefficients of POD modes by the total-least-squares method are likely to diverge. Meanwhile, the eigenvalues of $\mathbf{A}_\mathrm{fb}$ and $\mathbf{A}_\mathrm{ode}$ are mostly located on the unit circle as shown in Figs.~\ref{fig:eigenvalues}(a) and (b). Nevertheless, in some cases, the magnitudes of a few eigenvalues of $\mathbf{A}_\mathrm{fb}$ are greater or much less than unity, as shown in Fig.~\ref{fig:eigenvalues}(c). Besides, the arguments of the eigenvalues indicate that the total-least-squares and forward-backward models produce higher-frequency oscillations than the forward (standard) and ODE-based models. The features of the eigenvalue spectra are qualitatively consistent with the results obtained in the previous works \citep{kutz2016dynamic}.

Figure~\ref{fig:model_error} shows the averaged estimation error for each model with different $r$. The error of the forward (standard) model is smaller for all $r$ conditions shown here. The errors of three noise robust models works similarly for $r=10$. The error of the total-least-squares model is larger than the other two models for $r=100$, and finally the errors of the forward-backward and total-least squares models are very large from the beginning and cannot be displayed for $r=1000$. This illustrates that the ODE-based model works most robustly among three noise robust models.
{Here, the larger errors in the noise robust methods even without unstable modes (the ODE-based model) than the forward (standard) method are considered to be caused by the neutral oscillation with totally different phases. Once the phase is estimate to be different, the error finally becomes the $\sqrt{2}$ times larger than that of the forward (standard) model. This is because the error of the forward (standard) model is converged to the root-mean-squares of the true temporal coefficients of POD modes as discussed in Sec.~\ref{sec:evaluationmethod}, while the error for the noise-robust models is estimated to be the sum of the uncorrelated temporal coefficients (estimated and true) of POD modes of almost the same standard deviations (due to the neutral oscillation of the model in addition to the true temporal coefficient) as each other. In addition, the quicker increase in the error of noise robust methods might be explained as follows: the error caused by the phase difference of the neutral oscillation of the temporal coefficients in the noise-robust methods increases more quickly than the error caused by the amplitude difference of the temporal coefficients in the forward (standard) method.}

The initial error of all the models ($y$ intercepts of error curves) increases when $r$ increases, as shown in Fig.~\ref{fig:model_error}. This might be explained as follows:  If the number of modes increases, the high-order modes are accounted for the model, and the discrepancy in the high order modes between model and data are enhanced even immediately after one step. This might be because the high-order mode has relatively smaller amplitude and their temporal coefficients of POD modes are inaccurately estimated by relatively strong observation noises. This leads to the increase in $y$ intercepts of error curves. Similar discussion is given in Sec.~\ref{subsec:dis}

\begin{figure}[h]
\captionsetup{justification=raggedright}
    \subfigure[forward (standard) model]{\includegraphics[width=60mm]{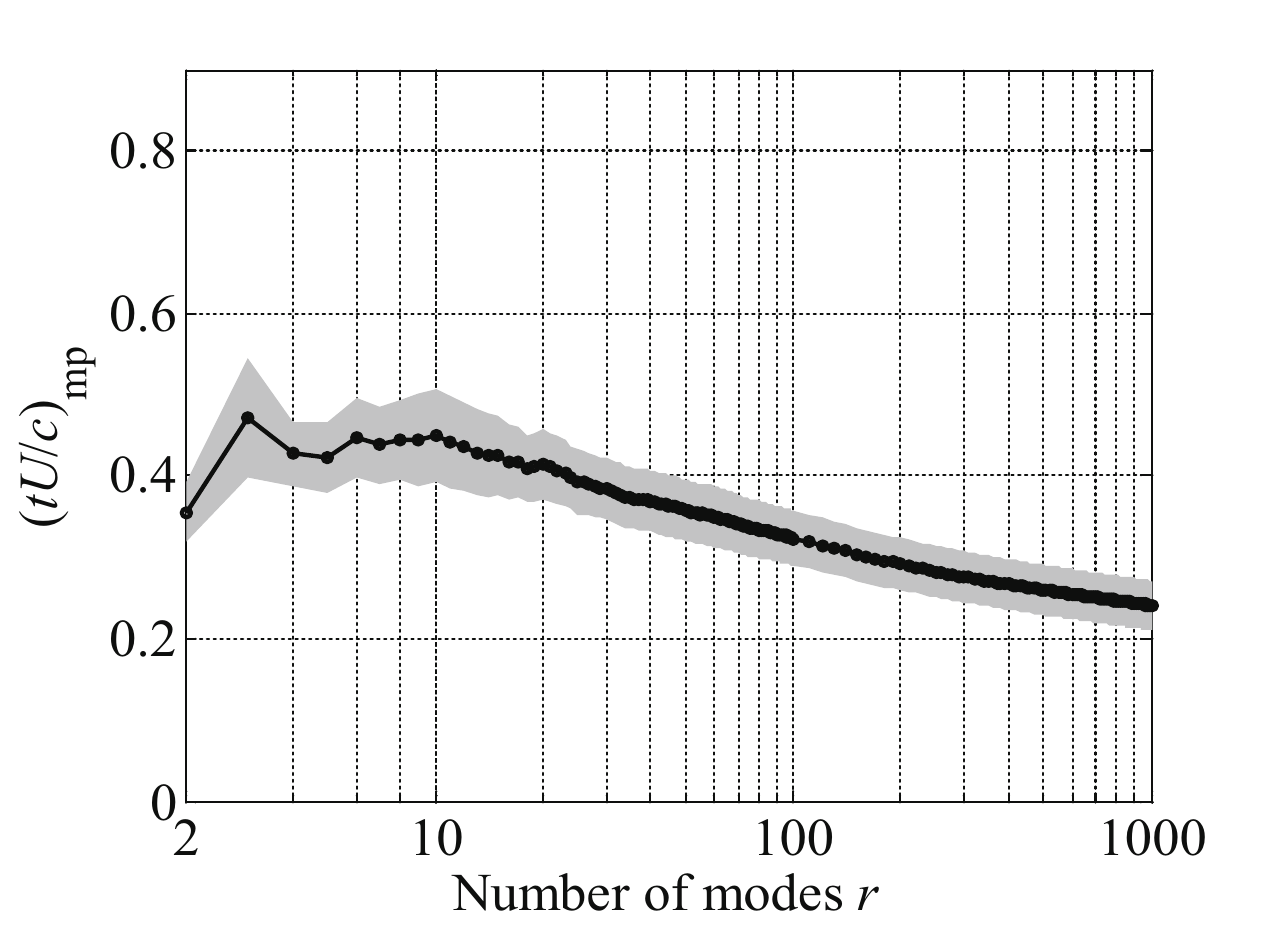}}
    \subfigure[forward-backward model]{\includegraphics[width=60mm]{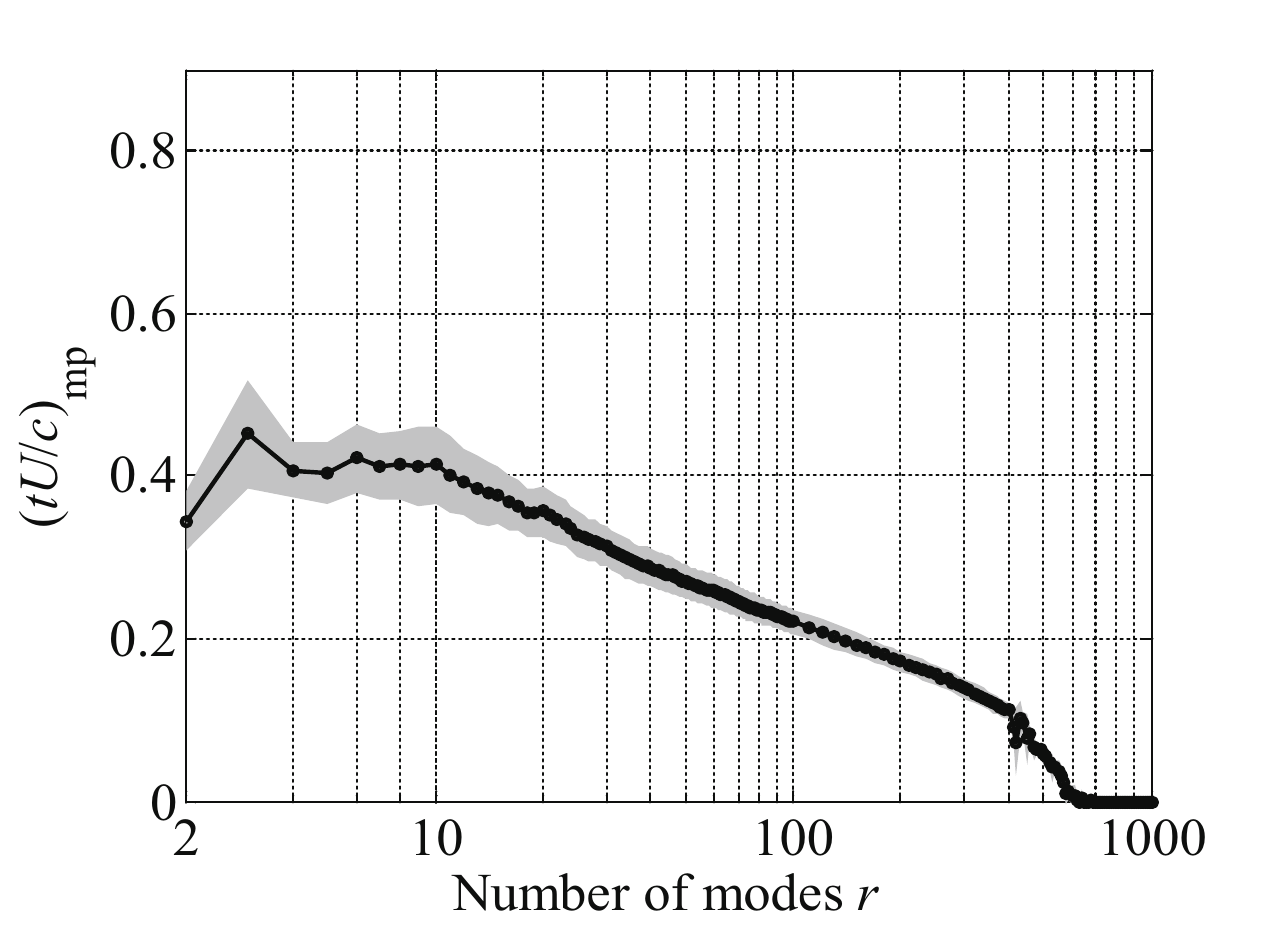}}
    \subfigure[total-least-squares model]{\includegraphics[width=60mm]{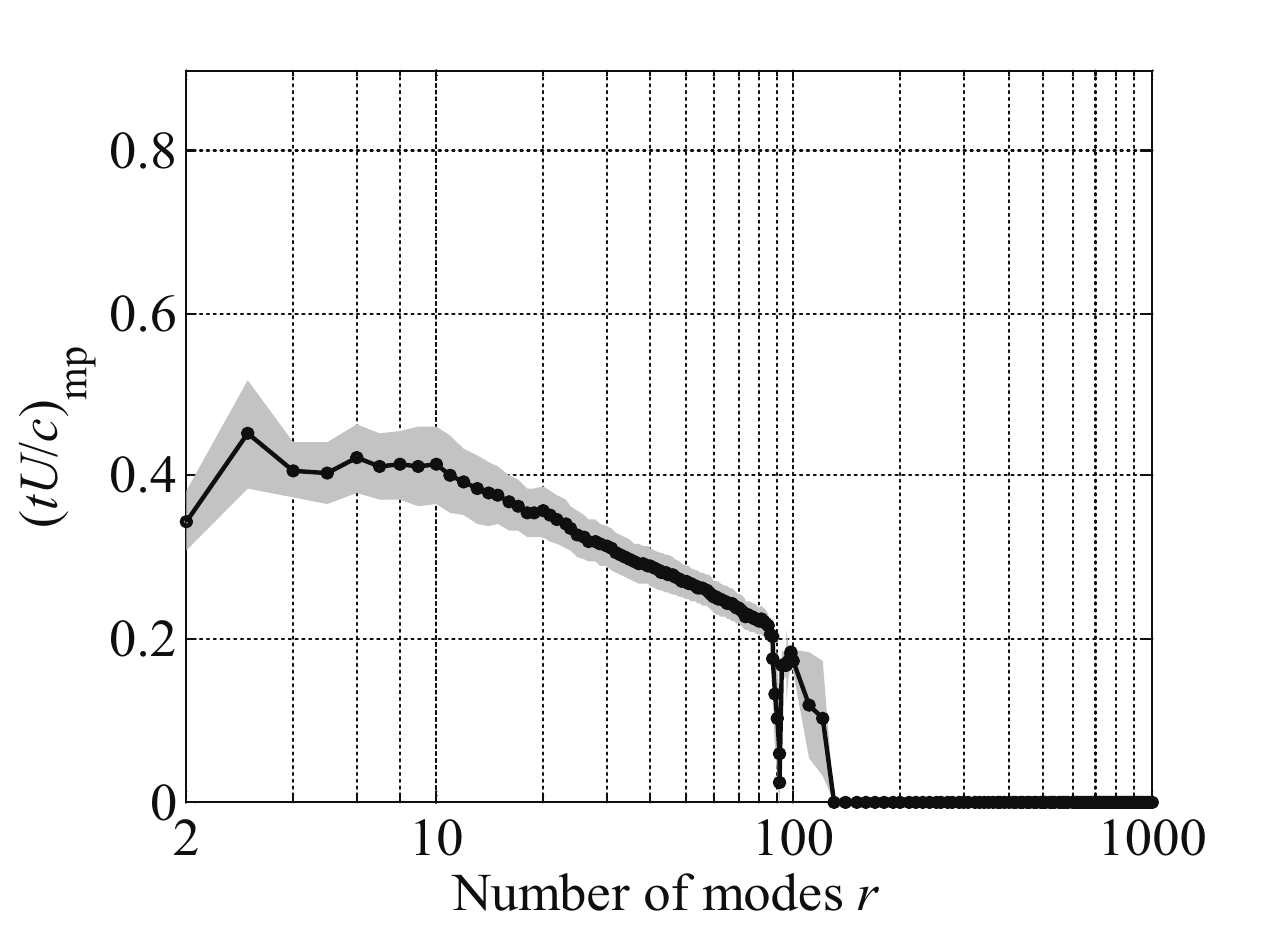}}
    \subfigure[ODE-based model]{\includegraphics[width=60mm]{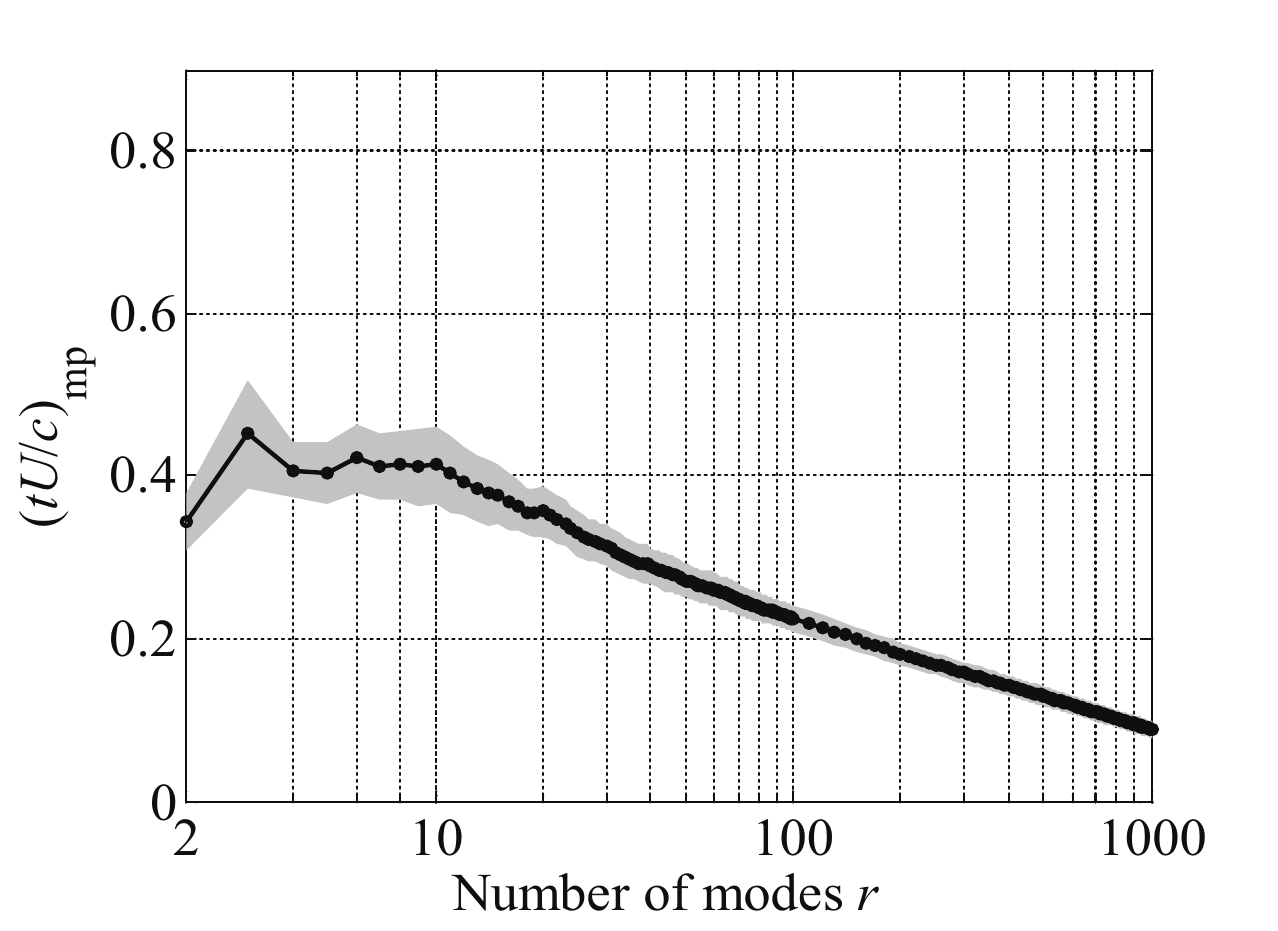}}
    \caption{cross-validation results of effects of $r$ on the model predictability. Shaded region corresponds to standard deviation estimated by cross-validation.}
    \label{fig:tperm_model_modes}
\end{figure}

\begin{figure}[h]
\captionsetup{justification=raggedright}
    \centering
    \includegraphics[width=60mm]{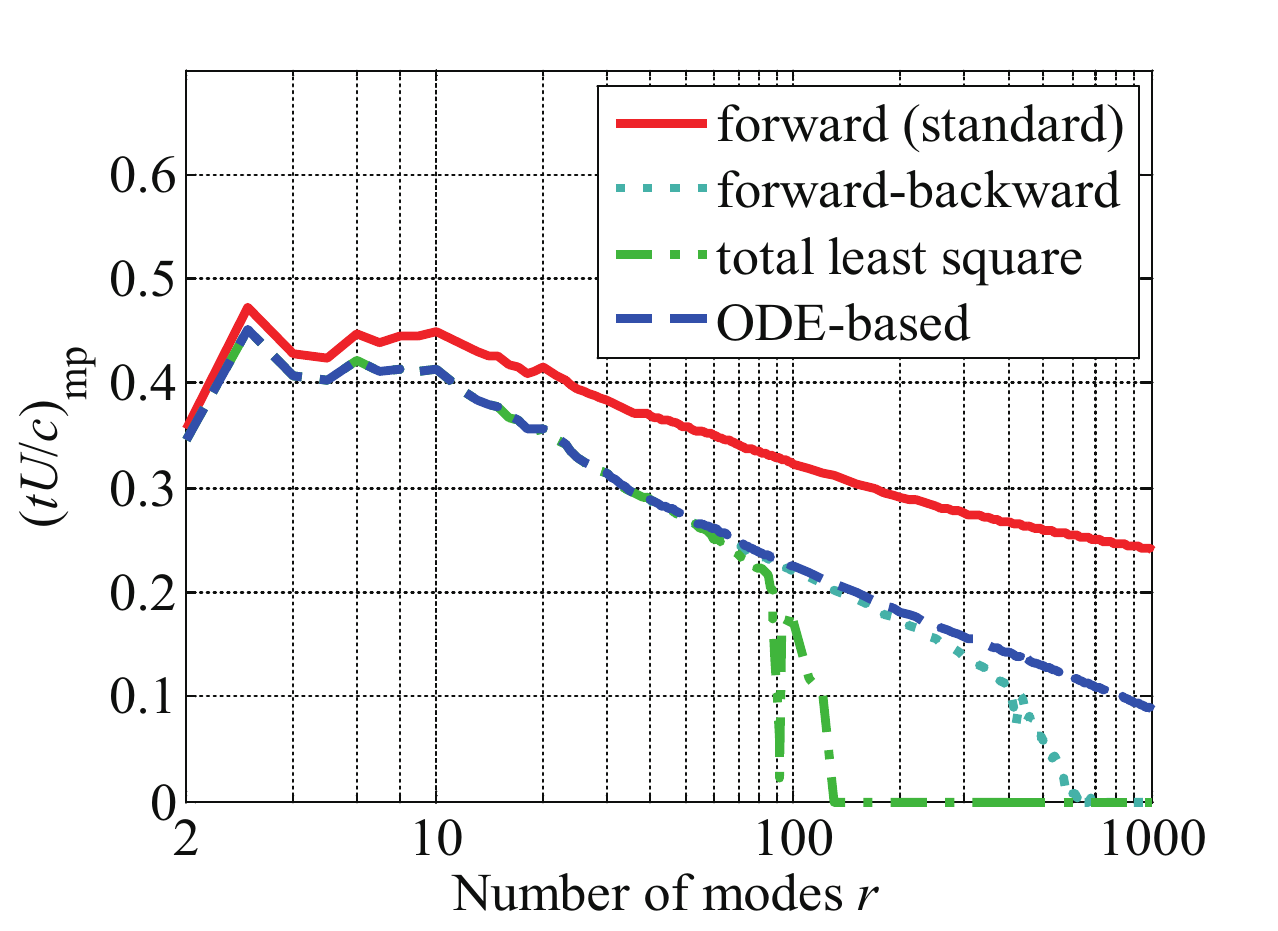}
    \caption{cross-validation results of effect of $r$ on the model predictability.}
    \label{fig:tperm_model_modes_allmodel}
\end{figure}

The effect of $r$ on the model prediction permissive time is shown in Figs.  \ref{fig:tperm_model_modes} and \ref{fig:tperm_model_modes_allmodel}. In the present paper, $r$ was changed up to one-thousand, which corresponds to 10\% of the total number of POD modes. The shaded region represents the standard deviation calculated from $k=10$ times iterations of the testing for the cross-validation in Fig.~\ref{fig:tperm_model_modes}. Figures~\ref{fig:tperm_model_modes_allmodel} shows that the predictability does not change monotonically with respect to $r$ and reaches a maximum at a specific value in a small-$r$ range. All the models have the first, second and third peaks at around $r=2$, $r=6$ and $r \approx 10$. It is noteworthy that the model shows high performance with small $r$ and increasing the POD-mode numbers of greater than 10 is not effective for improving the model performance. This result implies that the present linear model has difficulty expressing the complex wake dynamics which is considered to be represented by the high-order POD modes. The practical mode number should be chosen in the condition of $r\le10$ considering the trade-off of the model predictability and complexity of the model. 

The three noise-robust models show similar predictability in the low $r$ region (approximately $r<30$, corresponding to more than 70\% of the total energy) with differences in their predictability gradually increasing as $r$ increases as described in Fig.~\ref{fig:tperm_model_modes_allmodel}. The estimation results and the eigenvalue distributions shown in Figs.~\ref{fig:SV_estimate} and \ref{fig:eigenvalues} demonstrate that the total-least-squares model is likely to diverge because of its high amplification factors than unity, and its predictability was very low at large $r$. Furthermore, the predictability of the forward-backward model was lower than that of the ODE-based model at large $r$. This is considered to be because the eigenvalue distribution of $\mathbf{A}_\mathrm{fb}$ is more unstable than that of $\mathbf{A}_\mathrm{ode}$. The eigenvalues of the ODE-based model stably lie on the unit circle, and the predictability does not significantly drop even with increasing $r$. These results illustrate that the ODE-based model has the best performance in the noise-robust models. However, the predictability of the ODE-based model is lower than that of the forward (standard) model as shown in Fig.~\ref{fig:tperm_model_modes_allmodel}. This might be again because of difference between damping modes estimated by the forward (standard) model and neutral modes estimated by the ODE-based models. In conclusion, the forward (standard) model shows the best predictability of the present linear models. On the other hand, in terms of the attenuation of the model, i.e., the magnitudes of the eigenvalues of the coefficient matrices, the ODE-based model works the best. In addition, Fig.~\ref{fig:tperm_model_modes_allmodel} also demonstrates that the ODE-based model shows the best performance under the condition of the same $r$ among the three noise-robust  models.


\clearpage

\subsection{DMDsp-based Reduced-order Model}
Then, the results of DMDsp-based reduced-order model is considered. Here, $r_\textrm{POD}=100$ is chosen for the computation of the DMD modes. Therefore, the eigenvalues of all the DMD modes obtained are exactly corresponds to those obtained in the POD-based linear reduced-order model with $r_\textrm{POD}=100$, which are shown in Fig. \ref{fig:eigenvalues}(b). The sparse DMD modes are selected from the all modes calculated by the DMDsp algorithm. The DMDsp-based reduced-order models for on standard (forward) and ODE-based DMD modes are only presented in the present section because the forward (standard) method works the best in all the DMD implementations and the ODE-based DMD works the best in all the robust DMD implementations similar to discussion in Sec.~\ref{subsec:pod}. {It should be noted that $\mathbf{A}$ matrices and corresponding DMD modes are different for DMDsp-based reduced-order models using standard (forward) and ODE-based methods, even before the DMDsp mode selection.} Here, the weight $\gamma$ of regularization term in Eq.~\ref{eq:DMDsp} is changed from $10^{-3}$ to $10^6$.

\begin{figure}[h]
\captionsetup{justification=raggedright}
    \centering
    \includegraphics[width=70mm]{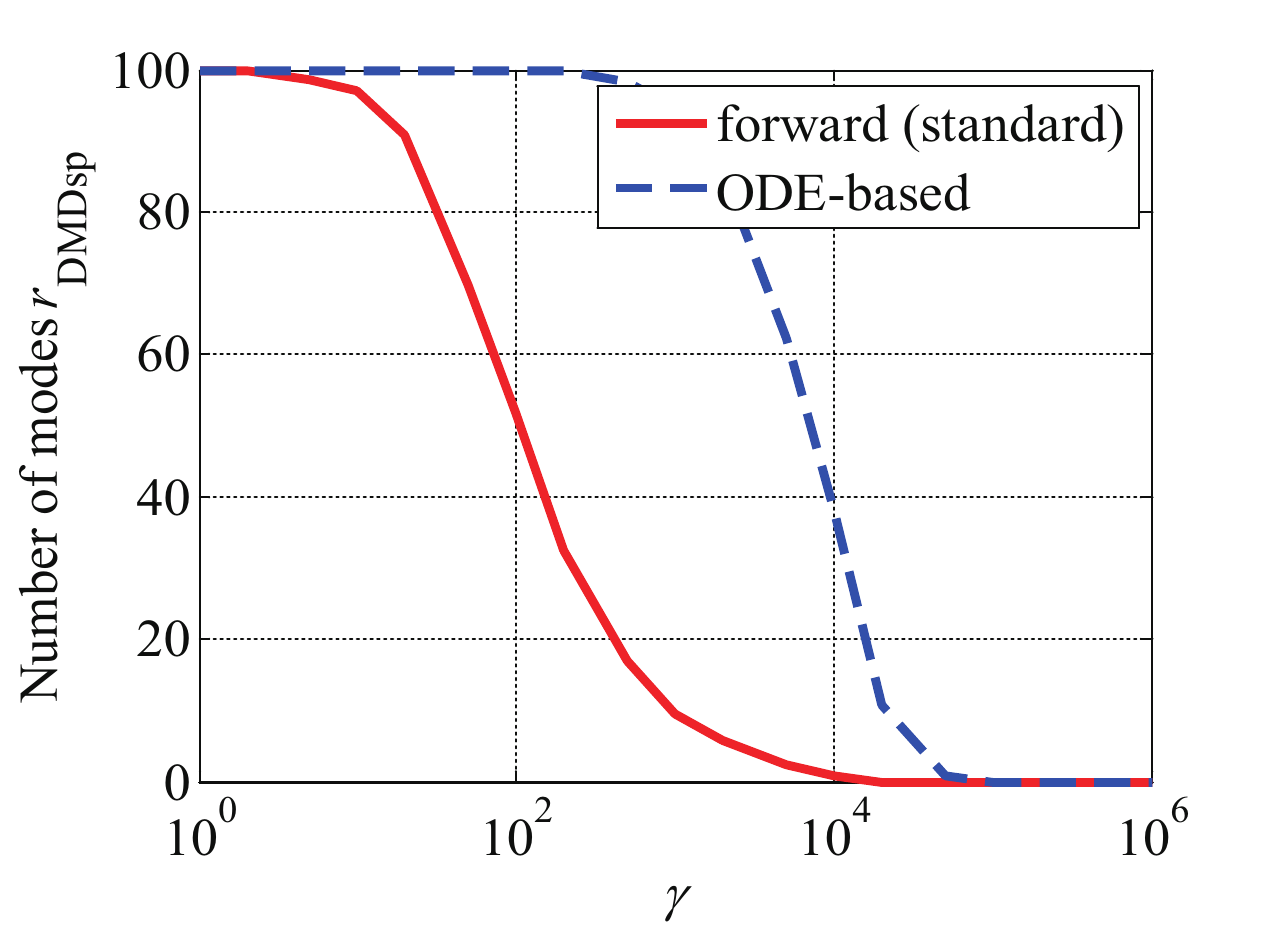}
    \caption{cross-validation results of effects of $\gamma$ on the predictability of the model}
    \label{fig:gamma_r}
\end{figure}

First, Fig. \ref{fig:gamma_r} shows a number of the selected modes with changing the weight $\gamma$ of a sparsity promoting (second) term in Eq.~\ref{eq:DMDsp}. Here, the numbers of selected modes by DMDsp decrease for both forward (standard) and ODE-based methods. The number of selected modes for the DMD modes constructed by the forward (standard) method decreases with smaller $\gamma$ than that by the ODE-based method. This might be because more damping modes are generated by the forward (standard) method than those by the ODE-based method. The damping modes could be discarded by DMDsp with the smaller regularization term coefficient $\gamma$ because the damping modes only contribute to the reconstruction of the early stage of the time-series data and does not play important role for the entire time-series data. Figure \ref{fig:gamma_r} shows that the slopes of curves of the ODE-based method is steeper than forward method and it implicates that the ODE-based method requires the fine tuning to obtain the favorable number of DMD modes for users.

\begin{figure}[h]
\captionsetup{justification=raggedright}
    \subfigure[forward (standard) with $\gamma=100$   ]{\includegraphics[width=60mm]{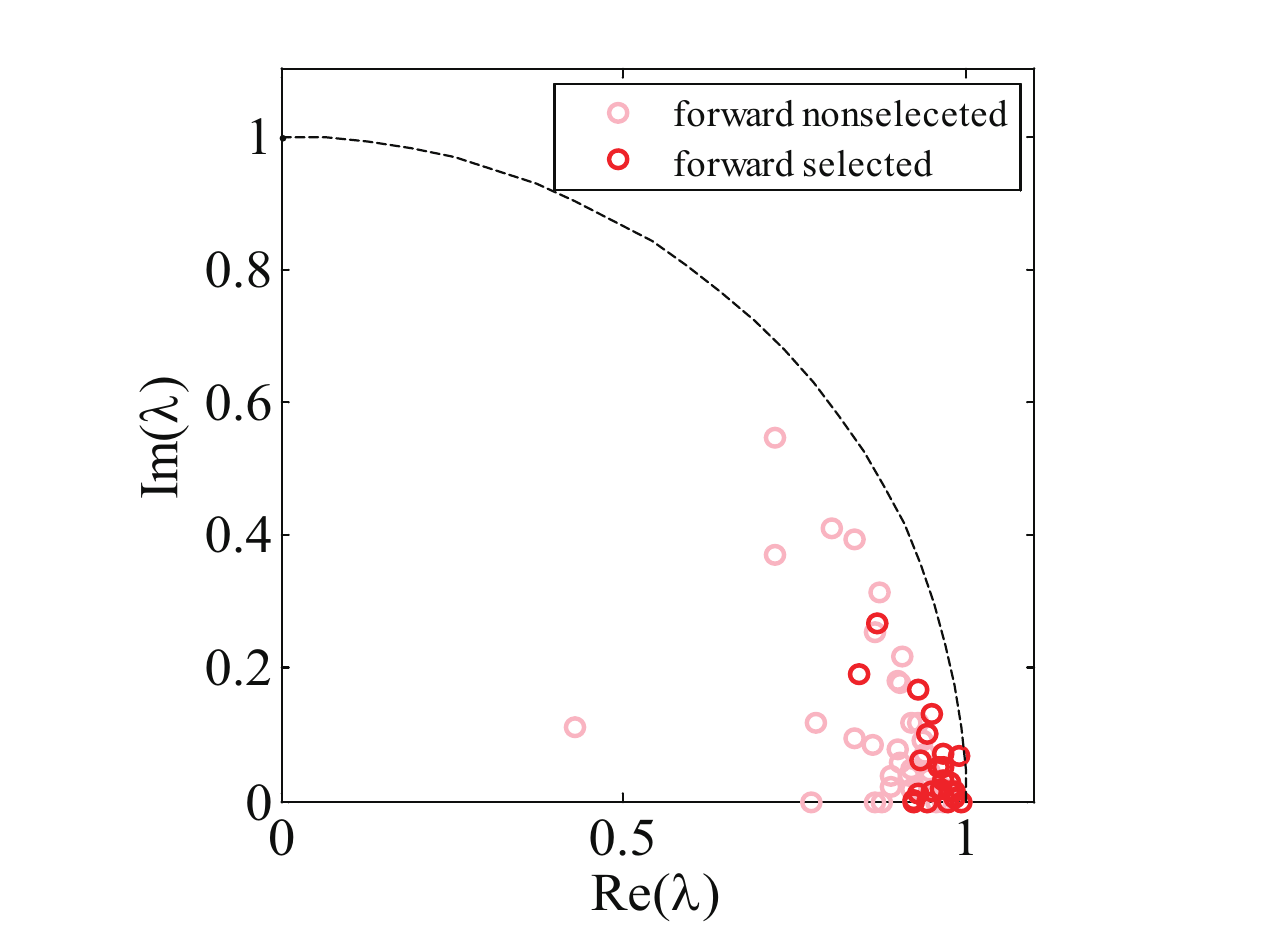}}
    \subfigure[forward (standard) with $\gamma=1,000$ ]{\includegraphics[width=60mm]{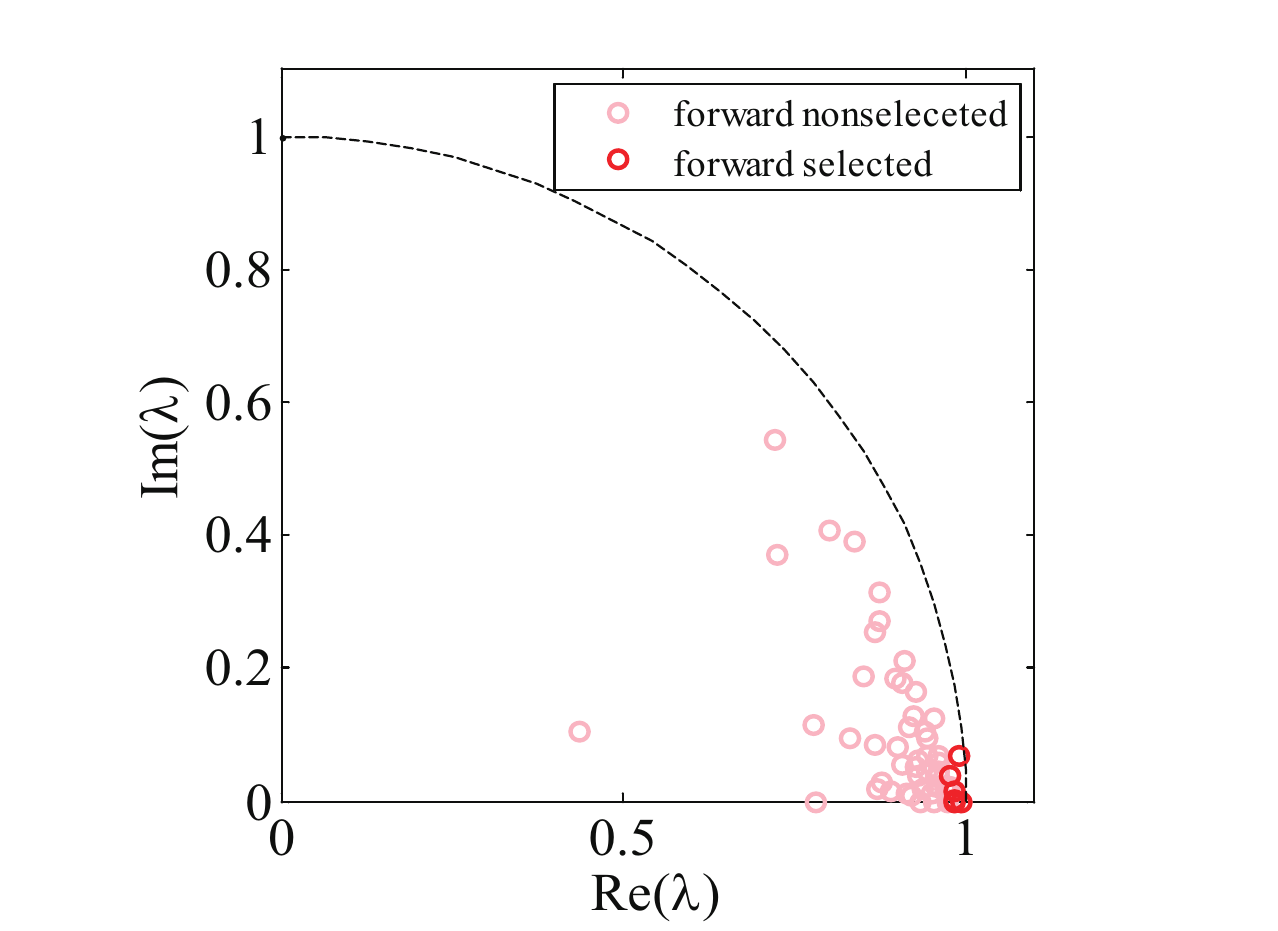}}\\
    \subfigure[         ODE-based with $\gamma=1,000$ ]{\includegraphics[width=60mm]{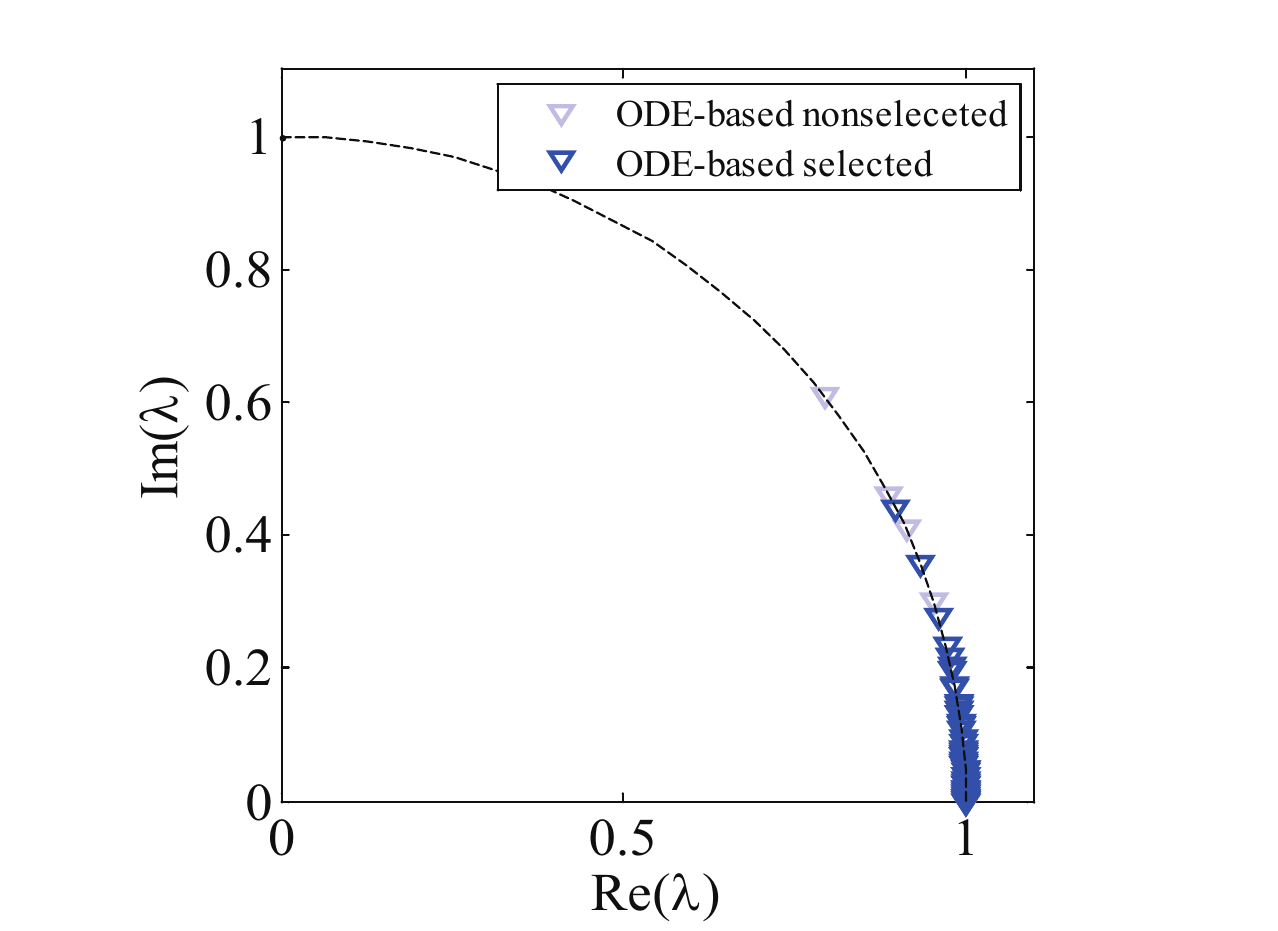}}
    \subfigure[         ODE-based with $\gamma=10,000$]{\includegraphics[width=60mm]{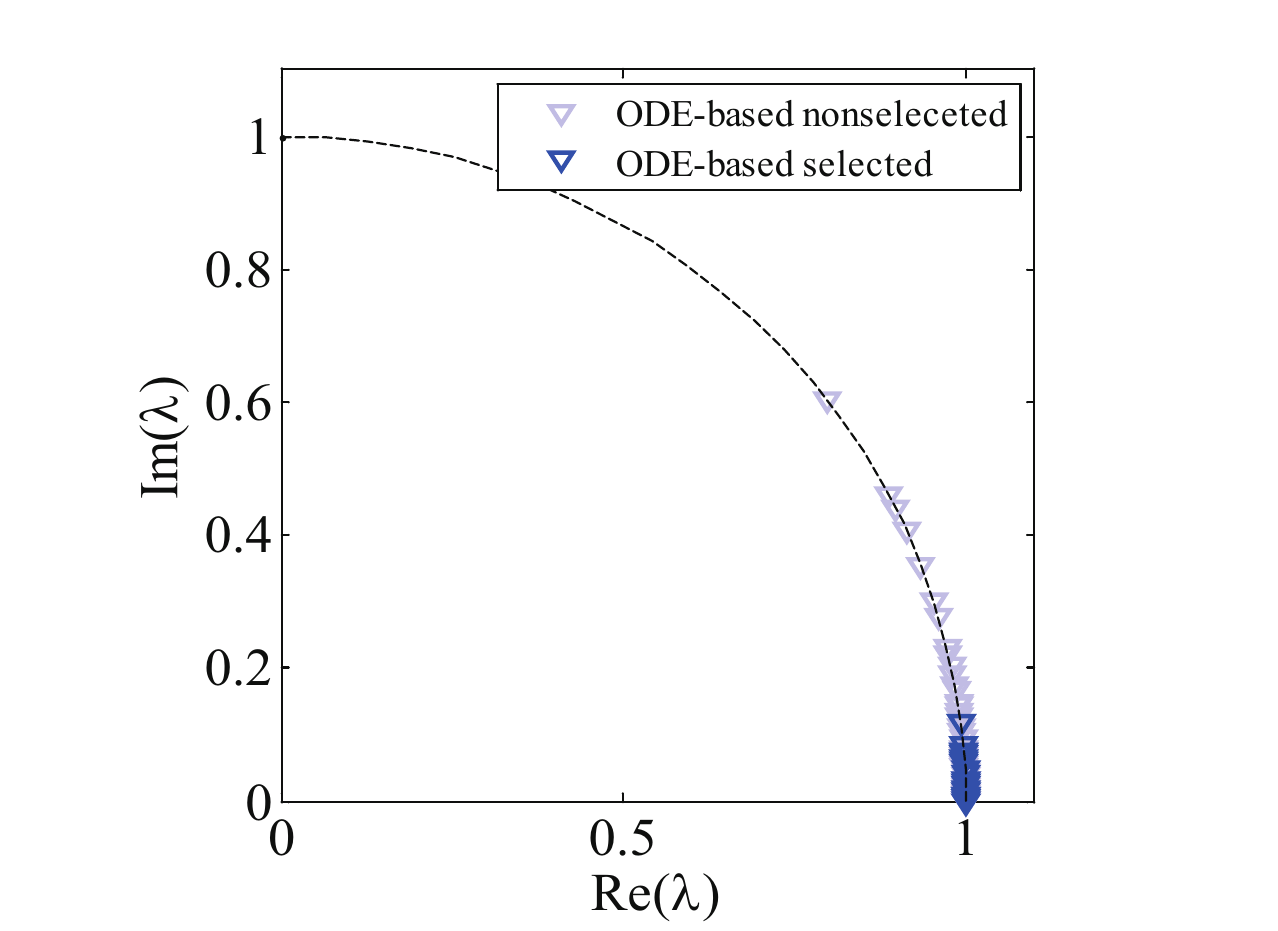}}\\
    \caption{The eigenvlaues of the selected and nonselected DMD modes of forward (standard) and ODE-based methods by DMDsp with different $\gamma$.}
    \label{fig:sel_eigenvalues}
\end{figure}

\begin{figure}[h]
\captionsetup{justification=raggedright}
    \subfigure[streamwise velocity field  ]{\includegraphics[width=110mm]{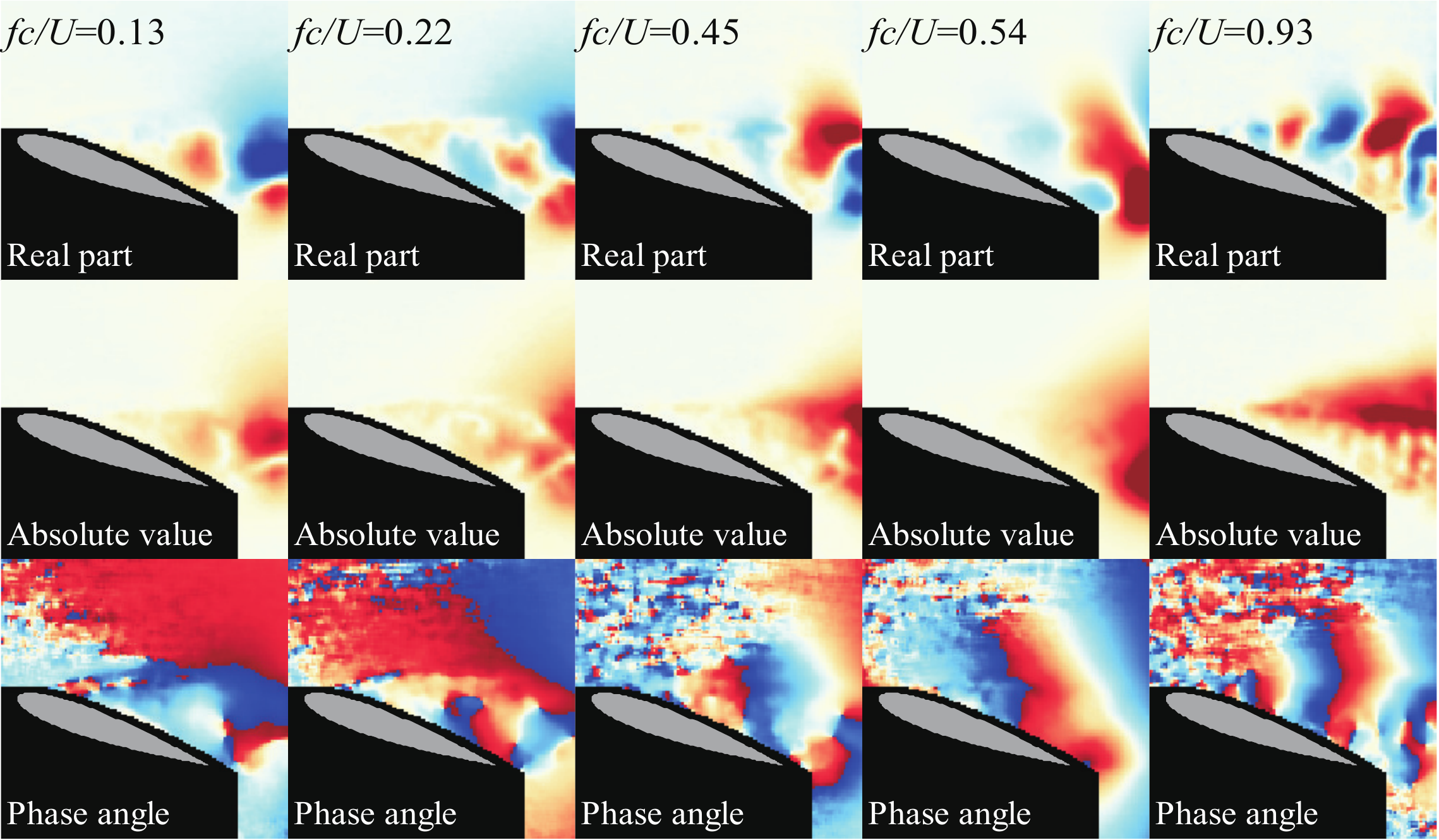}}
    \subfigure[transverse velocity field ]{\includegraphics[width=110mm]{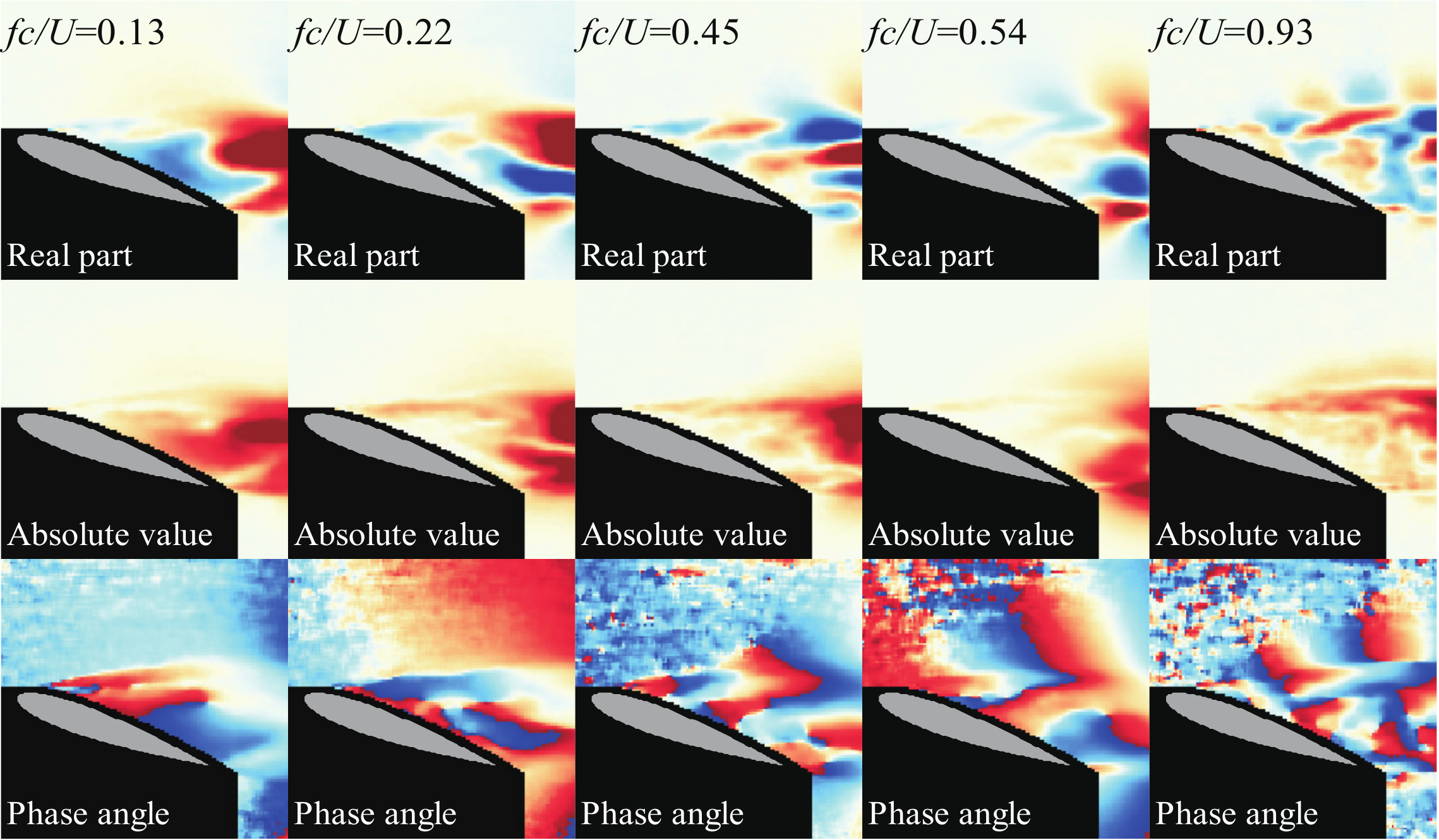}}
    \caption{Corresponding distributions of the selected DMD modes with $\gamma=100$ of the forward (standard) model.}
    \label{fig:dmdmodes}
\end{figure}

\begin{figure}[h]
\captionsetup{justification=raggedright}
    \subfigure[forward (standard) DMDsp with $\gamma=100$, POD mode 1]{\includegraphics[width=60mm]{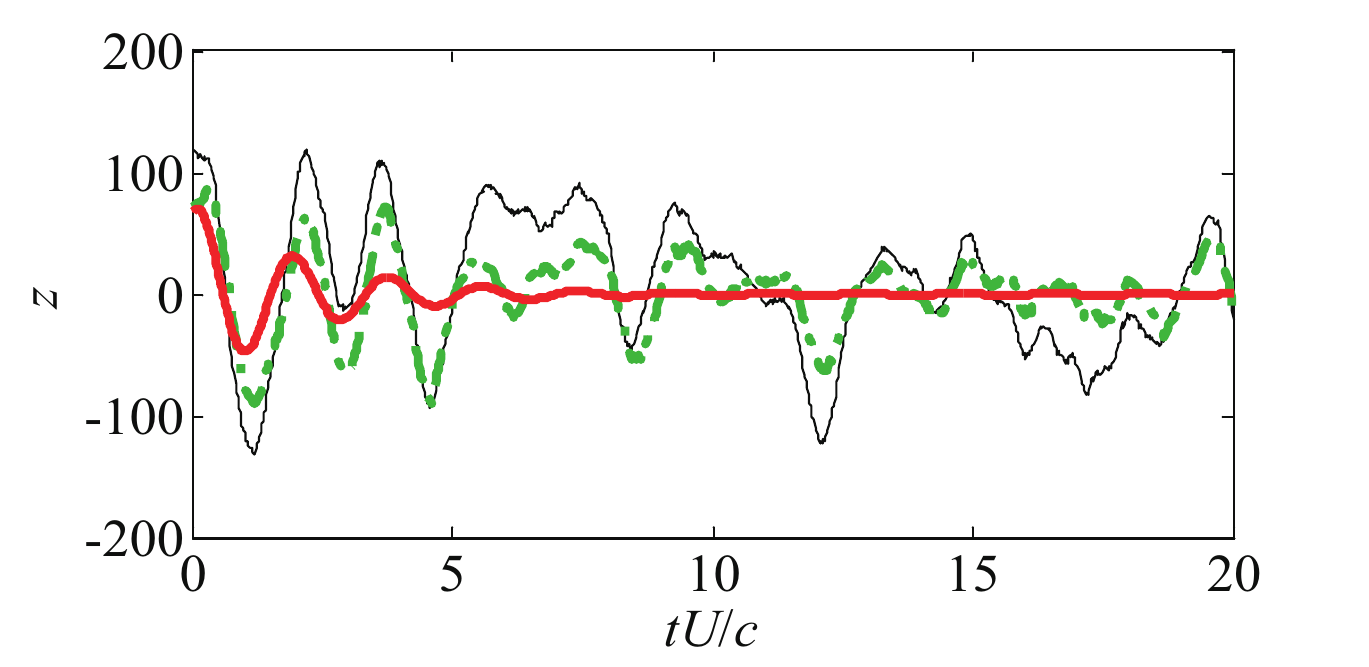}}
    \subfigure[forward (standard) DMDsp with $\gamma=100$, POD mode 2]{\includegraphics[width=60mm]{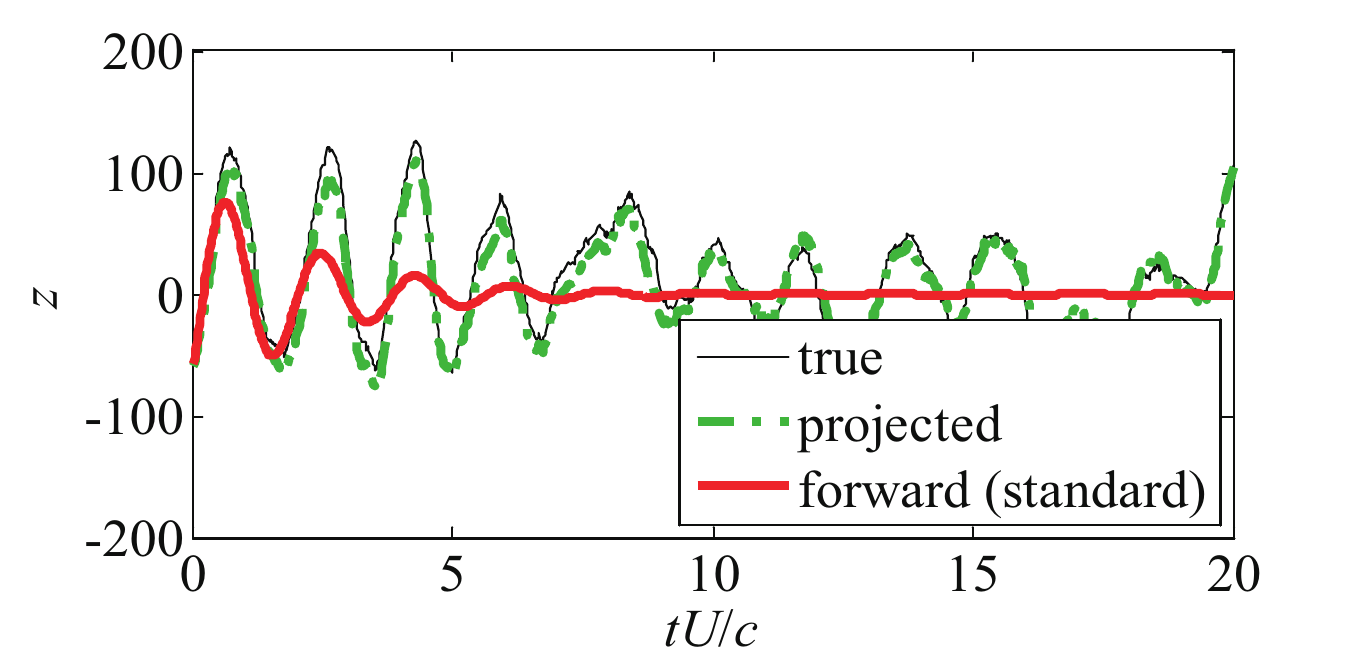}}
    \subfigure[ODE-based DMDsp with $\gamma=1000$, POD mode 1]{\includegraphics[width=60mm]{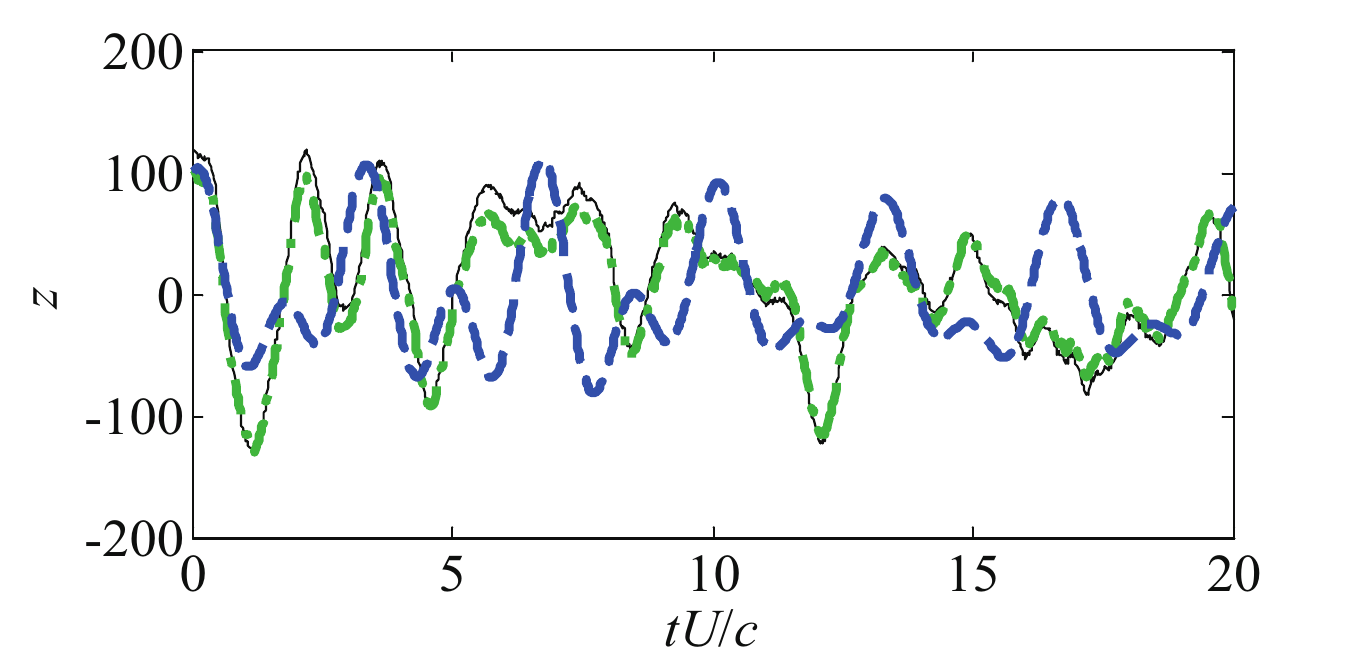}}
    \subfigure[ODE-based DMDsp with $\gamma=1000$,  POD mode 2]{\includegraphics[width=60mm]{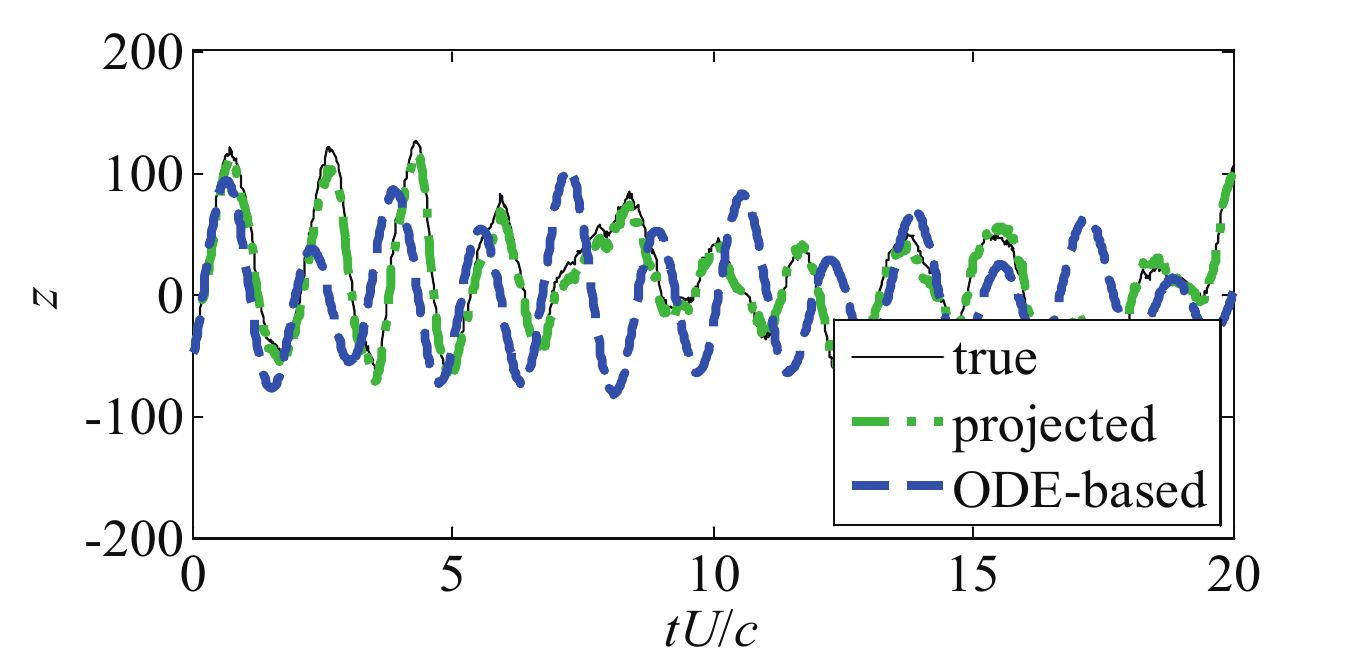}}
    \caption{Histories of estimated temporal coefficients of POD modes using the DMDsp-based reduced-order model. Results of the forward (standard) and ODE-based models are shown with the true POD mode amplitude and that projected to DMDsp subspace. }
    \label{fig:SV_estimates_DMDsp}
\end{figure}

Then, Fig. \ref{fig:sel_eigenvalues} illustrates the selected DMD eigenvalues. Here, $\gamma$ is set to be  100 and 1000 for the forward (standard) method and is set to be 1000 and 10,000 for the ODE-based method. Figure~\ref{fig:sel_eigenvalues} implicates that the lower frequency eigenvalues, that have smaller imaginary components, are chosen for the reconstruction. Figure~\ref{fig:dmdmodes} presents corresponding distributions of the DMD modes whereas the real part, the absolute value and the phase of complex value distribution of the modes are shown respectively. Similar to POD modes, ranges of the contours are set to be the same for all the mode, while the color bar is not shown because the DMD mode in the present study is normalized.
Figures~\ref{fig:SV_estimates_DMDsp} shows the estimated temporal coefficients of the POD modes using the DMDsp-based reduced-order model. {Here, the POD mode amplitudes projected to the DMDsp subspace should be the reference quantities for the model evaluation, because they are only the predictable components for the DMDsp-based reduced-order models. Here, the DMD modes themselves are different for the forward (standard) and ODE-based methods, and therefore, the POD mode amplitudes projected to the DMDsp subspace are different for those two methods. The less damped modes seem to be selected for both DMDsp-based reduced-order models based on those two methods. However, the estimated temporal coefficients by the forward (standard) model damp in the early time similar to that by the POD-based linear reduced-order model because all the modes of the forward (standard) method are estimated to have damping eigenvalues. In addition, although the estimated temporal coefficients by the ODE-based method are not considered to spuriously damp according to the DMDsp mode selection, the estimated neutral oscillations become to have different phases from those of the true temporal coefficients and the error grows larger in early time, again similar to the POD-based liner reduced-order model. Therefore, the model predictability does not seem to significantly change from that of the POD-based linear reduced-order model.} It should be noted that the data are only reconstructed by the initial POD mode amplitude and the model predictability is discussed. The way to reconstruct is totally different from the standard way of DMDsp while DMDsp is only used for the model selection in the present study as noted in Sec.~\ref{sec:LROMDMDsp}.

\begin{figure}[h]
\captionsetup{justification=raggedright}
    \subfigure[forward (standard) DMDsp-based reduced-order model ]{\includegraphics[width=60mm]{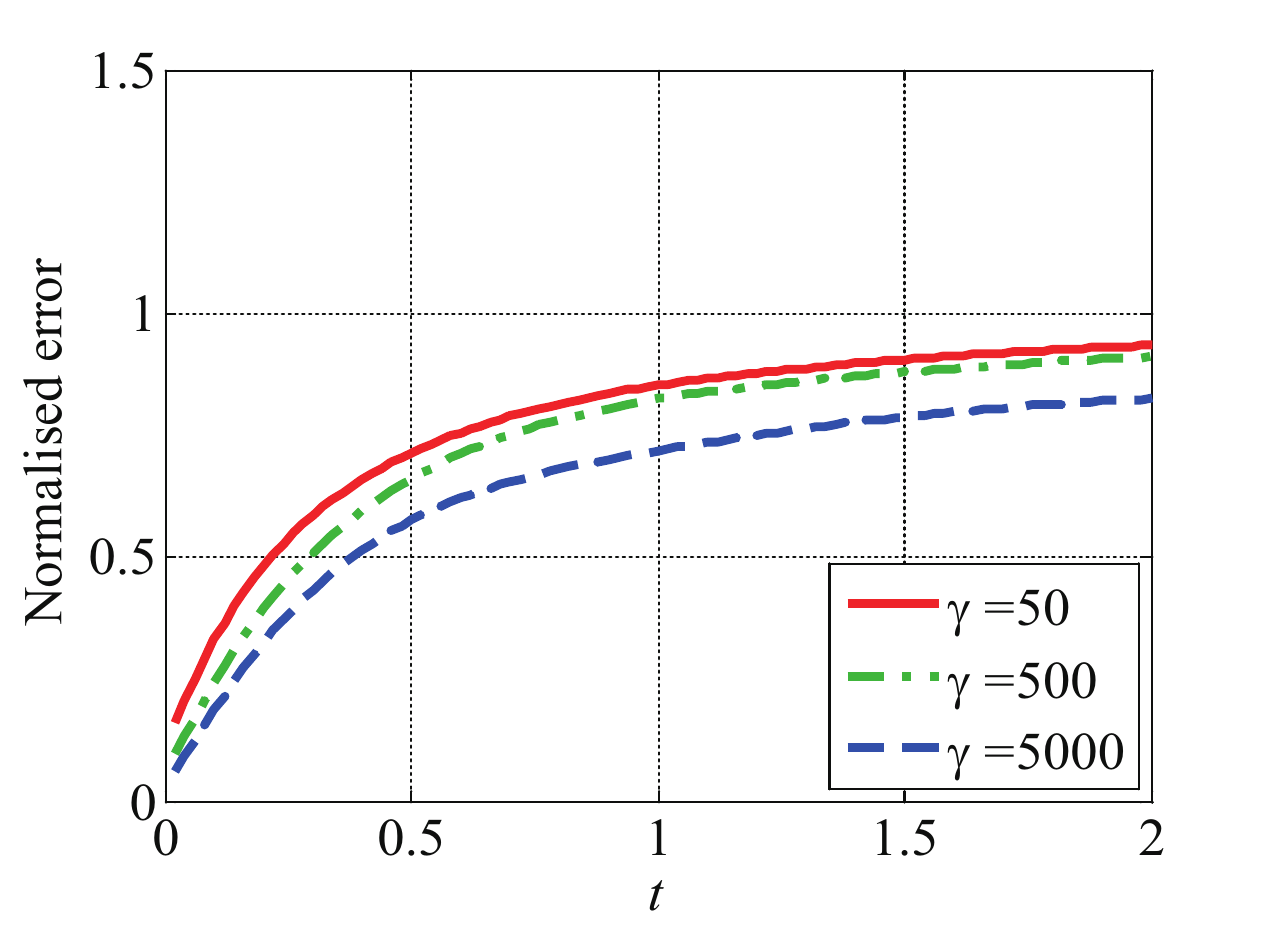}}
    \subfigure[ODE-based DMDsp-based reduced-order model]{\includegraphics[width=60mm]{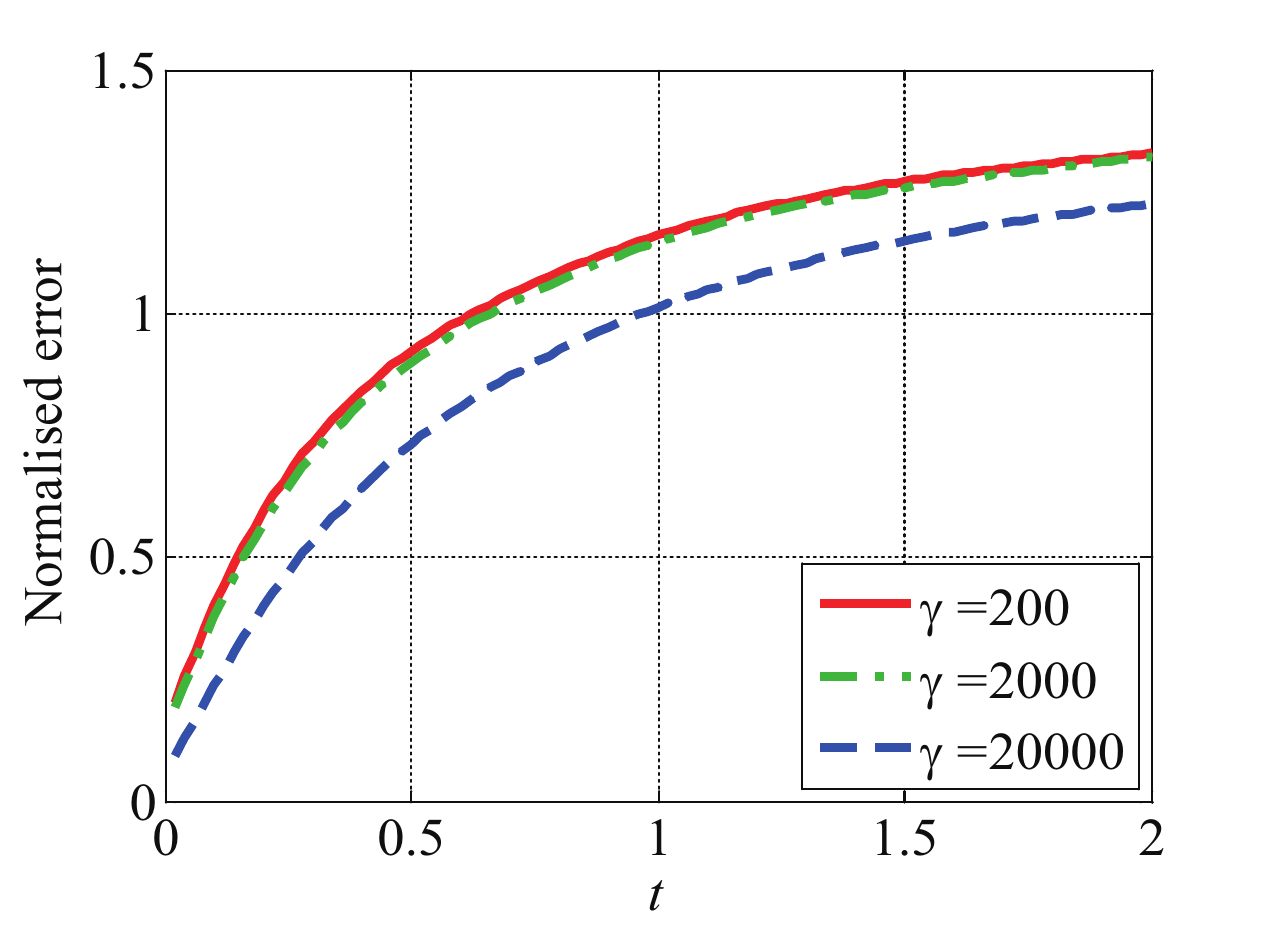}}
    \caption{Estimation error of the DMDsp-based reduced-order model of different $\gamma$.}
    \label{fig:model_error_DMDsp}
\end{figure}

\begin{figure}[h]
\captionsetup{justification=raggedright}
    \subfigure[forward (standard) DMDsp-based reduced-order model]{\includegraphics[width=60mm]{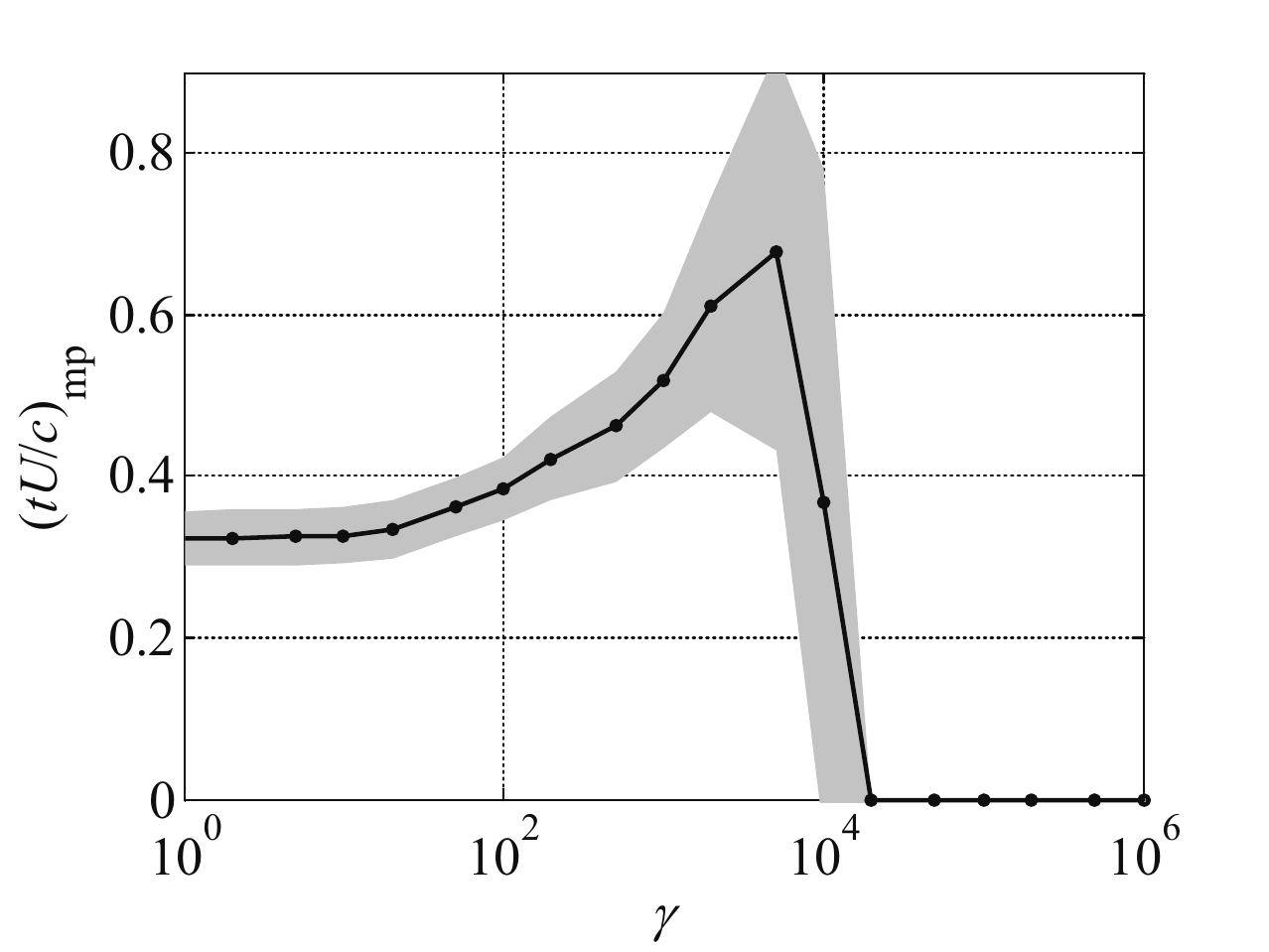}}
    \subfigure[ODE-based DMPsp model]{\includegraphics[width=60mm]{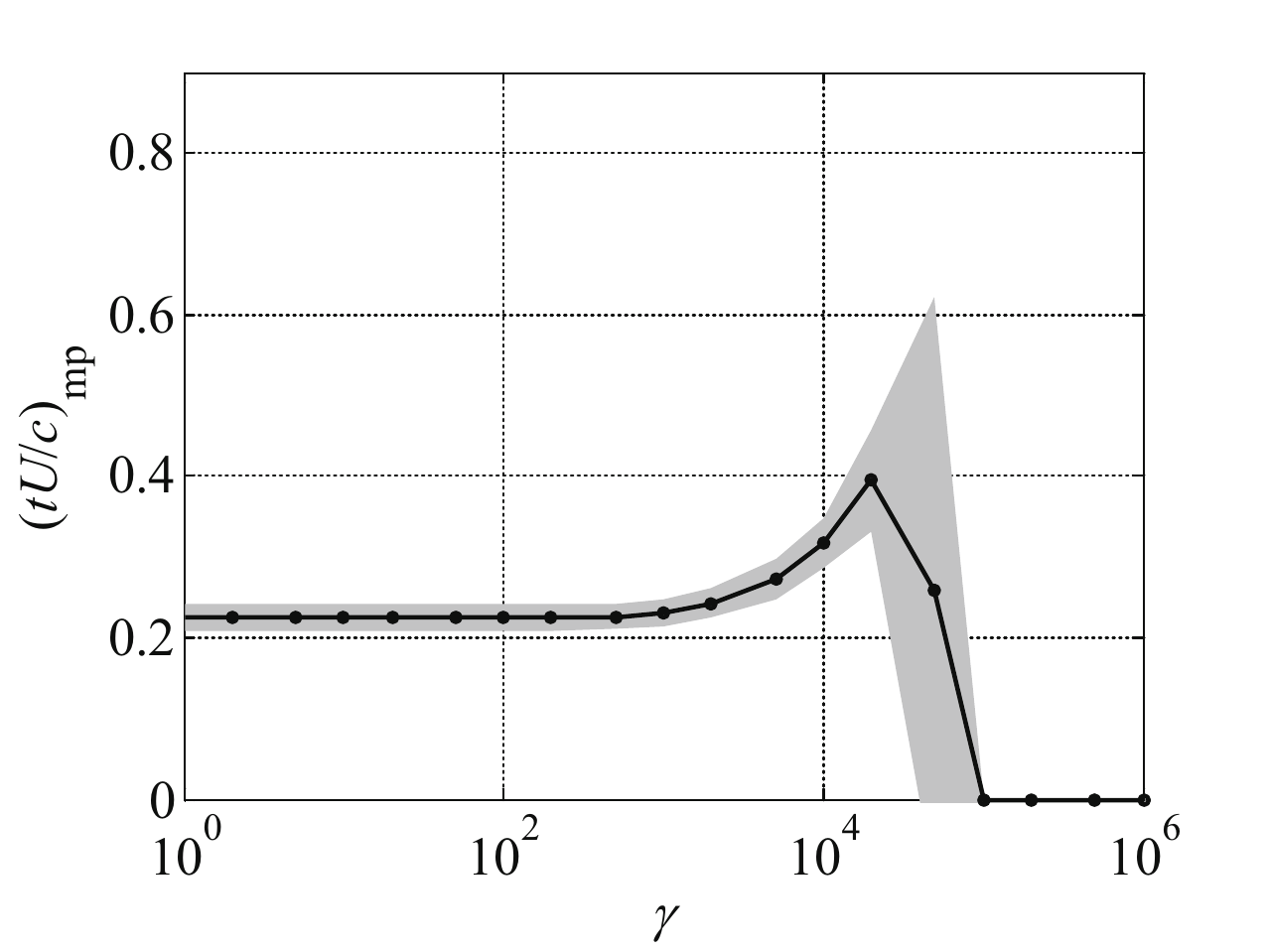}}
    \caption{cross-validation results of effects of $\gamma$ on the model predictability. Shaded regions show the standard deviation estimated by cross-validation.}
    \label{fig:tperm_model_gamma_modes}
\end{figure}

\begin{figure}[h]
\captionsetup{justification=raggedright}
    \centering
    \includegraphics[width=70mm]{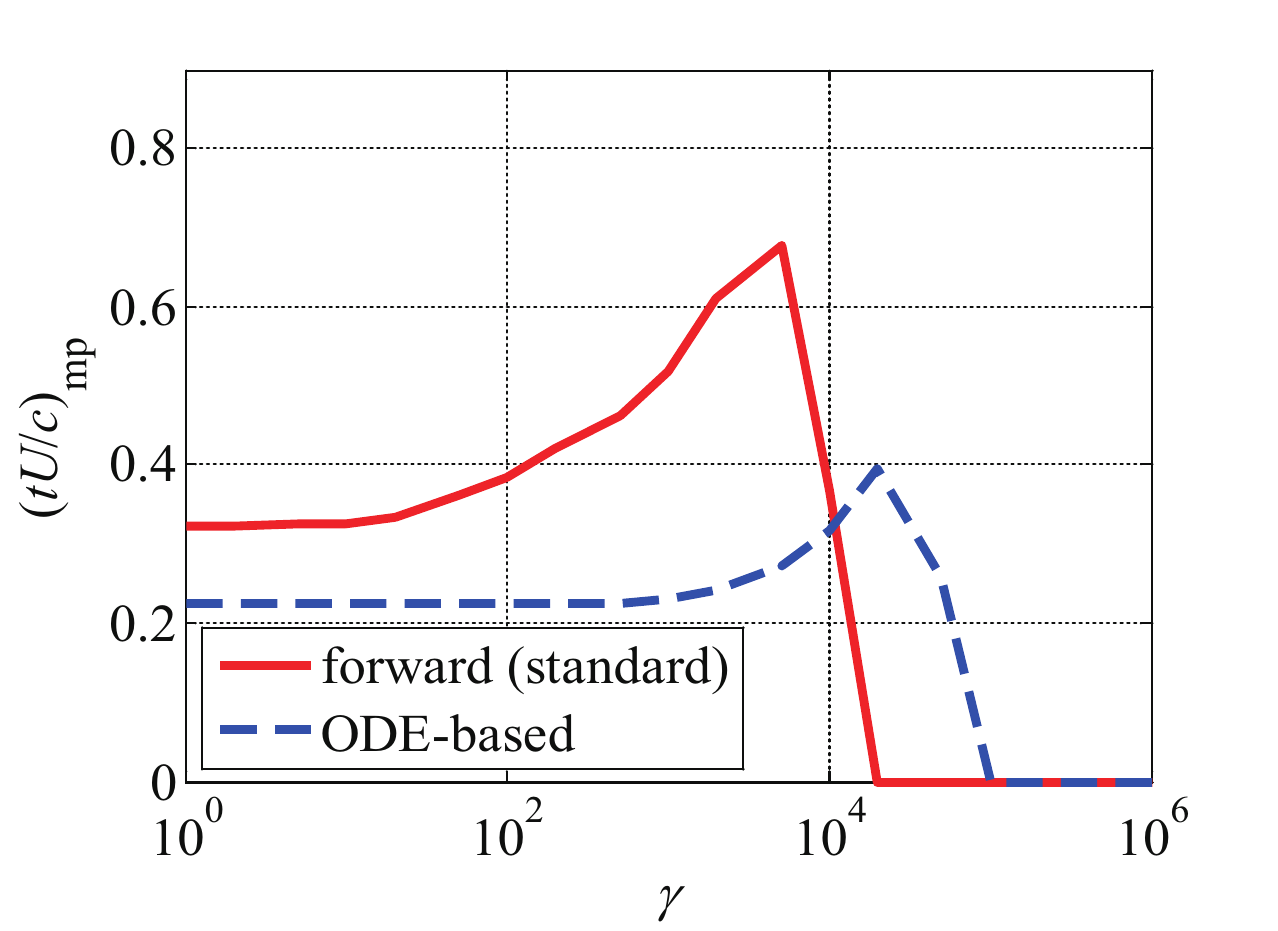}
    \caption{cross-validation results of effect of $\gamma$ on the model predictability in the DMDsp-based reduced-order model.}
    \label{fig:tperm_model_gamma_allmodel}
\end{figure}

\begin{figure}[h]
\captionsetup{justification=raggedright}
    \centering
    \includegraphics[width=70mm]{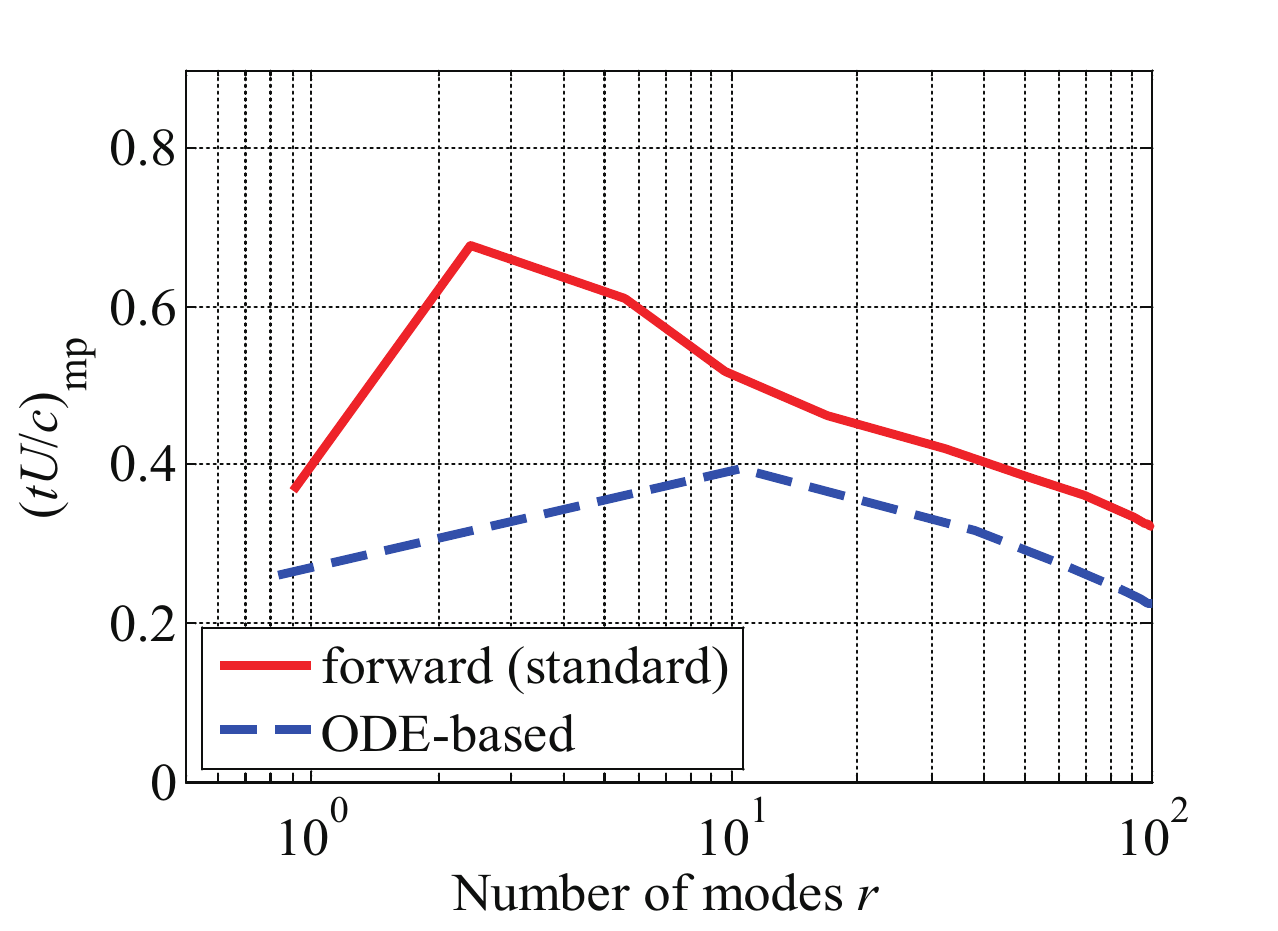}
    \caption{cross-validation results of effect of $\gamma$ on the model predictability in the DMDsp-based reduced-order model.}
    \label{fig:tperm_model_r_allmodel}
\end{figure}

\begin{figure}[h]
\captionsetup{justification=raggedright}
    \centering
    \includegraphics[width=70mm]{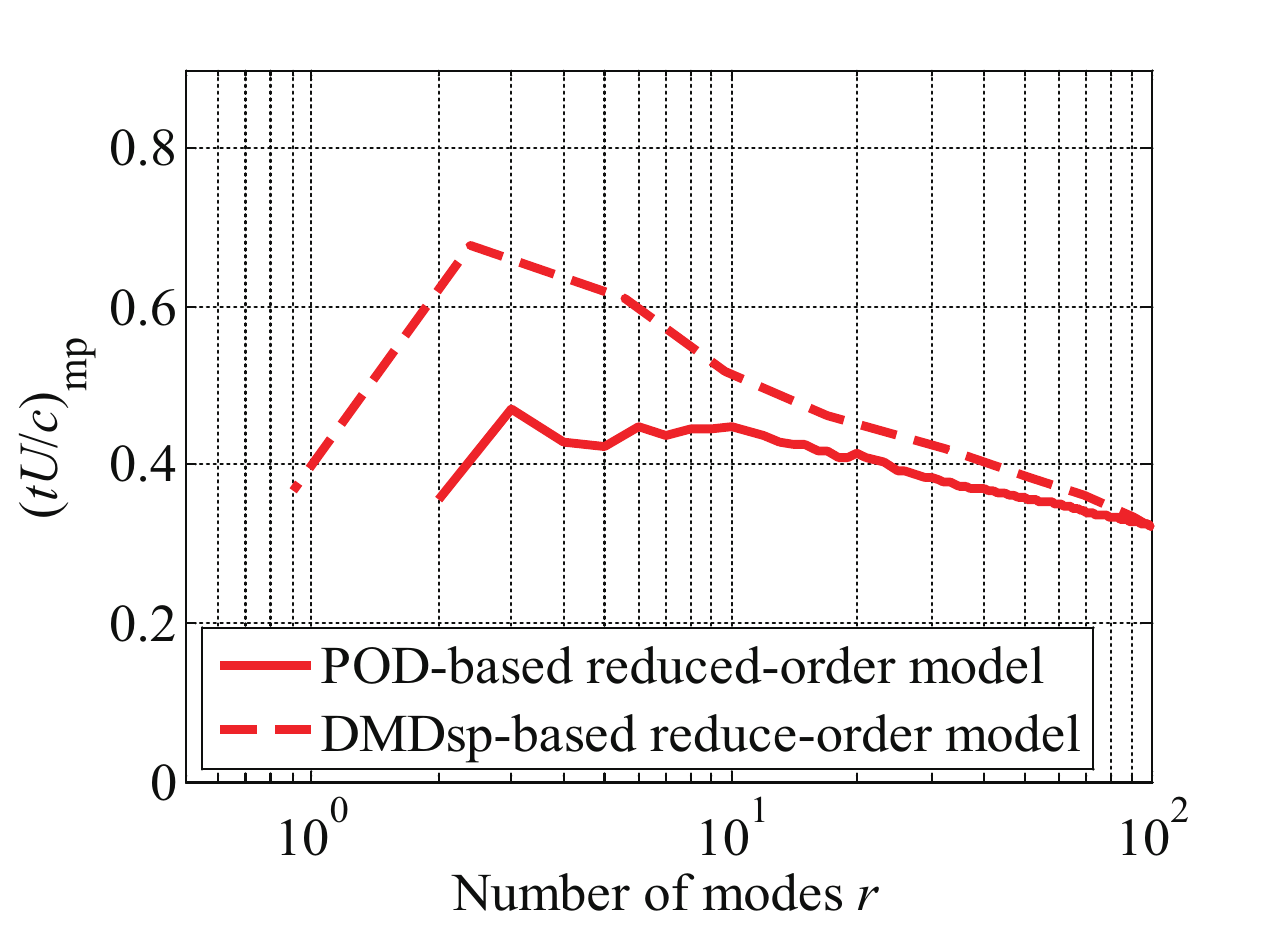}
    \caption{Comparison of the model predictability of the POD-based linear model and the DMDsp-based reduced-order model for prediction of the data.}
    \label{fig:compPODDMDsp}
\end{figure}

Finally, the model predictability using selected DMD modes above is discussed. Figure \ref{fig:model_error_DMDsp} shows the temporal growth of the model error which is difference between the temporal coefficients of POD modes predicted by DMDsp and the temporal coefficients of POD modes projected to the DMDsp subspace. This model error is employed for the evaluation of the model predictability. Figure~\ref{fig:tperm_model_gamma_modes},  \ref{fig:tperm_model_gamma_allmodel} and \ref{fig:tperm_model_r_allmodel} show the model prediction permissive times against  $\gamma$ and a resulting number of selected DMD modes $r_{\textrm{DMDsp}}$. Those results show that the DMDsp-based reduced-order model based on the forward (standard) method has long model prediction permissive time to follow the behavior of flow dynamics with $\gamma=5,000$ and resulting $r_{\textrm{DMDsp}}=2.3$, where the number of DMD modes selected becomes noninteger because the cross-validation for the constant $\gamma$ is considered and the number of those is averaged in the present study. If the coefficient of the regularization term is appropriately chosen for from two to six modes, selected DMD modes work well. On the other hand, the DMDsp-based reduced-order model based on the ODE-based method does not work better than those based on the forward (standard) method. This is because the ODE-based method is more sensitive to $\gamma$ as discussed before, and it selects no DMD modes in some cases in cross-validation depending on the training data and the resulting averaged model prediction permissive time does not become larger, whereas the model prediction permissive time is assumed to be zero when no DMD modes are selected. This happens when $\gamma$ is set to be from 20,000 to 50,000. Even if the sensitivity of DMDsp selection for the ODE-based method is relaxed by tuning the parameter, the model prediction permissive time for the ODE-based method is shorter than that for the forward (standard) method similar to the POD-based linear reduced-order models, and therefore, the further analysis has not been conducted.  The results of the POD-based linear reduced-order model and the DMDsp-based reduced-order model for the forward (standard) method are compared. Figure \ref{fig:compPODDMDsp} shows that the DMDsp-based reduced-order model with the forward (standard) method works approximately 1.5 times longer than the POD-based linear model when $r\approx 2$ while their difference becomes smaller at $r \ge 10$. This illustrates that the DMDsp-based reduced-order model can improve the model when the forward (standard) method is employed with the small number of degree of freedom. The insights obtained here will be utilized for the feedback control of flow fields using a reduced-order model in the future study. 


\subsection{Discussion}
\label{subsec:dis}
The present study illustrates that the linear reduced-order model works for approximately $0.6c/U$ at longest, and the performance is relatively poorer than the prediction of the K\'{a}rm\'{a}n vortex flow at the low-Reynolds number condition by the similar method in the previous study \citep{cammilleri2013pod}. The possible reasons that the present authors consider are as follows: 
\begin{enumerate}[i)]
\item data length is not sufficiently long for modeling of the low frequency dynamics, 
\item the observation noises in initial data degrade the model, and
\item the dynamics is aperiodic (e.g. chaotic or even more complex). 
\end{enumerate}

With regard to the first possible reason, although the present authors considers the data length of $200c/U$ in which the freestream goes through the airfoil chord 200 times is sufficiently long for modeling, the further longer dataset might improve the low frequency dynamics. However, the experimental data are limited by the memory of the high-speed camera, and the database size seems to be reasonable for the experimental setup. In addition, although the combination with the other point sensors might improve the prediction, such a combination is clearly out of scope of the present paper. Therefore, the possible improvement of the model using the much longer dataset or combination with the other sensors is left for the future study. 

\begin{figure}[h]
\captionsetup{justification=raggedright}
    \centering
    \includegraphics[width=70mm]{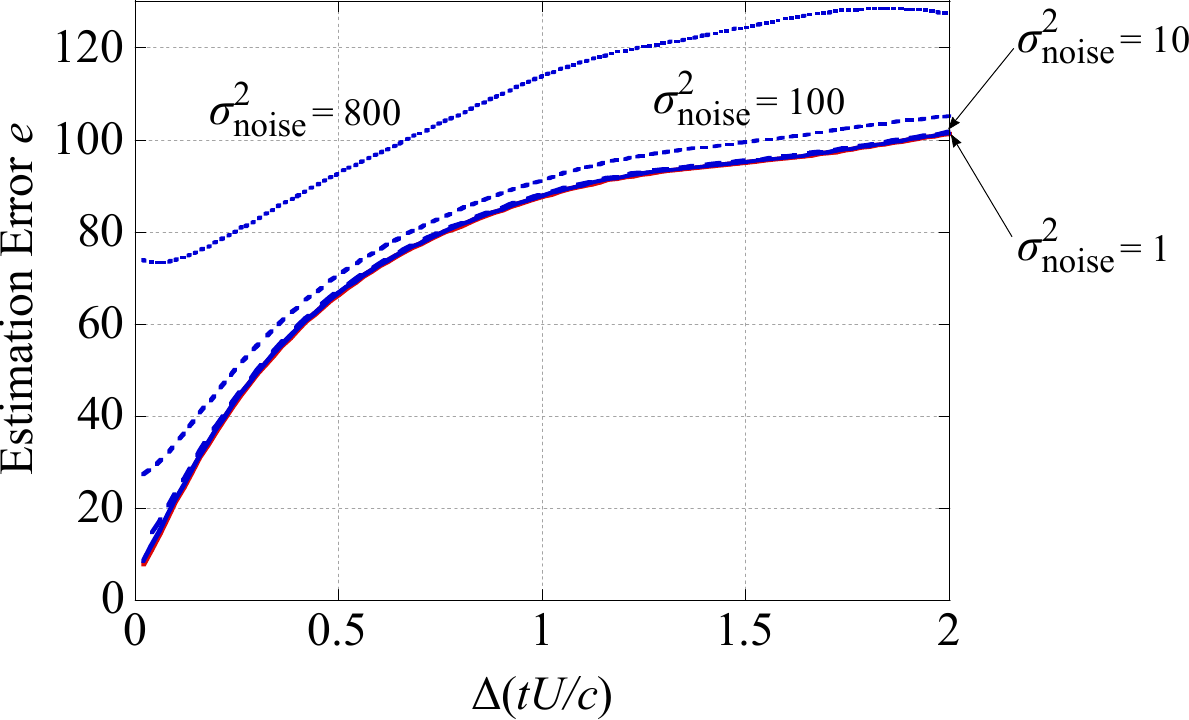}
    \caption{Estimation error of the model using the noisy initial value. Line colors: red, using the original initial value; blue, using the noisy initial value}
    \label{error_10modes_noisy}
\end{figure}

Then, the second possible reason is considered. Although the initial values of temporal coefficients of POD modes are given by the snapshot of velocity fields in this study, the initial value might be affected by observation noise. If this is very strong, the observation noise might affect the predictability of the model. However, this effect is explained to be weak as follows. The noises already contained in the data are difficult to be perfectly removed. Therefore, artificial white Gaussian noises are added and the change in the error behaviors is considered.
White noise with $N\left(0, \sigma_{\text {noise }}^{2}\right)$ is added to the data that are used as initial values for the temporal estimation and the initial value dependence of the estimation is investigated in the case of $\alpha=18$~deg. Adding the white noise on the POD mode amplitudes corresponds to the assumption that the velocity field data acquired by PIV measurements have the additional white noise \citep{nonomura2018dynamic}. The estimation is conducted using the forward (standard) model with $r=10$~modes. 
Figure \ref{error_10modes_noisy} displays the estimation errors, and it shows that the $y$ intercept of the error curve increases with ${\sigma^2}$ in the case of $\sigma^2 \ge 100$. This implicates that the $y$ intercept of the error curve is determined by the strength of the noise that contains in the initial data. On the other hand, the error curve in the case of $\sigma^2 \le 10$ does not change at all. This implies that the noise of the $\sigma^2=10$ level is already contained in the initial data and it leads to the constant $y$ intercept in the case of $\sigma^2 \le 10$. It should be noted that the error curve of the cases with smaller additional noises (including the case without noise) is almost the same as that of the case with $\sigma^2=1$. If the noise already included in the initial data can be perfectly removed, the error curve is slightly right shifted so that the $y$ intercept could be zero, but this leads to the change of less than $0.1c/U$ in the model prediction time based on the simple linear extrapolation of the curve and the $y$ intercept value. Therefore, although the initial value contains the observation noise, the observation noise does not significantly degrade the model predictability.  

Finally, the last possible reason is considered. The flow field is fully turbulent and three-dimensional while the two-dimensional cut plane is only observable. Although the turbulent flow field itself can be nonlinear and behave chaotically, the lack of the information of outside the plane might accelerate this. Therefore, the third reason seems to be the most reasonable in addition to the system error as the disturbance coming from outside the plane. Therefore, the system is assumed to behave chaotically, and the Lyapunov exponent which is inverse of the Lyapunov time horizon was tried to be estimated for the discussion of the longest time to be predicted. However, the orbits in phase space of POD modes does not seem to approach to any attractors, and therefore, the data-driven estimations (\citep{rosenstein1993practical,mehdizadeh2019robust} of the Lyapunov exponent which requires the very close solutions in the different period does not work at all. Those additional analyses show that the system to be predicted is more complex than the chaotic system. This might be because of the presence of the process noises coming from outside the plane as well as the nonlinearity of the system. Therefore, the model prediction should be improved by the data assimilation based on the real-time observation. The present authors are now trying to construct the real-time sparse processing PIV system \citep{kanda2021feasibility} using optimized sensing locations \citep{saito2019determinant,saito2020data} for this purpose.

\section{Conclusions}
The estimation performance of linear reduced-order models based on the PIV data of the flow field around an NACA0015 airfoil were quantitatively investigated in this study. A method for quantitative evaluation of the model predictability which is based on the time advancement of the estimation error was introduced. The effects of modeling parameters such as the number of POD modes for the order reduction and the method for computing the linear operator of the model on the predictability were investigated using the cross-validation technique. The two approaches are investigated for reduced-order modeling: the POD-based and DMDsp-based reduced-order modelings. In addition, the four approaches were tried for the linear regression of the coefficient matrix: the forward (standard) method, the forward-backward method and the total least-squares method that are developed in DMD, and the ODE-based method based on a previous study on POD-based reduced-order modeling of turbulent practical flows.

Here, the model characteristics are firstly summarized for the POD-based reduced-order model because the similar results are obtained for the DMDsp-based reduced-order model, and the additional results using the DMDsp-based reduced-order model follows after that. 
The evaluation results for the POD-based reduced-order model reveal that the predictability and the number of POD modes do not have a simple correlation. The predictability does not significantly increase with the number of POD modes; additionally, the predictability of the three noise-robust  models (forward-backward, total least-squares and ODE-based methods) remarkably worsens with increasing the number of POD modes. In other words, the increase in the number of POD modes does not greatly contribute to the improvement of the predictability. The reduced-order model of two to ten modes work the best in terms of predictability defined in the present study. This implies that two to ten modes are sufficient for the linear model construction of the practical flow fields and the reduced-order model with a larger number of degree of freedom cannot be expected to work well. The amplitude of the temporal coefficients of the POD modes estimated by the forward (standard) model is attenuated due to its damping eigenvalues, while the POD-mode coefficients estimated by the three noise-robust  models are not attenuated. The estimation errors of all the model are shown to increase as time goes and their model prediction permissive times are only $\approx 0.4c/U$ at longest. The forward (standard) model displayed the lowest growth rate of the error and the best predictability.  

Then, the additional results using the DMDsp-based reduced-order models are summarized. Although the basic characteristics of the model does not change from the POD-based linear reduced-order model, the DMDsp-based reduced-order models can predict 1.5 times longer times ($\approx 0.6 c/U$) at maximum for the criteria proposed in the present study than the POD-based linear reduced-order model. This illustrates that the prediction accuracy can be improved by DMDsp-based reduced-order models. The performance improvement by the DMDsp-based reduced-order models is large in the case of two to ten modes and it is not for the cases of more than ten modes.  

This study shows the linear reduced-order model construction of turbulent practical flow fields towards the future flow control. 
These results illustrate that the choices of the DMDsp-based reduced-order model, the forward (standard) method for the construction of the coefficient matrix, and the use of two to ten modes are better for the model predictability, though the predictable time is limited even in the best choice of methods as discussed finally. Those insights will be utilized for the practical reduced-order model which can be used with the modern control theory.

\appendix
\section{Full-data prediction error}
Similar to the model prediction error, the full-data prediction error is also considered for the reference. The number of reference POD mode truncation which is considered to be sufficient to represent the full data matrix is set to be $r_\textrm{ref}$ where as $r_\textrm{ref}>r$. The error vector ${\bm{\epsilon}_{\textrm{full}}}(t_{n})$ and the instantaneous error ${\epsilon_{\textrm{full}}}(t_{n})$ in the full-data prediction are defined as follows:
\begin{eqnarray}
\label{eq:error}
    {\bm{\epsilon}_{\textrm{full}}}(t_{n})={\mathbf{z}_{\textrm{full}}}(t_{n})-{\hat{\mathbf{z}}_{\textrm{full}}}(t_{n}),\\
\label{eq:error_allmodes}
    {\epsilon_{\textrm{full}}}(t_{n})=\| {\bm{\epsilon}_{\textrm{full}}}(t_{n}) \|_2,
\end{eqnarray}
where, ${\mathbf{z}_{\textrm{full}}}$ is defined to be the vector which consists of the first to $r_\textrm{ref}$ POD modes and  ${\hat{\mathbf{z}}_{\textrm{full}}}=[\begin{array}{cc}\hat{\mathbf{z}}^{\mathsf{T}} & \bm{0}^{\mathsf{T}}\end{array}]^{\mathsf{T}}$ is the estimated vector of ${\mathbf{z}_{\textrm{full}}}$. Here, the ($r_\textrm{POD}+1$)th to $r_\textrm{ref}$th components of ${\hat{\mathbf{z}}_{\textrm{full}}}$ is assumed to be zero due to the POD mode truncation.  Also, the asymptotic value of the error was similarly defined by RMS of the original POD-mode coefficients $\bar{z}_{\mathrm{fullRMS}}$ as 
\begin{equation}
\label{eq:zrms}
    {z_{\mathrm{fullRMS}}}=\sqrt{\frac{\sum_{n=1}^{N} \| {\mathbf{z}_\textrm{full}}(t_{n}) \|^2_2}{N}} .
\end{equation}
Here ``full-data prediction permissive time range'' $t_{\mathrm{dp}}$ is defined as the time at which the error reaches $1-e^{-1}$ of ${z}_{\mathrm{fullRMS}}$, i.e., $\bar{{\epsilon}_{\textrm{full}}}(t_{\mathrm{dp}})=0.632{z}_{\mathrm{fullRMS}}$, similar to the definition of the model prediction permissive time range. 
This error is a better index for the prediction of the full-data matrix, but it is not considered to be good compared with the model predictability because our target is building a reduced-order model of a limited number of degrees of freedom which can be used for the practical flow control. Therefore, the results using this index are shown as the reference in the present paper. 

\begin{figure}[h]
\captionsetup{justification=raggedright}
    \centering
    \includegraphics[width=60mm]{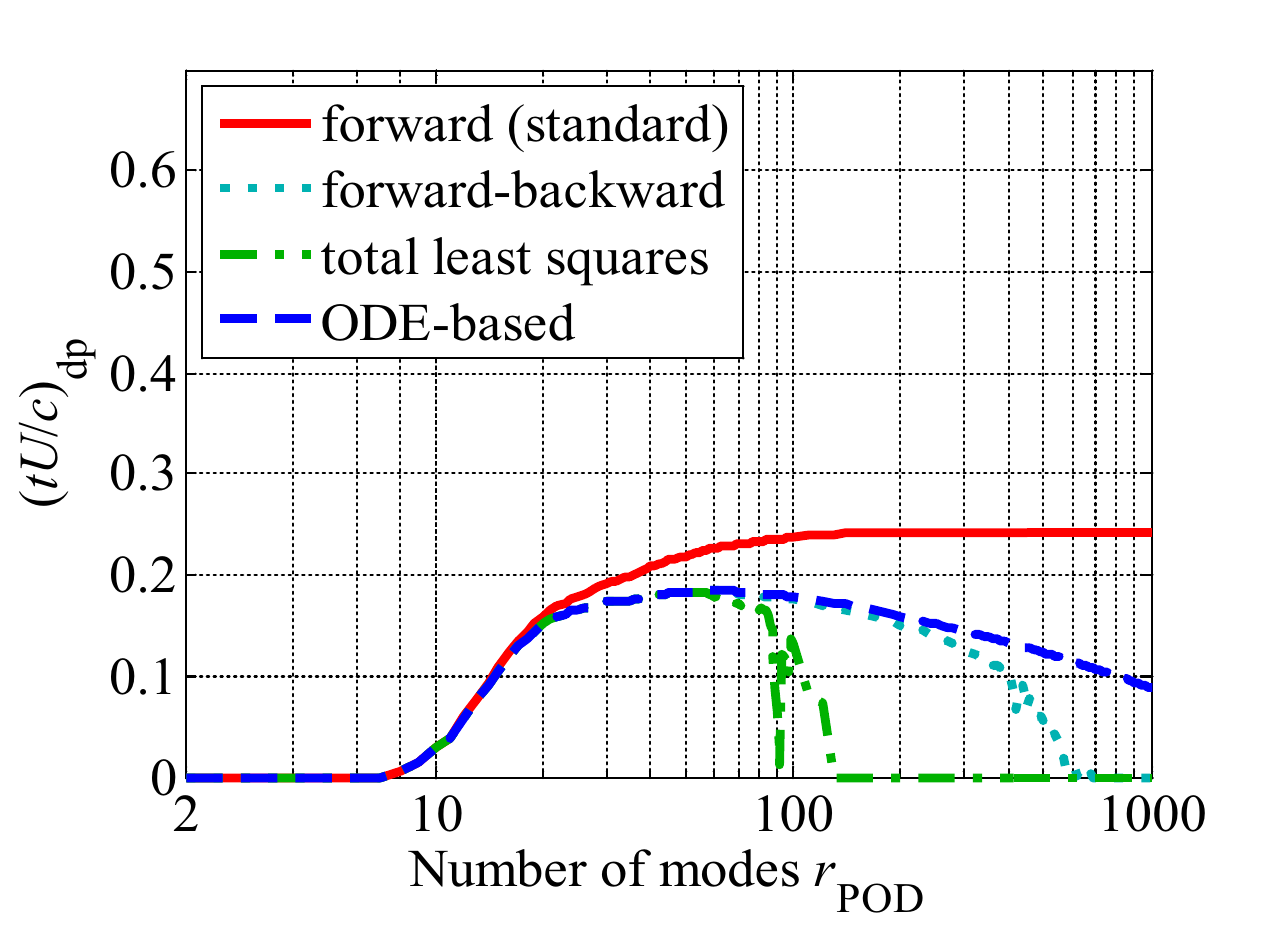}
    \caption{cross-validation results of effect of $r$ on the full-data predictability of the data by POD-based linear model.}
    \label{fig:tperm_rec_modes_allmodel}
\end{figure}

Here, the full-data prediction error is discussed as the reference. When the full-data prediction error is evaluated, the error of the reduced-order model becomes higher because the reduce-order model no longer recovers the information of higher truncated POD modes. The full 1,000 modes are chosen for the reference modes and the error becomes the difference between the raw data and the date reconstruction by the reduced-order model. Figure~\ref{fig:tperm_rec_modes_allmodel} shows the full-data prediction permissive times. The full-data prediction permissive times shown in Fig. \ref{fig:tperm_rec_modes_allmodel} are much shorter in all over the range than the model prediction permissive times shown in Fig. \ref{fig:tperm_model_modes_allmodel}. In addition, the maximum permissive time are observed at $r_{\textrm{POD}}\approx500$ for the forward (standard) model and $r_{\textrm{POD}}\approx70$ for the other models. This is because more of modes are required to represent the original data. However, the full-data reconstruction is not our objective but the extraction of low-dimensional dynamics for the flow control is our objective, and the model permissive times are more important in the present study. It should be noted that the ODE-based method works the best in the noise-robust implementation of linear model constructions also in this prediction tests. 

\bibliography{xaerolab}
\bibliographystyle{spbasic}
\end{document}